\documentclass[usenatbib,usegraphicx]{mn2e}
\usepackage{times}
\usepackage{graphicx, epsfig}
\usepackage[fleqn]{amsmath}
\usepackage{amsfonts}
\usepackage{amssymb}
\usepackage{deluxetable}
\usepackage{times}
\usepackage[totalwidth=17.8cm, totalheight=24.0cm]{geometry}

\title[Stellar discs in early-type galaxies]{The ATLAS$^{\rm 3D}$ project -- XVII. Linking photometric and kinematic signatures of stellar discs in early-type galaxies}

\author[Davor Krajnovi\'c et al.]  {Davor
   Krajnovi\'c,$^{1}$\thanks{E-mail: dkrajnov@eso.org},  Katherine Alatalo$^{2}$, Leo Blitz$^{2}$, Maxime Bois$^{3}$, Fr\'ed\'eric Bournaud$^{4}$,  \newauthor Martin Bureau$^{5}$, Michele Cappellari${^5}$,  Roger L. Davies$^{5}$,  Timothy A. Davis$^{1}$, \newauthor P. T. de Zeeuw$^{1,6}$, Pierre-Alain Duc$^{4}$, Eric Emsellem$^{1,7}$, Sadegh Khochfar$^{8}$,  \newauthor Harald Kuntschner$^{1}$, Richard M. McDermid$^{9}$,  Raffaella Morganti$^{10,11}$, Thorsten Naab$^{12}$,  \newauthor Tom Oosterloo$^{10,11}$, Marc Sarzi$^{13}$, Nicholas Scott$^{14}$,  Paolo Serra$^{10}$,   \newauthor Anne-Marie Weijmans$^{15}$\thanks{Dunlop Fellow}, and Lisa M. Young$^{16}$ 
    \\
   $^1$European Southern Observatory,Karl-Schwarzschild-Strasse 2, 85748 Garching bei M\"unchen, Germany\\
   $^{2}$  Department of Astronomy and Radio Astronomy Laboratory, University of California, Berkeley, CA 94720, USA  \\
   $^{3}$ Observatoire de Paris, LERMA and CNRS, 61 Av. de l'Observatoire, F-75014 Paris, France\\
   $^{4}$  Laboratoire AIM Paris-Saclay, CEA/IRFU/SAp Ð CNRS Ð Universit\'e Paris Diderot, 91191 Gif-sur-Yvette Cedex, France\\
   $^{5}$  Sub-department of Astrophysics, Department of Physics, University of Oxford, Denys Wilkinson Building, Keble Road, Oxford OX1 3RH\\
   $^{6}$  Sterrewacht Leiden, Leiden University, Postbus 9513, 2300 RA Leiden, the Netherlands \\
   $^{7}$  Universit\'e Lyon 1, Observatoire de Lyon, Centre de Recherche Astrophysique de Lyon and Ecole Normale Sup\'erieure de Lyon, 9 avenue Charles Andr\'e, F-69230 Saint-Genis Laval, France\\
   $^{8}$  Max Planck Institut f\"ur extraterrestrische Physik, PO Box 1312, D-85478 Garching, Germany\\
   $^{9}$  Gemini Observatory, Northern Operations Centre, 670 N. A'ohoku Place, Hilo, Hawaii 96720, USA \\
   $^{10}$ Netherlands Institute for Radio Astronomy (ASTRON), Postbus 2, 7990 AA Dwingeloo, The Netherlands\\
   $^{11}$ Kapteyn Astronomical Institute, University of Groningen Postbus 800, 9700 AV Groningen, The Netherlands\\
   $^{12}$ Max-Planck-Institute for Astrophysics, Karl-Schwarzschild-strasse 1, 85741 Garching, Germany \\
   $^{13}$ Centre for Astrophysics Research, University of Hertfordshire, Hatfield, Herts AL1 09AB, UK \\
   $^{14}$  Centre for Astrophysics and Supercomputing, Swinburne University of Technology, Hawthorn, Vic, 3122, Australia\\
   $^{15}$ Dunlap Institute for Astronomy \& Astrophysics, University of Toronto, 50 St. George Street, Toronto, Canada \\
   $^{16}$ Department of Physics, New Mexico Institute of Mining and Technology, Socorro, NM 87801, USA \\
}

\pagerange{\pageref{firstpage}--\pageref{lastpage}}
\pubyear{2011}

\date{Accepted 2012 October 30.  Received 2012 October 30; in original form 2012 June 29}

\newcommand{\kinemetry}{{\it kinemetry}}
\newcommand{\atlas}{ATLAS$^{\rm 3D}$ }

\def\aj{AJ}             
\def\araa{ARA\&A}       
\def\apj{ApJ}           
\def\apjl{ApJ}          
\def\apjs{ApJS}         
\def\aap{A\&A}          
\def\aaps{A\&AS}        
\def\mnras{MNRAS}       
\def\pasa{PASA}         

\begin{document}
\label{firstpage}

\maketitle

\clearpage

\begin{abstract}
We analyse the morphological structures in galaxies of the \atlas sample by fitting a single S\'ersic profile and decomposing all non-barred objects (180 of 260 objects) in two components parameterised by an exponential and a general S\'ersic function. The aim of this analysis is to look for signatures of discs in light distributions of nearby early-type galaxies and compare them to kinematic properties. Using S\'ersic index from single component fits for a distinction between slow and fast rotators, or even late- and early-type galaxies, is not recommended. Assuming that objects with $n>3$ are slow rotators (or ellipticals), there is only a 22 per cent probability to correctly classify objects as slow rotators (or 37 per cent of previously classified as ellipticals). We show that exponential sub-components, as well as light profiles fitted with only a single component of a low S\'ersic index, can be linked with the kinematic evidence for discs in early-type galaxies. The median disk-to-total light ratio for fast and slow rotators is 0.41 and 0.0, respectively. Similarly, the median S\'ersic indices of the bulge (general S\'ersic component) are 1.7 and 4.8 for fast and slow rotators, respectively. Overall, discs or disc-like structures, are present in 83 per cent of early-type galaxies which do not have bars, and they show a full range of disk-to-total light ratios. Discs in early-type galaxies contribute with about 40 per cent to the total mass of the analysed (non-barred) objects. The decomposition into discs and bulges can be used as a rough approximation for the separation between fast and slow rotators, but it is not a substitute, as there is only a 59 per cent probability to correctly recognise slow rotators. We find trends between the angular momentum and the disc-to-total light ratios and the S\'ersic index of the bulge, in the sense that high angular momentum galaxies have large disc-to-total light ratios and small bulge indices, but there is none between the angular momentum and the global S\'ersic index. We investigate the inclination effects on the decomposition results and confirm that strong exponential profiles can be distinguished even at low inclinations, but medium size discs are difficult to quantify using photometry alone at inclinations lower than $\sim50\degr$.  Kinematics (i.e. projected angular momentum) remains the best approach to mitigate the influence of the inclination effects. We also find weak trends with mass and environmental density, where disc dominated galaxies are typically less massive and found at all densities, including the densest region sampled by the \atlas sample.

\end{abstract}


%
%

\section{Introduction}
\label{s:intro}

Excluding those unsettled systems undergoing mergers, bright galaxies  come in two flavours: with and without discs. This was recognised in the early part of the twentieth century \citep{1920MNRAS..80..746R, 1922ApJ....56..162H, 1926ApJ....64..321H, 1929Jeans,1936RNeb..........H} and today is characterised as the Hubble sequence of galaxies \citep[][for a review]{2005ARA&A..43..581S}. Recognising where discs disappear on the sequence, however, is a much more difficult task as projection effects play a key role in our (in)ability to quantify their incidence. This is evident in the fact that the idea of S0 galaxies actually being similar to spirals, while present in the works of \citet{1951ApJ...113..413S} and \citet{1970ApJ...160..831S}, waited some forty years after the appearance of the Hubble tuning fork to be qualitatively presented \citep{1976ApJ...206..883V}. The importance of the parallelism between the two sequences of late- and early-type galaxies for the understanding of galaxy structure was nearly ignored for decades. The parallelism between the two classes of galaxies was recently revived by our project, thanks to the use of integral-field stellar kinematics \citep[][hereafter Paper VII]{2011MNRAS.416.1680C}, which allowed us to recognise discs even at low inclinations. This was followed a few months later by two independent photometric studies reaching the same conclusion  \citep{2011AdAst2011E..18L,2012ApJS..198....2K}.

In practice, there are three ways to look for discs in galaxies: by means of photometric or kinematic analysis, or by constructing dynamical models using both types of information. Dynamical models are often complex and typically rely on certain assumptions. One of these is an assumption on the shape, which could be a limitation if we are interested in quantifying structural components such as discs. 

The photometric analysis is based on recognising structural components of galaxies in their light distributions, while the kinematic analysis is based on recognising features in the higher moments of the line-of-sight velocity distribution (i.e. the mean velocity, velocity dispersion). Stellar discs, which are the main topic of this study, are flattened structures in which stars move on orbits of high angular momentum, hence they should leave both photometric and kinematic traces. Next to their flattened shape, which is clearly recognisable only when viewed directly from a side, or edge-on, discs could be expected to have a specific distribution of light. Indeed, discs of late-type spirals were found to have exponential light profiles \citep{1970ApJ...160..811F}. By contrast, ellipticals and bulges of spirals were first fitted with an $R^{1/4}$ profile \citep{1959HDP....53..311D,1977ApJ...217..406K}, but since the early 1990s the paradigm shifted towards describing these structures with a more general \citet{1968adga.book.....S} $R^{1/n}$ law which provided a continuous parameter applicable across the Hubble sequence \citep[e.g.][]{1993MNRAS.265.1013C,1995MNRAS.275..874A,1996ApJ...465..534G,1996A&AS..118..557D}. 

Early-type galaxies, traditionally divided into ellipticals and S0s, are particularly interesting as among them the separation into objects with and without discs is ambiguous. Photometric analysis of their isophotes revealed that some do contain non-obvious discs \citep{1989A&A...217...35B}, that these might be very common \citep{1990ApJ...362...52R}, and that inclination effects misclassify S0s as ellipticals \citep{1994ApJ...433..553J}. A new way of searching for discs in early-type galaxies was found in the so-called {\it bulge-disc} decompositions \citep[e.g.][]{1985ApJS...59..115K,1997ApJS..109...79S,1998A&AS..131..265S,2001MNRAS.326.1517D}. The essence of these techniques is that they attempt to separate the light contribution from a bulge (having an $R^{1/4}$ or an $R^{1/n}$ light profile) and a disc (having an exponential light profile). As disc dominated galaxies are frequently made of more than just a bulge and a disc, and contain also bars, rings, ovals, nuclear discs and nuclear clusters, as well as of bulges which are not necessary similar to elliptical galaxies \citep[e.g][]{2004ARA&A..42..603K}, recent decomposition techniques allow for a more general description of sub-components \citep[e.g.][]{2003ApJ...582..689M, 2004MNRAS.355.1155D,2009ApJ...692L..34L,2009ApJ...696..411W,2010MNRAS.405.1089L,2012ApJS..198....2K}, as well as applying it on two-dimensional spectra \citep{2012MNRAS.422.2590J}.

The other way of looking for discs is by observing the kinematics of galaxies. As stars in discs rotate at large velocities, and as their motion is typically ordered, observing regular rotation similar to those expected from ideal thin discs, implies those systems are discs, contain discs, or are related to discs by evolution. Elliptical galaxies, or bulges that are similar to them, should not exhibit such ordered and simple rotations \citep[e.g.][]{1991AJ....102..882S,1994MNRAS.271..924A}. Early studies of kinematics of early-type galaxies indeed pointed out there are differences between them \citep{1983ApJ...266...41D,1994MNRAS.269..785B}, but to bring kinematic and photometric analysis to a comparable level it was necessary to wait for integral-field spectrographs (IFS) and two-dimensional maps of stellar kinematics.

The benefits of such observations were clearly pointed out by the SAURON Survey \citep{2002MNRAS.329..513D} and \atlas project \citep[][hereafter Paper I]{2011MNRAS.413..813C}. Using velocity and velocity dispersion maps \citep[e.g.][]{2004MNRAS.352..721E}, it is possible to robustly classify early-type galaxies according to their global angular momentum, even though it is still a projected quantity \citep{2007MNRAS.379..401E,2007MNRAS.379..418C}. This study proposed a separation of early-type galaxies into fast and slow rotators based on a physical property more robust to the effects of the inclination, instead of the traditional elliptical/S0 separation which is based on the apparent shape. This point was taken further with the \atlas data, which comprise observations of a sample of nearby ETGs, volume limited and complete down to a magnitude of -21.5 in the K-band. Using this statistical sample, \citet[][hereafter Paper III]{2011MNRAS.414..888E} showed that $86\pm2$ per cent of ETGs are fast and $14\pm2$ per cent are slow rotators. This separation agrees closely with a quantitative separation of the morphology of the kinematics maps  \citet[][hereafter Paper II]{2011MNRAS.414.2923K}, supporting the robustness of the distinction between the two classes. 

Furthermore, utilising kinemetry \citep{2006MNRAS.366..787K}, it is possible to quantify how well the velocity maps of early-type galaxies agree with those of ideal discs. \citet{2008MNRAS.390...93K} and Paper II found that differences of only 2-4 per cent, between observed stellar velocity maps of early-type galaxies and maps of inclined discs, are typical for fast rotators, while velocity maps of slow rotators simply can not be represented by those of ideal discs. This suggest that fast rotators as a class are indeed discs or at least disc-like objects, and this is the essence of the fast-slow rotators separation used in \citet[][hereafter Paper VII]{2011MNRAS.416.1680C} to set apart objects with and without discs and update the Hubble sequence accordingly. The fact that the presence, or lack of, discs differentiates fast from slow rotators is also confirmed though semi-analytical modelling. In \citet[][hereafter Paper VIII]{2011MNRAS.417..845K}, we show that selecting galaxies by disc fraction, where fast rotators are selected to have more than 10 per cent of mass in discs, semi-analytic model is able to reproduce the observed abundance of fast and slow rotators as a function of mass or luminosity. 

Armed with these results on galaxies' internal kinematics, we now turn our attention to the photometric analysis of \atlas galaxies. We fit single S\'ersic profiles to all \atlas galaxies and attempt to separate the light contributions into a general S\'ersic and an exponential profiles. It is generally assumed that exponential profiles can be associated with discs. This is applicable to spiral and edge-on S0s galaxies, where discs are obvious, but for a general early-type galaxy, seen at a random orientation, where a disc might be masked due to the projection, it is not obvious that the exponential profile is really related to a (hidden) disc. Put in another way, the existence of an exponential profile does not necessary prove that the galaxy contains a disc. This was pointed out by \citet{2004MNRAS.355.1155D} and \citet{2006MNRAS.369..625N}, who suggest that the kinematic information is crucial for determining the disc nature of early-type galaxies. The purpose of this work is to quantify the incidence of exponential light profiles, make a link with the observed kinematics and investigate the difference between fast and slow rotators from the point of view of their light distributions. 

In Section~\ref{s:samp} we briefly outline the \atlas sample, relevant observations and define samples of galaxies used in this work. In Section~\ref{s:1d} we present the method used for the parametrisation of the light distributions and for the disc/bulge decomposition. In Section~\ref{s:1comp} we outline our global fits with a single S\'ersic function. In Section~\ref{s:res} we show and discus the results, while in Section~\ref{s:conc} we summarise the main conclusions of this work. A further discussion on the merits of the chosen method is presented in Appenidx~\ref{A:choice}, a comparison of our results with literature is in Appendix~\ref{C:litcomp} and Table with the results is in Appendix~\ref{B:master}.

%
%

\section{Sample and observations}
\label{s:samp}

The ATLAS$^{\rm 3D}$ sample and its selection are described in detail in Paper I. Briefly, ETGs were visually selected from a parent sample of objects in the Northern hemisphere ($|\delta - 29\degr| < 35\degr$, where $\delta$ is the sky declination), brighter than $M_K < -21.5$ mag and within a local volume of radius of $D = 42$ Mpc. The final sample contains 260 nearby early-type galaxies, which were observed with the SAURON IFS \citep{2001MNRAS.326...23B} mounted on the William Herschel Telescope (WHT). The SAURON kinematics was introduced in Paper I, and we refer to that paper for details on the extraction, while the stellar velocities maps used here were presented in Paper II. 

Photometric data of 258 galaxies were assembled from the Sloan Digital Sky Survey (SDSS) DR7 \citep{2009ApJS..182..543A} and from our own imaging with the Wide-Field Camera (WFC) mounted on the 2.5m Isaac Newton Telescope (INT). These data, their reduction and photometric calibrations are presented in \citet{2012Scott}. In this study we use the {\it r}-band imaging. We exclude two galaxies without SDSS or INT imaging from further analysis.  We used the same zero points and the photometric calibration as \citet{2012Scott}.

In Paper II we showed that at least 30\% of galaxies in \atlas sample contain bars and/or rings. These systems obviously have more than two components, comprising at least: a bulge, a bar, a ring (alone or in addition to the bar), and a disc. A two component fit will not describe these systems well. Crucially, bars (and rings) are disc phenomena; they happen only if there is a disc in the first place. Therefore, we removed from the sample all galaxies showing clear bars (and/or large scale rings), according to classification in Paper II. This reduced the number of galaxies for the decomposition analysis to 180. Included are 34 of 36 slow rotators (two slow rotators are actually barred galaxies), and 146 of 224 fast rotators, as classified in Paper III. It is, however, still possible that among the remaining galaxies there are barred systems or galaxies with more than two components. The global one component fits, however, we do on all \atlas galaxies (258 galaxies with the SDSS or INT imaging). We caution the reader that in all statistical consideration throughout the paper we use the limited sample of 180 galaxies (no barred galaxies), unless stated otherwise. Specifically, in Section~\ref{ss:1compres}, which deals with the one components S\'ersic fits, we use the 258 galaxies of the \atlas sample.

%
%

\section{Decomposition of one dimensional profiles}
\label{s:1d}

\subsection{One or two dimensional decomposition?}
\label{ss:1or2}

Parametric decomposition of light into various structural components is often done in two dimensions \citep[e.g][]{2003ApJ...582..689M, 2004MNRAS.355.1155D, 2006MNRAS.371....2A, 2007MNRAS.379..841B,2009MNRAS.393.1531G,2009A&A...508.1141S,2009ApJ...696..411W,2010MNRAS.405.1089L,2011ApJS..196...11S}, as more information is available to constrain the parameters of the components. The extra information held in the original images (e.g. on ellipticy and position angle) may be diluted when deriving a one-dimensional profile, and the analysis of one-dimensional profiles may not use changes in the other properties to constrain the model parameters. This is important because, for example, while position angle can remain unchanged between the components, the ellipticity will generally differ; if a systems is composed of a spheroidal bulge and a thin disc, there will be a marked change in the ellipticity as one of the components starts dominating over the other \citep[e.g.][p 217]{1998gaas.book.....B}. 

Based on simulations, \citet{1995ApJ...448..563B}, \cite{1996A&AS..118..557D} and \citet{2002ApJS..142....1S} argued that two dimensional decompositions are superior to those done in one dimension, and several algorithms, of which some are publicly available, have been developed with that purpose, such as GIM2D \citep{2002ApJS..142....1S}, GALFIT \citep{2002AJ....124..266P}, BUDDA \citep{2004ApJS..153..411D,2008MNRAS.384..420G}, GASPHOT \citep[][using a hybrid 1D/2D approach]{2006A&A...446..373P}, GASP2D \citep{2008A&A...478..353M} and GALPHAT \citep{2011MNRAS.414.1625Y}. A number of authors, however, continue to work in one dimension \citep[e.g.][]{2001AJ....121..820G, 2002MNRAS.333..633A,2003ApJ...582L..79B,2003ApJ...594..186B, 2006MNRAS.369..625N, 2008AJ....136..773F, 2012ApJS..198....2K,2012arXiv1204.5188F}, while \citet{1996ApJ...457L..73C} and \citet{2003ApJ...582..689M} argued that one dimensional decompositions should not be disfavoured as they give similar results as two dimensional fits, provided the data have high signal-to-noise ratios. 

Our purpose here is to attempt to decompose and look for discs in a robust and homogenous way in both fast and slow rotators. To do this, we limit ourselves to considering only simple one- or two-component models. We therefore consider that the additional information gained in fitting two-dimensional images is offering a negligible improvement while introducing significant additional complexity and computational effort. The high signal-to-noise images and the large size of the \atlas galaxies ensures that extraction of the profiles can be done robustly. In the next section we present our method in detail, and in Appendix~\ref{A:choice} we present additional considerations regarding the choice of our methods.

\subsection{Method}
\label{ss:method}

One dimensional light profiles were extracted by azimuthally averaging the light along the best fitting ellipses obtained by means of an isophotal analysis (for an overview of other possibilities see Appendix~\ref{A:choice}). The best fitting ellipses were found using the method of \kinemetry\footnote{An IDL implementation of kinemetry is available at this address: http://www.eso.org/$\sim$dkrajnov/idl} \citep{2006MNRAS.366..787K}, run in the {\it even} mode optimised for images. It this case, kinemetry reduces to the analysis of even moments of the line-of-sight velocity distribution (e.g. light distributions) and the methodology is similar to \citet{1987MNRAS.226..747J} and the {\it iraf} task ELLIPSE. For a given ring of radius r (semi-major axis length) and thickness $\Delta r$ (which is a geometric function of $r$ such that rings at larger radii are wider), the intensity $I(r)$ is sampled at equal intervals in the eccentric anomaly $\theta$ along a trial ellipse defined by the position angle {\it PA}, flattening $Q =b/a$, where $a$ and $b$ are the lengths of the semi-major and semi-minor axis, respectively, and the centre ($X_0$,$Y_0$). The intensity $I(r, \theta)$ is expanded into a Fourier series and the amplitudes of the Fourier coefficients are minimised until a fit as close as possible to $I(r, \theta)=const.$ is achieved. 

In practice, the centre of a galaxy was pre-determined as the centroid of the light distributions, obtained in the same way as the global photometric position angle and ellipticity in Paper II, and kept fixed during the analysis. Bright stars and companion galaxies were masked prior to the fit. Dust is not often seen in our galaxies, and we masked or excluded from fitting the most contaminated regions. Sky levels were estimated and subtracted from the images using a routine {\tt sky.pro} available from the IDL Astronomy Library \citep{1993adass...2..246L}. 

In addition to extracting along the best fitting ellipses where {\it PA} and {\it Q} were allowed to vary freely, we also extracted a second set of profiles for which {\it PA} and {\it Q} were fixed to the global values from Paper II. These two sets of light profiles are used for different purposes: the set from the fixed ellipses for a global single component fit (see Section~\ref{s:1comp}) and the set from free ellipses for the decompositions as outlined below. 

We use two different forms of the \citet{1968adga.book.....S} fitting function to describe the components in the light profiles. The first one is a general $r^{1/n}$ model, often used to describe the surface brightness profiles (and images) of bulges or whole galaxies \citep[e.g][]{1993MNRAS.265.1013C,2001AJ....121..820G,2004MNRAS.355.1155D,2009ApJ...696..411W,2011MNRAS.411.2439H}:
\begin{equation}
\label{e:free}
I(r) = I_e \exp \left\{ -b_n \left[ \left(\frac{r}{R_{e}}\right)^{1/n} - 1 \right] \right\}
\end{equation}
\noindent where $I_e$ is the intensity at the effective radius $R_{e}$ that encloses half of the light of the component, $n$ is the parameter which describes the shape of the function, while $b_n$ is dependent on $n$, and not an additional free parameter. It can be obtained by solving the equation $\Gamma(2n)=2\gamma(2n,b_n)$, where $\Gamma$ is the gamma function and $\gamma(2n,b_n)$ is the incomplete gamma function \citep{1991A&A...249...99C}. We use an accurate numerical approximation of $b_n=2n - 1/3 + 4/(405n) + 46/(25515n^2)$ given in \citet{1999A&A...352..447C}. A number of useful mathematical expressions related to the S\'ersic model are given in \citet{2005PASA...22..118G}. 

The other function is a special case of the S\'ersic model when $n=1$. In this case the model simplifies to an exponential function: 
\begin{equation}
\label{e:disc}
I{_d}(r) = I_0 \exp \left(-\frac{r}{R_d}\right)
\end{equation}
\noindent where $I_0=I_e e^{b_n}$ is the central surface brightness, $R_d=R_{e}/b_n$ is the scale length and $b_n=1.678$ for $n=1$. This exponential form is usually used to define a disc component, as it reproduces well the outer light profiles of disc galaxies \citep{1970ApJ...160..811F}. 

In this work we use two sets of parameters linked with eq.~(\ref{e:free}), one for a single component fit to the light profile, where the S\'ersic function describes the total light, and a two components fit to the light profile, where the general S\'ersic function describes the bulge light (more precisely, light not belonging to the exponential component). In the former case, the parameters of the eq.~(\ref{e:free}) are: $I_{e, tot}$, $R_{e,tot}$ and $n_{tot}$, and in the latter case: $I_{e,b}$, $R_{e,b}$ and $n_{b}$ . As will be seen later, after the decomposition of some galaxies it is evident that a sufficiently good fit is obtained using the general S\'ersic component only (i.e the decomposition and the exponential component are not necessary). In these cases, we will still refer to the parameters of the fit as the bulge parameters (e.g $n_b$), even though they describe the full galaxy, to differentiate if from the direct single component fit. In spite of both being results of single component fits, they are not necessary equal, as will become apparent in Section~\ref{s:1comp}. 

\begin{figure*}
\begin{minipage}[b]{\textwidth}
\centering
        \includegraphics[width=0.495\columnwidth]{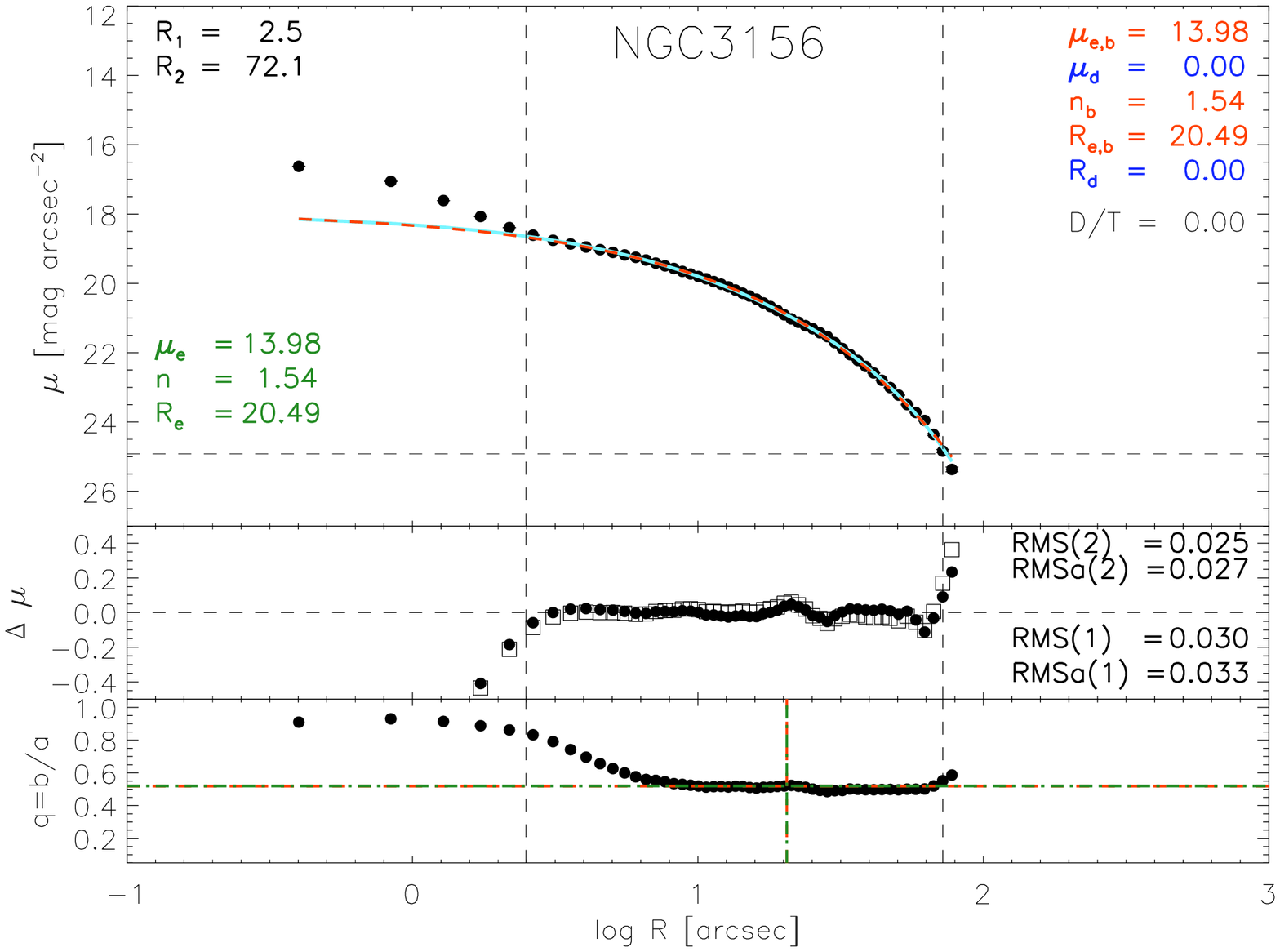}
        \includegraphics[width=0.495\columnwidth]{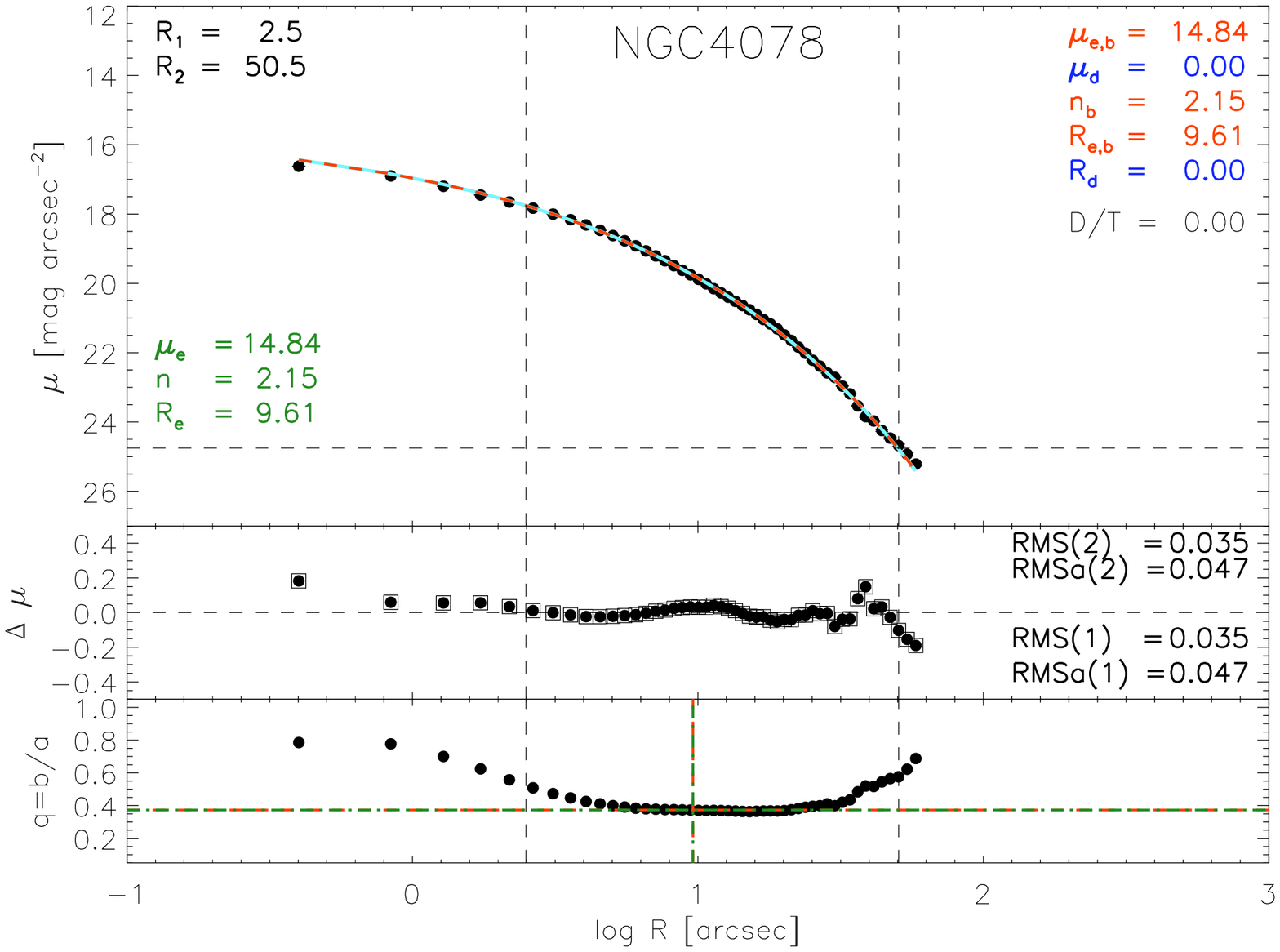}
\end{minipage}

\begin{minipage}[b]{\textwidth}
\centering
        \includegraphics[width=0.495\columnwidth]{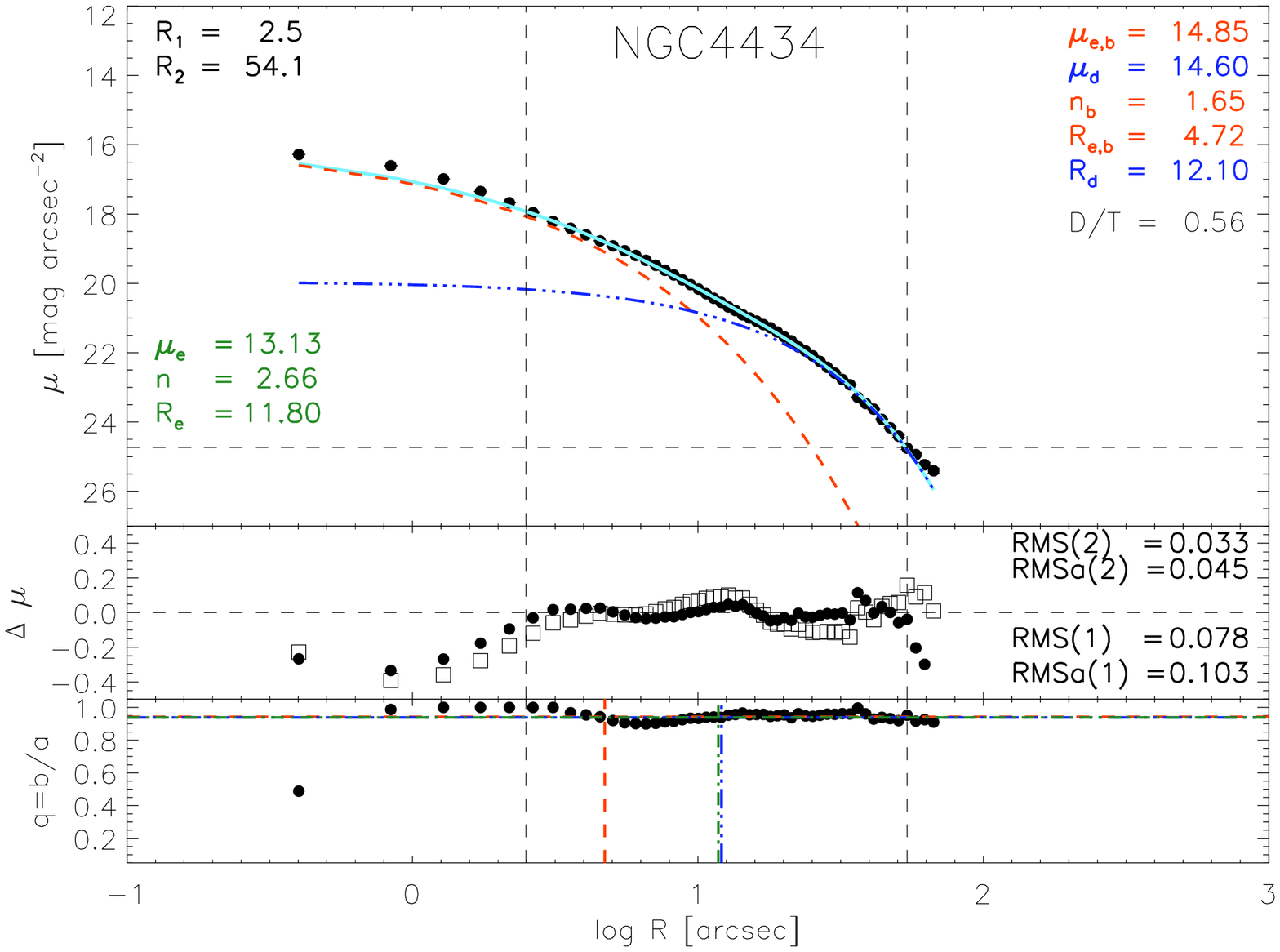}
        \includegraphics[width=0.495\columnwidth]{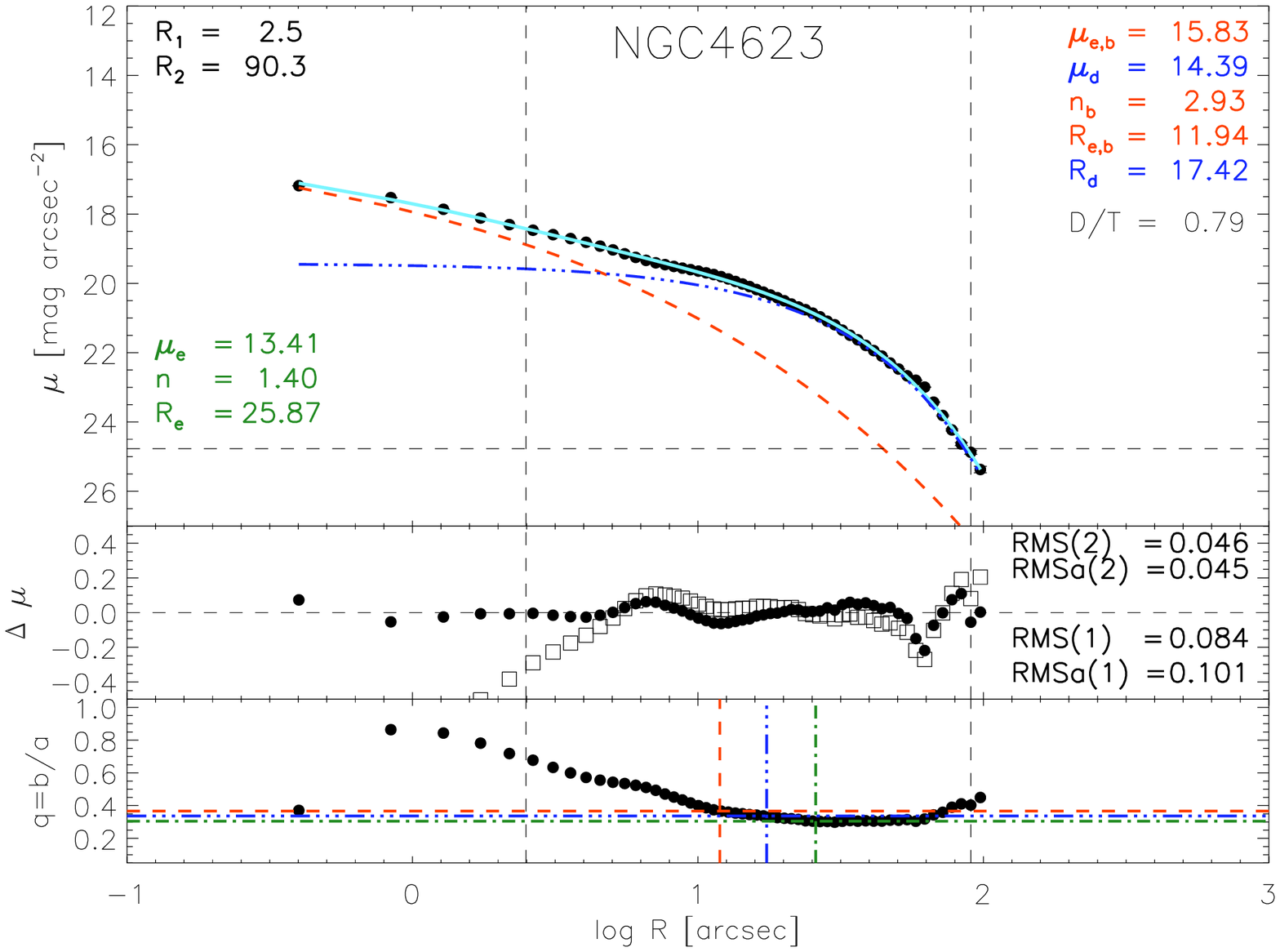}
\end{minipage}

\begin{minipage}[b]{\textwidth}
\centering
        \includegraphics[width=0.495\columnwidth]{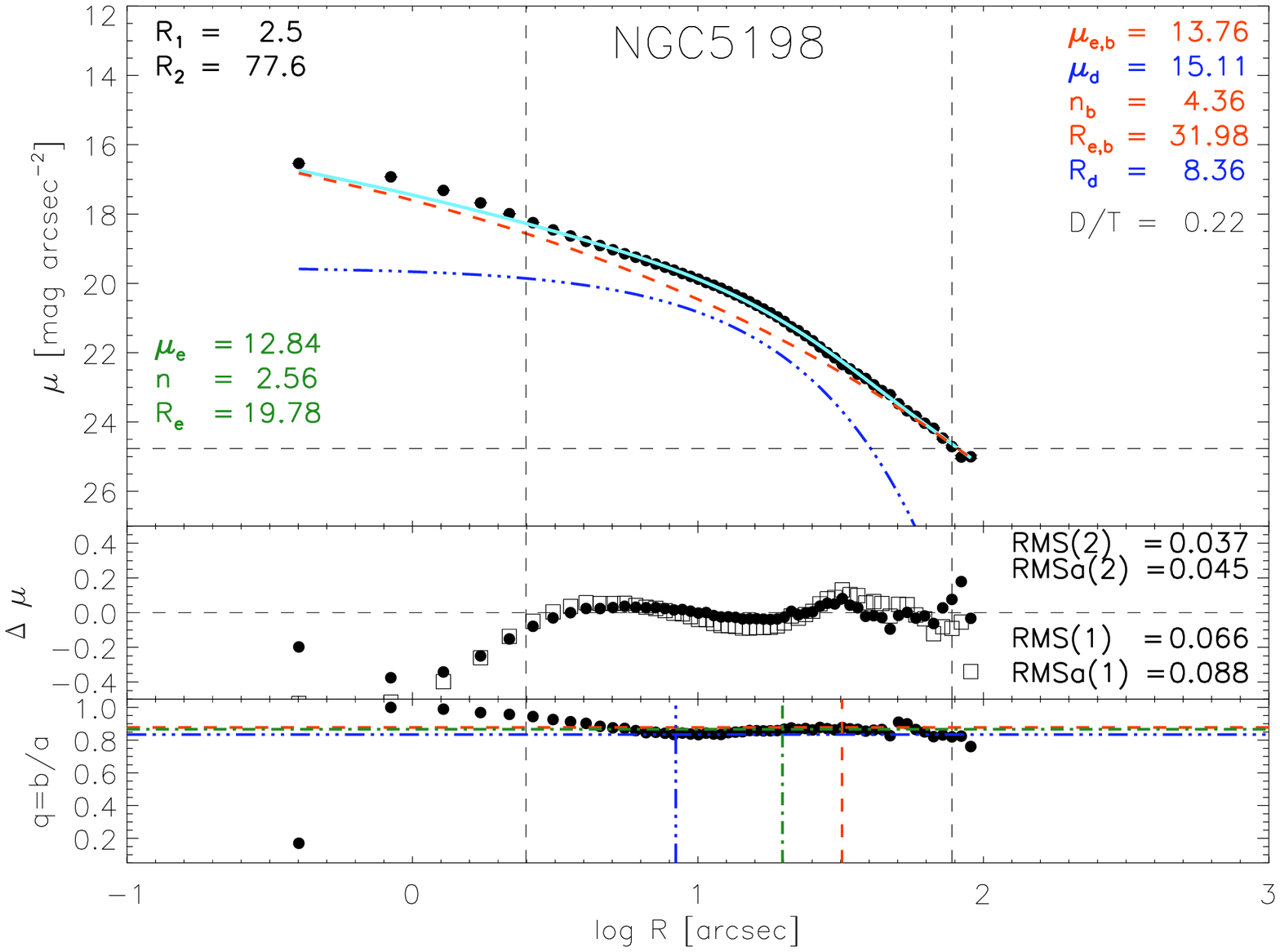}
        \includegraphics[width=0.495\columnwidth]{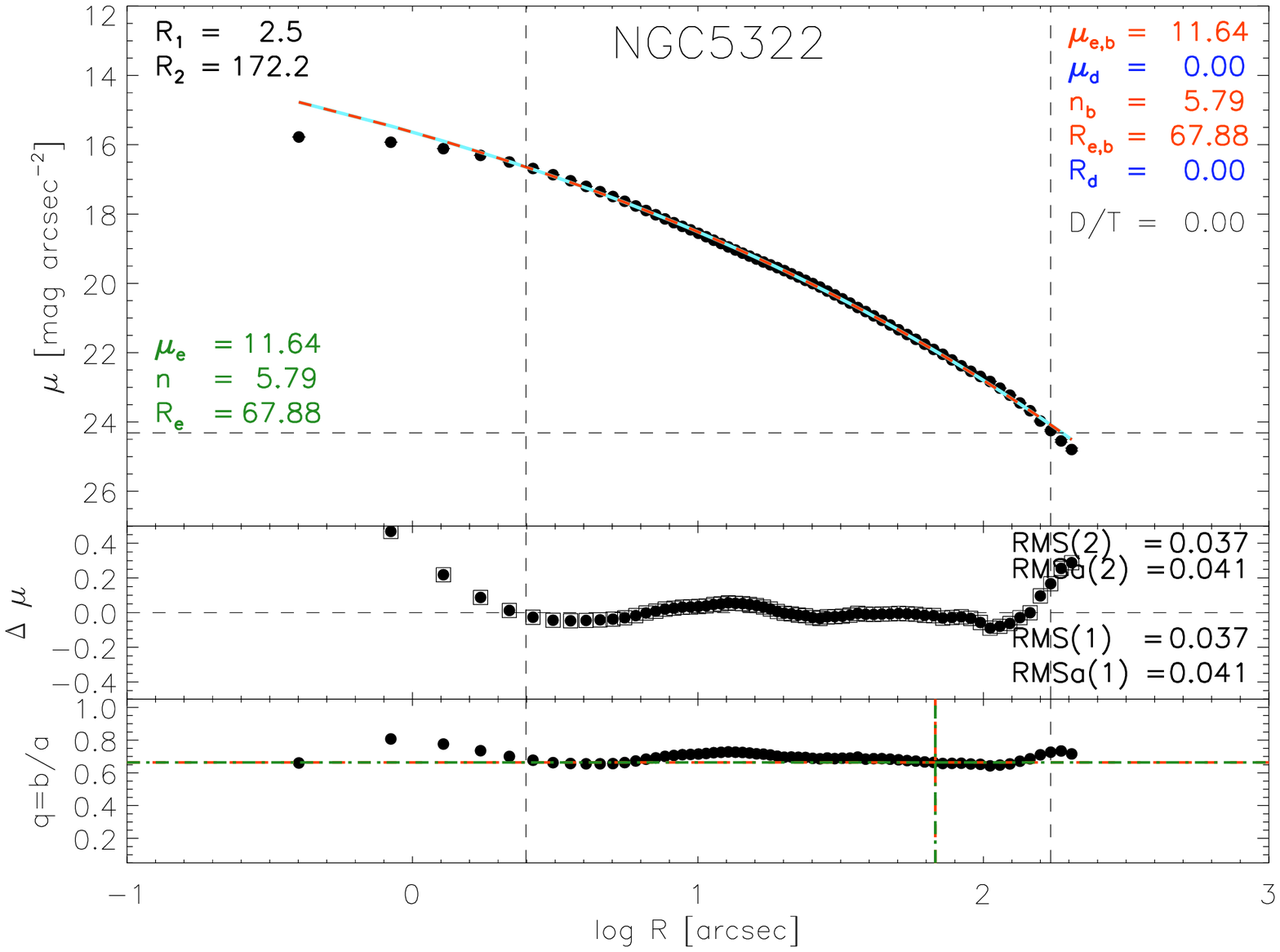}
\end{minipage}

  \caption{\label{f:example} Decomposition examples. Each galaxy is represented by three panels, where top panel shows the extracted light profile, the middle panel show the residuals (data - best fit model) in mag/$\arcsec^2$, and the bottom panel shows the flattening (q=$1-\epsilon$) profile extracted at the same time as the light profile. On the top panel the data are shown with solid symbols. Results of the two component fit (the effective radius $R_{e,b}$ and the bulge S\'ersic index $n_b$, disc scale height $R_d$, the total light for both components, $\mu_{e,b}$ and $\mu_{d}$, and the disc-to-total light ratio) are given in the upper right corner. The results of the one component fit (total light $\mu$, S\'ersic index $n$ and effective radius $R$) are shown in the lower left corner.  Vertical dashed lines indicate the region used in the fit. The actual values in seconds of arc are given in the upper left corner. These lines are also shown in the middle and bottom panels. The horizontal dashed line is our estimate of the sigma of the sky level. Light profiles of the different components are shown with lines: red dashed for the bulge model, blue tripple-dot-dashed for the exponential model and solid cyan for the combined fit. We do not show the one component fit.  On the middle panel solid symbols show residuals for the two component fit and open squares for the one component fit. The root-mean-square values for the fitted ({\it RMS}) and the full ({\it RMSa}) data range are shown in the upper and lower right corners for two and one component fits, respectively. On the bottom panel vertical red (dashed) and blue (triple-dot-dashed) lines correspond to the sizes of the bulge (R$_{e,b}$) and the exponential (R$_d$) components, respectively, and green (dot-dashed) line to the one fit component effective radius ($R_e$). The horizontal red and blue lines give the values of q used in eqs.~(\ref{e:b}) and~(\ref{e:d}), respectively.}
\end{figure*}

We decompose the light profiles $I(r)$ of \atlas galaxies by assuming that $I(r)=I_{e,b}(r) + I_d(r)$, with $I_{e,b}$, $R_{e,b}$, $n_b$, $I_0$ and $R_d$ as free parameters. The fit is performed using {\tt mpfit} \citep{2009ASPC..411..251M}, an IDL implementation of the MINPACK algorithm \citep{1980ANL.....80.74M} of the Levenberg-Marquardt method. As more parameters will always provide a better fit to the data, to decide on whether a one component model is sufficient to describe the galaxy, we used the following method. The same light profiles were fitted also using only the general $r^{1/n}$ S\'ersic model eq.~(\ref{e:free}), within the same radial range. The root-mean-square (rms) of the residuals (within the fitting range) of these single component fits ({\it rms$_1$}) were then compared with the rms of the residuals of the two component fits ({\it rms$_2$}). If {\it rms$_1$} $> 1.5\times${\it rms$_2$} then the two components fit was deemed better than the one component fit, and its parameters were adopted. It is important to note that we visually inspected all residuals (both one and two components) as it is not only the rms what should be considered, but also the systematic changes in the correlated residuals visible as wiggles. In this respect, adopting a higher threshold value (e.g. {\it rms$_1$} $> 2\times${\it rms$_2$}) does not change the results significantly, as long as one considers that the disappearance of the correlated wiggles is the prime evidence for the existence of multiple components (see Section~\ref{ss:exm} and Fig.~\ref{f:example} for more details and examples).

The total luminosity of the individual sub-components can be estimated by integrating:
\begin{equation}
\label{e:b}
B(r) = \int_0^\infty I_{e,b}(r) 2 \pi q_b r dr = \frac{2\pi I_{e,b} R_{e,b}^2 e^{b_n} n_b  q_b}{b_n^{2n_b}} \Gamma(2n_b)
\end{equation}
\noindent and for the case of an exponential disc:
\begin{equation}
\label{e:d}
D(r) = \int_0^\infty I_d(r) 2 \pi q_d r dr = 2\pi I_0 R_d^2 q_d
\end{equation}
\noindent where we assumed that the flattening of the sub-component $q_b$ and $q_d$ does not change with radius. The flattening of a sub-component was determined as the flattening at the representative radius of the sub-component. For the sub-component described with an $R^{1/n}$ model this means $q_b=q(R_{e,b})$ and for the exponential  $q_d=q(R_d)$. Finally, we want to know what is the relative fraction of light contained in the exponential sub-component and we calculate "disc-to-total" (D/T) ratio\footnote{At this moment we call the exponential components a disc component without proof that this is applicable for all early-type galaxies. This is done by convention, but in Section~\ref{ss:lesson} we address this issue in detail justifying our choice.}, with this expression: D/T= D/(B+D), where D and B are the expressions from eqs.~(\ref{e:b}) and (\ref{e:d}).

We also estimated the total luminosity within the radius, $R_{\rm max}$ which corresponds to the largest coverage of our IFU observations (matching the coverage of our kinematics). This was done by integrating the integrals in eqs.~(\ref{e:b}) and (\ref{e:d}) from $r=0$ to $r=R_{\rm max}$ to estimate the bulge and disc light within this regions, respectively. In practice, for the bulge component we use eq.~(2) from \citet{2005PASA...22..118G} and apply the tabulated form of the integral in eq~(\ref{e:d}) \citep[e.g.][page 357]{2000tisp.book.....G} for the exponential component. Depending on the coverage of the individual objects there are some modifications to D/T ratios, but non of the conclusions of this work change if we consider this limited luminosity instead of the  (standard) total luminosity. The main reason why this is the case comes from the fact that our IFU coverage is on average twice as large as $R_b$ and  $R_d$ estimated in this study. In the rest of the paper we only consider the total luminosities defined by eqs.~(\ref{e:b}) and (\ref{e:d}). 

A number of studies discuss the robustness of the decomposition parameters \citep{1987AJ.....93...60S, 1996A&AS..118..557D,2003ApJ...582..689M,2009ApJS..182..216K}. We found that the crucial step of our fitting procedure is an adoption of the radial range within which the fit is done, and partially the initial conditions for the fit. We use one continuous range excluding the central parts influenced by the effects of seeing and running until the sky level.  \citet{2012Scott} estimate that the average point spread function (PSF) of our data has full-width-half-maximum of 1.25\arcsec and we as a rule exclude a region twice as big (the fitted region starts at $\sim 2.5\arcsec$, or $\sim300$ pc assuming the average distance to \atlas galaxies). If necessary, and in a limited number of cases, both inner and outer radii for the fits were adapted for each galaxy individually (see Section~\ref{ss:exm}). 

\subsection{Decomposition examples}
\label{ss:exm}

In Fig.~\ref{f:example} we show six example fits to light profiles extracted along the best fitting ellipses. These include three profiles which can be reproduced with a single component of a low S\'ersic index, and three light profiles which are reproduced with two components of various relative fractions. We also show residuals of both one and two component fits for comparison. These examples are representative of the fits to other galaxies in the sense of their quality, types of residuals, fitting ranges and types of models that reproduce the observed light profiles. 

The residuals within the fitted range are generally small indicating good model fits; a median of the rms deviation is 0.05 mag/$\arcsec^2$ and its standard deviation is 0.03 mag/$\arcsec^2$. On the top left panel (NGC\,3156), we show an example of a galaxy for which residuals of the two component fit are not significantly smaller than the one component fit residuals. Hence, the one component fit was deemed sufficient, and the decomposition results were discarded. Contrary examples, when a two component fit was considered necessary, are shown for NGC\,4434, NGC\,4623 and NGC\,5198. 

After carrying out similar comparisons for all galaxies and choosing if the decomposition is necessary, we examined all galaxies with rms $> 0.1$ mag (29 objects) to understand the reasons for the deviations. In only one case (NGC\,4753), residuals could be connected with dust features, with a characteristically jagged distribution of values. In all other cases, the distribution of residuals was monotonically varying. These kind of features suggest there are possible additional components in the light profile, which can not be described by the assumed decomposition in two components only. 

Among the galaxies with high residuals, we found both those fitted with one (16 objects), and with two components (13 objects). The majority (9/13) of galaxies fitted with two components have $\epsilon > 0.6$, and are often seen in disc dominated systems close to edge on. NGC\,4623 from Fig.~\ref{f:example} is an example.  We tested these cases by decomposing their light profiles obtained as major axis cuts, but there were no significant improvements to the two components fits, nor large difference in the parameters of the best fitting components. The cause for the poor fits can be fully attributed to the existence of additional components, which could be interpreted as manifestations of instabilities (e.g. bars, rings) induced by secular evolution and hard to recognise due to the inclination angle.

On the other hand, systematic variations of residuals in galaxies with only one component might suggest that these galaxies are actually better fit with two components and that our threshold criterion should not apply here. However, for 9 (of 16) objects the fitting algorithm actually automatically excluded the two components solutions and this result was robust to changes in both the initial conditions and fitting ranges. Additionally, only 1 (of 16) objects has $n>3$, while for the majority (12/16) objects S\'ersic index ranges from 0.8 to 1.2. These single components, near exponential galaxies have additional structures, often seen in the shape of correlated wiggles in the residuals, but a two component fit is not sufficient to describe them. 

Inwards of the inner fitting range point ($2.5\arcsec$), one can often detect departures from the fitted and the observed light profiles. This trend is particularly visible in NGC\,3156 and NGC\,5322 of Fig.~\ref{f:example}. The models either over- or under-predict the light in the centres of the galaxies. In some cases, these can be directly associated with the excess/deficit observed within ETGs with the HST \citep{1994AJ....108.1598F,1997AJ....114.1771F, 2003AJ....125.2951G, 2006ApJS..164..334F,2009ApJS..182..216K}, or small nuclear components, but we do not attempt to quantify the effects as one generally needs higher spatial resolution for this analysis (e.g. the Hubble Space Telescope data) to allow fits that extend to smaller radii. 

Finally, we note that our decomposition was performed on relatively shallow SDSS images focusing on morphological structures within a few effective radii. Deeper images are likely to show more varied structures at larger radii introducing a need for more than just two components to describe the light distributions of galaxies \citep[e.g.][]{2011MNRAS.417..863D}. 

\begin{figure*}
        \includegraphics[width=\textwidth]{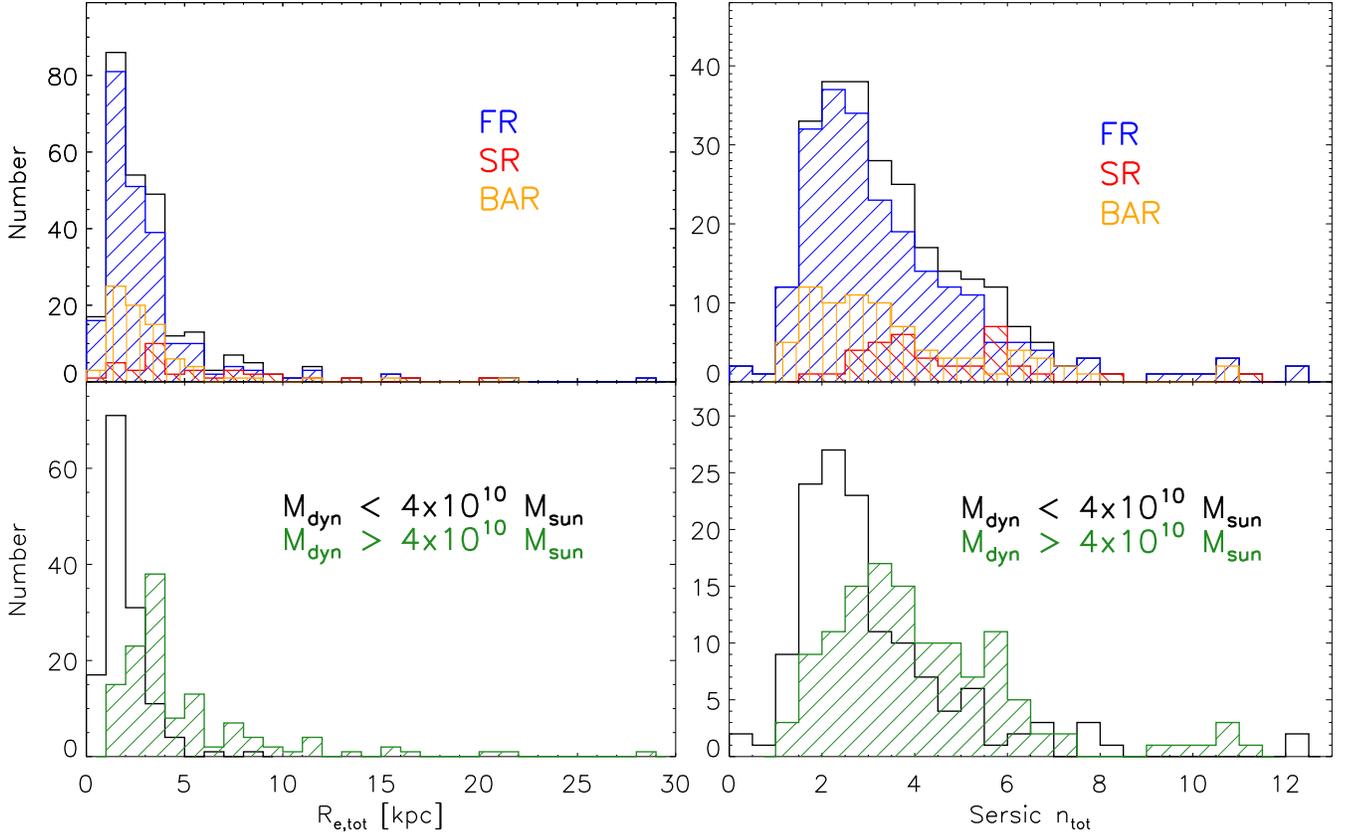}
\caption{\label{f:hist1c} Distribution of the effective radius $R_{e,tot}$ ({\it left column}) and the global S\'ersic index $n_{tot}$ ({\it right column}) of single S\'ersic fits to light profiles obtained averaging along fixed ellipses, for 258 \atlas galaxies. In the top row galaxies are divided in fast (blue histogram hatched to the left), slow (red histogram hatched to the right) rotators, and barred objects (orange histogram with vertical lines), while the open histogram is for all galaxies. In the bottom row, galaxies are divided by mass into less (open histogram) and more massive (green hatched histogram) than $4\times10^{10}$ M$_{\odot}$, which splits the sample in two roughly equal halves. }.

\end{figure*}

\subsection{Uncertainties}
\label{ss:error}

As mentioned above, we obtain the best fit parameters by doing a linear least-squares fit with the {\tt mpfit} routine. In doing so we assume constant relative errors, which ensures equal weighting to all points on our light profiles. To estimate the uncertainties to S\'ersic parameters we perform Monte Carlo simulations based on the {\it rms} scatter of the residuals to the fit. We perturb original light profiles, fit them again 100 times and estimate the uncertainties as the standard deviation of the simulations. These are only statistical estimates of the uncertainties, and they do not properly represent the systematic ones coming from the choice of the method, initial condition, sky levels and, in particular, the choice of the fitting range. In Appendix~\ref{A:choice} we discuss the systematic effects when using different methods outlined above. We caution the reader that these sources of the systematic uncertainties are what could drive the difference between our and literature results. 

In Appendix~\ref{C:litcomp} we present a comparison of our results (focusing on the S\'ersic index and the D/T ratio) with the results of other studies. We compare our results both directly and in a statistical sense: firstly, with studies that analyse samples which overlap with our own (i.e comparison of individual galaxies), and, secondly, with studies that analyse large numbers of galaxies. The reason for this approach is in the presence of large systematics (e.g. definition of the sample and fitting technicalities such as the fitting range or choice of one over two component fits) and absence of a similar to our own data set for which calculations were done in a comparable way (e.g. decomposition into free S\'ersic and exponential components for a significant number of galaxies in common with this study). Our conclusion is, based on comparing individual cases, that there is a sufficiently good agreement with previous work, but that different types of above mentioned systematics are the dominant factor for uncertainties. 

%
%

\section{S\'ersic fits to one dimensional profiles}
\label{s:1comp}

We also fitted a single component  S\'ersic function to the light profiles of all \atlas galaxies with SDSS and INT imaging, in order to derive their global structural parameters, as it is often done with early-type galaxies \citep[e.g.][]{1993MNRAS.265.1013C,1996ApJ...465..534G,2004AJ....127.1917T,2006ApJS..164..334F}. After some testing, and contrary to our choice for the decomposition, we decided to fit azimuthally averaged light profiles obtained along fixed ellipses. Note that in Section~\ref{ss:method}, when we outlined the method for choosing whether a profile needs to be decomposed or not, we stated that we fitted both one and two components to the same light profile extracted along the best-fitting ellipses. We, however, do not think these profiles are best suited for determination of the global parameters, and, hence, use profiles extracted along the fixed ellipse. 

Our choice for fixed ellipse profiles is motivated by our wish to parameterise the whole galaxy with a single component. As shown by  \citet{2008AJ....135...20E}, multicomponent systems will have different light profiles depending whether they are extracted along fixed or free ellipses. Our choice of fixing {\it PA} and {\it Q} is justifiable as we are fitting a single function to objects which are predominantly two or more component systems (see Section~\ref{s:res}).  For some objects, such as massive, triaxial slow rotators, the change in ellipticity or position angle is most likely not an indication of multiple components but of triaxiality or smoothly varying orbital structure. For these objects an approach with free ellipses could also be preferred. As there are, however, only a handful of such objects, we choose to fit a constant in {\it PA} and {\it Q} model, as for all other galaxies. As these galaxies typically do not warrant a decomposition (see Section~\ref{ss:dt}), an interested reader can find in Appendix~\ref{B:master} values for single component fits obtained on light profiles extracted from free ellipses. Our choice is similar to what a typical 2D fitting algorithm does: the component used to fit the galaxy image has a fixed shape and orientation. We support our decision with a discussion in Appendix~\ref{A:choice}. 

The parameters of the ellipses (PA, Q) were taken from Paper II, which are global and measured at large radii (typically around 2-3 effective radii). As another difference from the approach outlined in Section~\ref{s:1d}, we performed the fits on all galaxies, including objects with bars and/or rings. Note that the {\it PA} and {\it Q} used are not related to bars, because in Paper II we took care to obtain them at radii beyond these structures and, hence, in barred systems they describe the shape and orientation of host discs. 

We fitted the light profiles in the same radial range as for the two component fits with the general $r^{1/n}$ profile of eq.~(\ref{e:free}). The results of the fits are the global S\'ersic index $n_{tot}$, effective radius $R_{e,tiot}$ and the intensity $I_{tot}$ at the effective radius. As can be expected, one component fits have somewhat larger residuals than two component fits. The median rms is 0.08 mag/$\arcsec^2$, while the standard deviation is 0.05 mag/$\arcsec^2$. If we exclude barred galaxies and compare the rms for only those objects for which we also performed the disc/bulge decompositions, the median rms drops to 0.06 and its standard deviation to 0.04 mag/$\arcsec^2$. 

%
%

\section{Results}
\label{s:res}

\subsection{Global structural parameters of ETGs}
\label{ss:1compres}

Results of the single S\'ersic fits to all galaxies are presented in Fig.~\ref{f:hist1c} and given in Table~\ref{t:master}. In addition to division into slow and fast rotators (top panels), we split the sample by mass in two subsets similar in number using M$_{dyn}=4\times10^{10}$ M$_{\odot}$ as the divider (bottom panels), a value similar to the characteristic mass derived by \citet{2003MNRAS.343..978S}. 

The mass is constrained by the \atlas integral-field kinematics, images used in this paper and the Jeans Anisotropic Models \citep{2008MNRAS.390...71C}. It is defined as $M_{dyn}=L\times (M/L)_{dyn}$, where $L$ is the galaxy total luminosity and the mass to light-ratio was obtained via dynamical models. This mass represents $M_{dyn}\approx2\times M_{1/2}$ where $M_{1/2}$ is the total dynamical mass within a sphere containing half of the galaxy light. Given that the stellar mass dominates the mass inside $M_{dyn}(r=r_{1/2})$, $M_{dyn}$ provides a very good approximation (in median within 10\%) to the galaxy stellar mass \citep[][hereafter Paper XIX]{2012arXiv1208.3522C}. 

When mass is used as a proxy, there are clear trends in size (global effective radius of the S\'ersic profiles) and the S\'ersic index: high mass galaxies are typically larger and have larger $n_{tot}$. However, when using this particular mass pivot point, the overlap between the values of the two samples is large. 

When dividing galaxies into slow and fast rotators, there is a significant difference between the two classes based on these two parameters. A Kolmogorov-Smirnov (K-S) test gives a probability of $10^{-5}$ and $10^{-4}$ that sizes and S\'ersic $n$ of fast and slow rotators are drawn from the same distribution, respectively. On the other hand, barred galaxies (Paper II) show a very similar distribution of sizes and S\'ersic indices as other fast rotators. A K-S test gives a 98 per cent probability that bars are drawn from the distribution of fast rotators, implying that a typical non-barred fast rotator will have the same size or S\'ersic index as a barred galaxy. 

Detailed comparisons with literature data are difficult due to various ways samples of early-type galaxies are selected (e.g. morphology, magnitude cuts or colour properties). However, in terms of the distribution of the S\'ersic index, our results are in a reasonable agreement with previous studies of early-type galaxies, \citep[e.g.][]{1993MNRAS.265.1013C}, who found a large fraction of galaxies with $n_{tot}<4$. A more detailed comparison can be found in Appenidix~\ref{C:litcomp}.

The main differences between slow and fast rotators is that distributions of both $R_{tot}$ and $n_{tot}$ are flatter for slow than for fast rotators. The latter show a peak in size at about $R_{e,tot}=1.5$ kpc and a peak for S\'ersic index at about $n_{tot}=2$. Slow rotators do not display any specific peak, but their distributions are somewhat limited in the sense that there are no small galaxies (e.g. less than 1 kpc in effective radius) and the smallest $n_{tot}$ is about 2. Furthermore, slow rotators are also found at the upper extremes of the size and S\'ersic index distributions. Noteworthy is to mention that the low values in $R_{tot}$ and $n_{tot}$ among slow rotators occur for special kinematics, such as for galaxies with counter-rotating components. 

The distribution of the S\'ersic index $n_{tot}$ in this sample of ETGs is of special importance. Various authors use the S\'ersic index to separate galaxies into discs and spheroids, or late- and early-type galaxies \citep[e.g.][]{2003MNRAS.343..978S, 2005ApJ...632..191M, 2005ApJ...635..959B}. The typical divide is taken to be $n_{tot}=2$ or $n_{tot}=2.5$, but some authors separate galaxies into an exponential ($n_{tot}<1.5$) and a concentrated ($n{tot}> 3$) group\footnote{In the rest of the paper we will similarly use $n_{tot}=3$ (or $n_b=3$) to distinguish between galaxies with concentrated and non-concentrated S\'ersic profiles.} \citep[e.g.][]{2003ApJ...594..186B}, or use S\'ersic indices as part of their classifications \citep[e.g.][]{2007ApJS..172..406S}. If these values are adopted, about  21 per cent (using $n_{tot}<2$), 34 per cent (using $n_{tot}<2.5$), or 48 per cent (using $n_{tot}<3$) of the \atlas galaxies, would not be considered early-type galaxies. As shown in Paper I, none of the \atlas galaxies have spiral arms or large dust lanes (across the full body when seen edge on). However, as we argued in Papers II, III and VII, and show below, it is a fact that the majority of early-type galaxies are discs or strongly related to discs. 

Furthermore, parameterising with a single S\'ersic function, and using any values of S\'ersic index, is not sufficient to separate slow from fast rotators. It is true that only a few slow rotators have low $n_{tot}$ values (and none of them has $n_{tot}<2$), and these might be special cases. However, there is a large number of fast rotators with S\'ersic index value as high as that of more typical slow-rotators. There are 6 slow rotators with $n_{tot}<3$ (out of 124 objects) and 104 fast rotators with $n_{tot}>3$ (out of 134 objects). These fractions give a probability to classify an object as a slow rotator if its $n_{tot}>3$ is only 0.22. If we use the Hubble classification (data from HyperLeda, \citep{2003A&A...412...45P}, see Section~\ref{ss:outlier}), one gets that a probability for classifying an elliptical if its $n_{tot}>3$ is  37 per cent (there are 50 of 134 galaxies with $n_{tot}>3$ classified as ellipticals).

S\'ersic index alone can not distinguish between slow and fast rotators (beyond saying that objects with $n_{tot}<3$ are most likely fast rotators), and hence does not sufficiently distinguish between two dynamically different classes of objects with likely different formation histories. This is an important caveat which should be kept in mind in all studies of large number of galaxies, or samples at large redshifts.

\begin{figure}
        \includegraphics[width=\columnwidth]{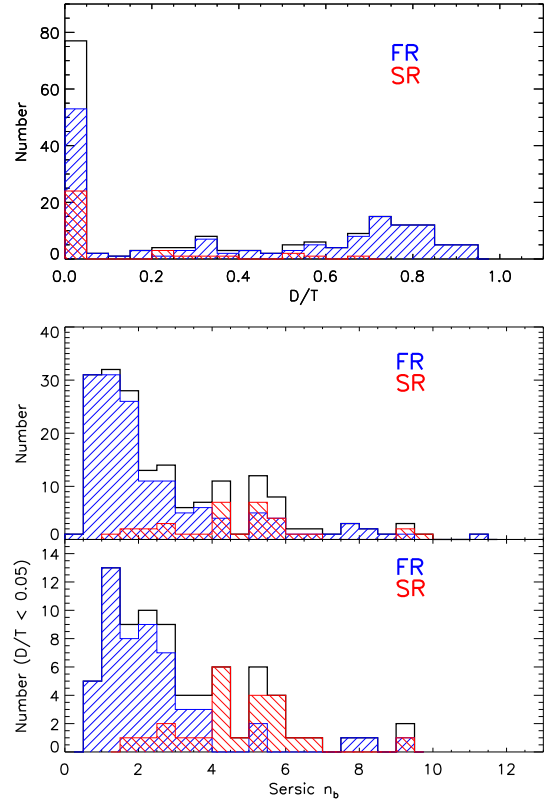}
\caption{\label{f:hist} Distribution of disc-to-total light (D/T) ratios ({\it top panel}) and S\'ersic $n$ indices ({\it middle and bottom panels}) for non-barred \atlas galaxies. In all panels blue (right slanted) hatched histograms are for fast rotators and red (left slanted) hatched histograms are for slow rotators. The bottom histogram is made of galaxies in the first bin of the top panel (galaxies with D/T$<0.05$)}
\end{figure}

\subsection{The decomposition results}
\label{ss:dt}

In Fig.~\ref{f:hist} we plot the results of our decompositions for non-barred \atlas galaxies following the procedure outlined in Section~\ref{ss:method}. The values are tabulated in Table~\ref{t:master}. The top panel shows D/T light ratios. Using Monte-Carlo simulations we estimate the errors to D/T light ratios and find that a median uncertainty is 0.08 for cases where D/T$ > 0$. Three main features are obvious: {\it (i)} 43 per cent of the analysed galaxies are in the first bin with D/T $<0.05$, {\it (ii)} early-type galaxies show a full range of D/T ratios, and {\it (iii)} there is an increase of galaxies around D/T $\sim 0.8$. We consider that the first bin (D/T $<0.05$) contains galaxies with no exponential sub-components, hence, it is remarkable that more than half of all non-barred ETGs contain at least some evidence, and typically a significant amount, of light parameterised with an exponential component. This is perhaps not so surprising when considering the finding of \citet{2009A&A...508.1141S} that visually selected early-type galaxies can have low B/T ratios (or high D/T ratios in our notation). 

Separating galaxies according to their angular momentum into fast and slow rotators reveals that the majority of slow rotators (71 per cent, or 24 of 34) actually have no exponential component, but six slow rotators (18 per cent, or 6 of 34 objects) have D/T $ >0.3$, and ten (29 per cent) have D/T $>0.1$. The latter value confirms the choice in Paper VIII to separate fast and slow rotators. In  conclusion, the majority of slow rotators are early-type galaxies with no exponential components, while those that have an exponential component typically also have specific signatures of rotation. We  will return to this issue in Section~\ref{ss:outlier}.

The middle panel of Fig.~\ref{f:hist} shows the distribution of S\'ersic indices of the bulge. There is a strong peak at low S\'ersic indices and a long tail at larger values, and a bump between $n_b \sim4-6$. This protuberance is obviously caused by slow rotators, which predominantly lie between 4-6, and 76 per cent (26 of 34 objects) of slow rotators have $n_b>3$. 
 
While the distribution of S\'ersic indices for slow rotators is as expected ($n_b$ is typically large), the distribution of $n_b$ for fast rotators is more surprising. There are galaxies with large indices (about a quarter of fast rotators have $n_b>3$), and a fast rotator can have as large a S\'ersic index as a slow rotator. The majority of fast rotators (61 per cent, or 89 of 146 objects), however, have small indices ($n_b<2$) and the large indices are distributed in a long tail of the distribution. This comparison is only partially proper, as more than two thirds of slow rotators are single components systems, while this is true only for a third of fast rotators. 

In the bottom panel of Fig.~\ref{f:hist} we show the distribution of the S\'ersic indices for all galaxies in the first bin (D/T $<0.05$) of the top panel. We consider these galaxies to be made of a single component; the decomposition did not improve on the one component fit significantly. There are 53 and 24 such fast and slow rotators, respectively. The distribution of $n_b$ is again asymmetric with a peak at low values of the S\'ersic index ($n_b=1-3$) and two peaks at larger values ($n_b=4-6$). As on the plot above, fast rotators make up the first peak and slow rotators the secondary bumps, with an overlap of a few galaxies in both directions, suggesting a clear difference in the structure of these two classes of early-type galaxies.

A most likely S\'ersic index for a single component fast rotator is between 1 and 2. This is remarkable, as not only more than half of fast rotators have a significant amount of light in an exponential component (e.g. 59 per cent, or 86 of 146, of fast rotators have D/T $> 0.2$),  but the majority of fast rotators which can be described as single component systems have $n_b<3$ (79 per cent, or 42 of 53, of single component fast rotators) and a profile similar to that of the exponential. There are 11 single component fast rotators with $n_b>3$, of which 4 show prominent shells and tidal tails, and one is actually a prolate rotator. We will discuss these galaxies in more detail below.  

\subsection{Correlation between single S\'ersic fits, the decomposition parameters and angular momentum}
\label{ss:corr}

\begin{figure}
        \includegraphics[width=\columnwidth]{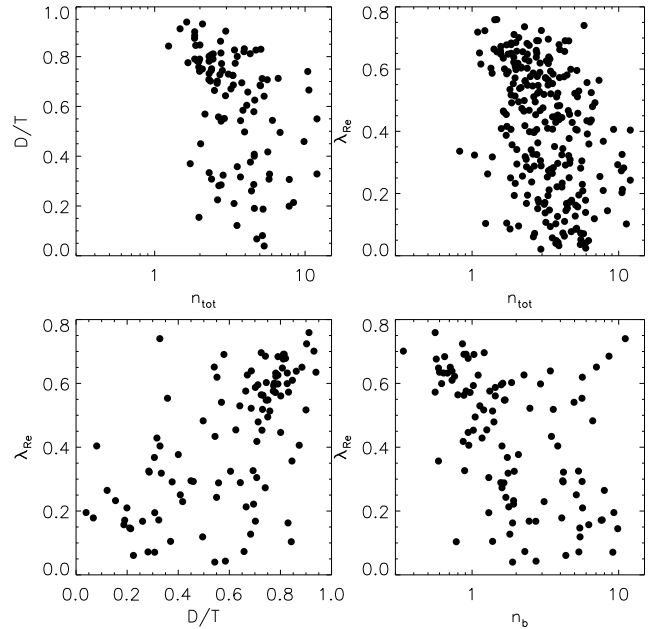}
\caption{\label{f:corr} From left to right, top to bottom: correlations between D/T ratio and S\'ersic index of the single component fits, $\lambda_R$ and S\'ersic index of the single component fits,  $\lambda_R$ and D/T ratio, and $\lambda_R$ and S\'ersic index of the bulge sub-component. In panels with D/T ratios, we show only those galaxies that required two components fits (e.g. D/T$>0$.) }
\end{figure}

In Fig.~\ref{f:corr} we show four diagrams with S\'ersic index of the single component fits, S\'ersic index of the bulge sub-components, D/T ratio, and angular momentum, $\lambda_R$, plotted against each other. The general conclusion is that there are no strong trends, except a general relation between D/T and $\lambda_R$. As it was reported previously \citep[e.g.][]{2009MNRAS.393.1531G, 2012MNRAS.421.2277L}, D/T (or rather bulge-to-total ratio\footnote{Note that B/T = 1- D/T only if the decomposition was done into two components like here and, hence, a comparison with other studies that decompose galaxies into, for example, bulge, bar and discs might not be straightforward. We prefer to use D/T ratio, where D is associated with the exponential component, while bulges are an in-homogenous set of objects with a range of S\'ersic indices \citep[for definitions of various types of bulges see][]{2005MNRAS.358.1477A}.}) ratio correlates poorly with the S\'ersic index, of both global and of the bulge sub-component. We will discuss further the relations between D/T and $n_b$ with $\lambda_R$ in the next section. There is a weak correlation between D/T and $\lambda_R$, which is tighter for larger values of $\lambda_R$ and high D/T ratios. On a contrary, there is no significant correlation between $\lambda_R$ and the S\'ersic index of single component fits, which confirms the finding of Section~\ref{ss:1compres}.

\begin{figure*}
        \includegraphics[width=\textwidth]{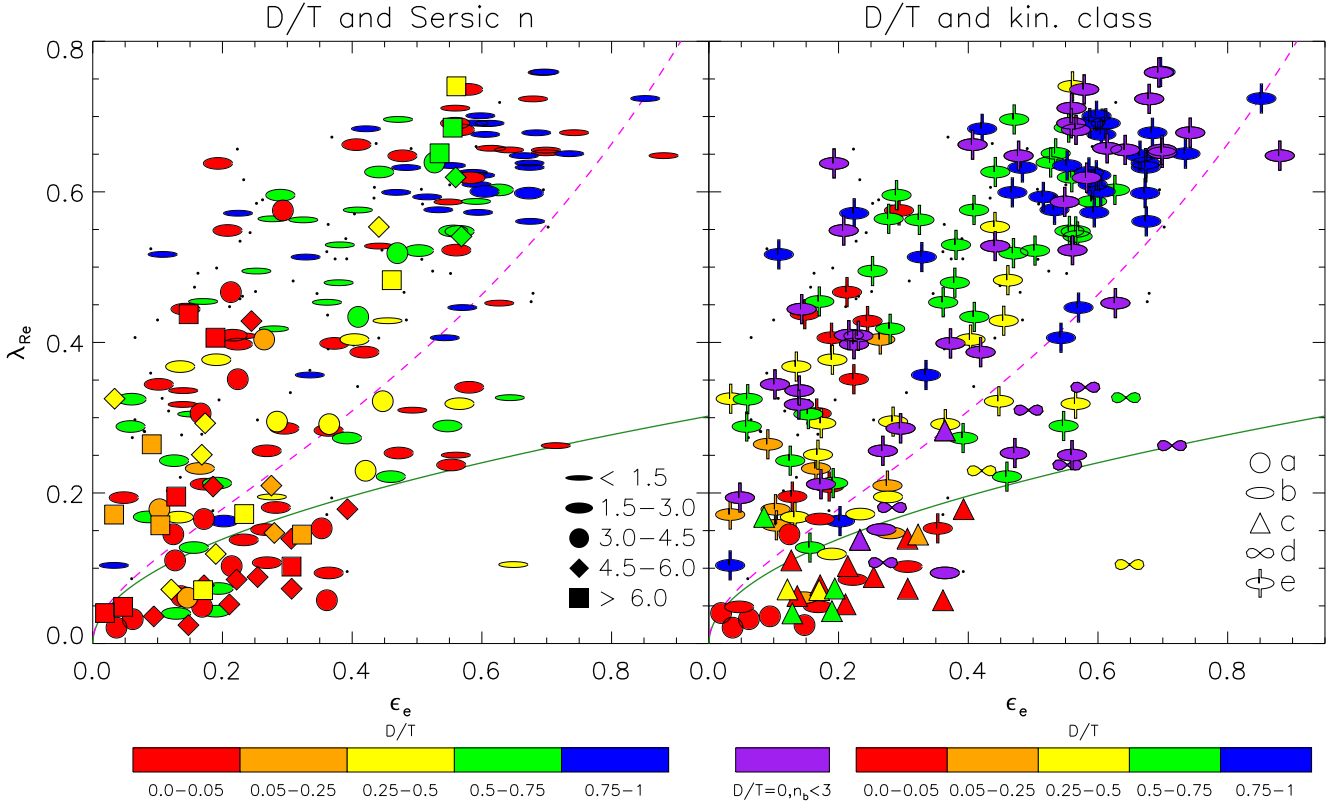}
\caption{\label{f:lambdaR} $\lambda_R$ versus $\epsilon$ for \atlas galaxies. Barred galaxies not used for the decomposition are shown as small dots for completeness. {\it Left:} Symbols represent S\'ersic indices as shown on the legend, while colour coding quantifies the D/T ratio, as shown on the colour bar under the diagram. {\it Right:} Symbols show different types of kinematics from Paper II and are described in the legend: $a$ - non rotating galaxies, $b$ - featureless non-regular rotators, $c$ - KDC, $d$ - $2\sigma$ and $e$ - regular rotators. Colours again quantify D/T ratios, as shown on the colour bar, but now we also highlight those galaxies which do not have an exponential component, but have $n_b<3$ (purple). The green line separates slow (below the line) from fast (above the line) rotators (Paper III). The dashed magenta line shows the edge-on view for ellipsoidal galaxies with anisotropy $\beta = 0.7 \times \epsilon$, from Cappellari et al. (2007). }
\end{figure*}

\subsection{Exponential profiles in ETGs are discs}
\label{ss:lesson}

\subsubsection{Morphological properties and angular momentum of early-type galaxies}
\label{sss:struct}

As pointed out by \citet{2004MNRAS.355.1155D} and \citet{2006MNRAS.369..625N}, finding exponential components in the light profiles of ETGs does not imply they correspond to discs. Combining the bulge/disc decomposition results with the stellar kinematics analysis, however, can elucidate the true nature of structural components of ETGs. Judging from Fig.~\ref{f:hist} there is a clear separation between slow and fast rotators in their structural properties. To investigate in greater detail the relationship between kinematics and photometric structures we present in Fig.~\ref{f:lambdaR} two $\lambda_R$ vs $\epsilon$ diagrams. In the left hand panel we compare the amount of light in the exponential component, as quantified by the D/T ratio, and the S\'ersic index $n_b$ of the bulge component. In the right hand panel we correlate the types of rotation found in our galaxies with the amount of light in the exponential component. 

Looking at the left hand panel of Fig.~\ref{f:lambdaR}, and as seen in Fig.~\ref{f:corr}, galaxies with low S\'ersic indices are typically found at high $\lambda_R$, while the fraction of galaxies with low D/T ratios is higher at low $\lambda_R$. There are some outliers, especially that galaxies with $D/T<0.05$ can be found also at larger $\lambda_R$. These objects, however, typically have a low S\'ersic index, typically $n_b <3$ (shown as ellipses). On the contrary, objects with $D/T<0.05$ at low  $\lambda_R$ (e.g. slow rotators), have typically higher S\'ersic indices ($>3$). This division sets two extremes of early-type galaxies: those with low angular momentum and that are best described with a single S\'ersic component of a high index, and those with high angular momentum, best described with two S\'ersic components of a similar index or with a single S\'ersic component of a low index.

Until this point we did not consider the detailed kinematic properties of our galaxies, except their global angular momentum. In Paper II we analysed our integral-field data by means of {\it kinemetry}, optimised for the mean velocity maps, and divided the galaxies in five groups depending on their complexity. We plot these on the right hand panel of Fig.~\ref{f:lambdaR}, colour coding with the D/T ratios. Here we also separate galaxies best parameterised with single components of low S\'ersic indices. This allows us to recognise that galaxies classified as non-rotators (Group {\it a}) are single component systems with high S\'ersic indices. Galaxies showing featureless but non-regular rotatation (group {\it b}) and kinematically distinct cores (KDCs; Group {\it c}), are typically made of a single component with a high index, but in some cases low fractions of the exponential components can be attributed to their light profiles. Finally, galaxies made of two-counter rotating discs ($2\sigma$ galaxies or Group {\it d}) are mostly single component systems of low S\'ersic index, or have large D/T ($>0.25$) and low $n_b$ ($<3$). In that respect they are structurally similar to Group {\it e}, or galaxies with regular and most disc-like rotation, which are also characterised with low S\'ersic indices and a range of D/T values. These include both single component systems (of low S\'ersic index) and systems with the highest contributions of the exponential light profiles.

\subsubsection{$V/\sigma - h_3$ correlation}
\label{sss:h3}
\begin{figure}
        \includegraphics[width=\columnwidth]{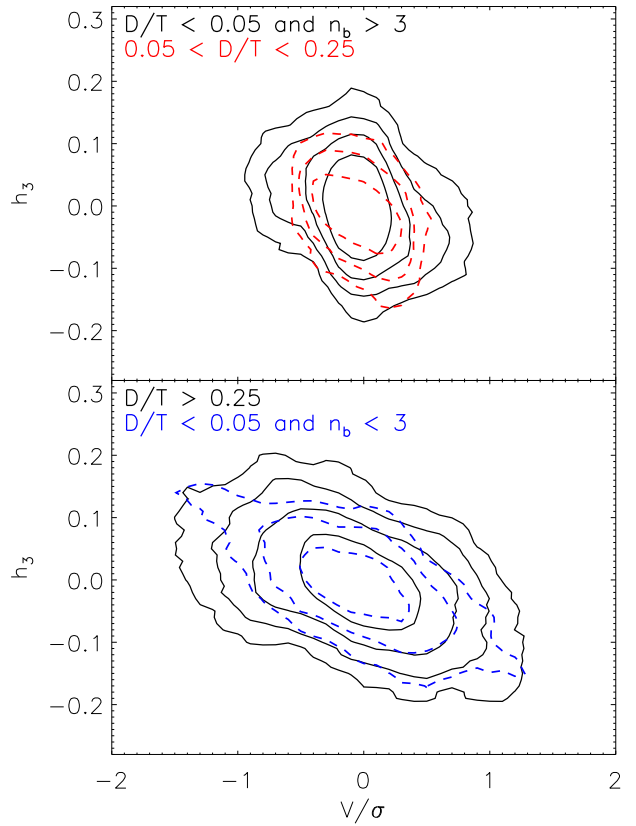}
\caption{\label{f:dt_h3} Local $h_3 - V/\sigma$ relation for every spectrum in galaxies with $\sigma > 120$ kms$^{-1}$ and an error on $h_3 < 0.05$. The contours show distribution of values in bins of 0.1 in $V/\sigma$ and 0.01 in $h_3$, smoothed with a boxcar filter of a window of 2 pixels in both dimensions. The contour levels decrease in step of 0.5 in log from 2 for the smallest contours. {\it Top:} solid contours show the distribution of values for galaxies described by a single component of a high S\'ersic index and dashed (red) contours show galaxies with low D/T fraction. {\it Bottom:}  solid contours show the distribution for galaxies with substantial disc fractions, while dashed (blue) contours show values for galaxies described by single components of a low S\'ersic index.}

\end{figure}

Next to kinematic information presented in Fig.~\ref{f:lambdaR} based on the angular momentum content and kinemetric analysis of the disc-like rotation in \atlas galaxies, we now use the information found in $h_3$, analogous to the skewness, the higher order moment of the line-of-sight velocity distribution \citep{1993ApJ...407..525V, 1993MNRAS.265..213G}. In Fig.~\ref{f:dt_h3} we show $h_3$ values against  $V/\sigma$ for all \atlas galaxies which we decomposed and for which we were able to measure this moment on individual spectra. We divided galaxies in those that are characterised by a single component of a large S\'ersic index, those that have a low contribution of exponential components, those with a high contribution of the exponential components and galaxies of single components with small S\'ersic indices. The first two classes are shown on the top panel (solid and dashed contours, respectively) and the second two on the bottom panel (solid contours) of Fig.~\ref{f:dt_h3}. 

There is an evident difference between the distributions on the two panels. Galaxies with high contribution of the exponential components show strong anti-correlation between $h_3$ and $V/\sigma$, which is often used as a kinematic manifestation of stellar disc kinematics, or at least evidence for stars at high rotational speeds \citep[e.g.][]{1994MNRAS.269..785B}. There is also a small difference between the two distributions on the top panel, as galaxies with single components (and large S\'ersic indices) are dominated by $V/\sigma\sim0$ values. On the bottom panel of this figure one can see that the tightest anti-correlation of $h_3 - V/\sigma$ is seen in single component galaxies of small S\'ersic indices.

The combination of various kinematic information and the decomposition results allows us to conclude that the rotation in early-type galaxies is typically associated with the presence of the exponential components in the light profiles. More specifically, the exponential profiles are only present when there is at least some indication of rotation, and galaxies in which the light is dominated by the exponential profiles are all galaxies with high stellar angular momentum. Furthermore, in cases where fits did not warrant the existence of exponential sub-components, but regular disc-like rotations is present and $h_3$ is anti-correlated with $V/\sigma$, the profiles are described by a single component of a small ($<3$) S\'ersic index. This leads to a conclusion that any component with a S\'ersic index less than about three can be associated with a disc, or is at least closely related to discs. The inverse is also true as galaxies with no detected rotation are typically single component systems of high S\'ersic indices. 

\subsubsection{Similarities of fast rotators galaxies and spirals}
\label{sss:spiral}

The existence of bulges of low $n_b$, a large range of D/T ratios, and a substantial fraction of objects with large D/T ratios in fast rotators  confirms their similarity with spirals \citep[e.g.][]{2001AJ....121..820G, 2003ApJ...582..689M,2009ApJ...696..411W,2010MNRAS.405.1089L}, and strongly suggest an evolutionary link. Our results support the revision of the Hubble diagram put forward initially by \citet{1976ApJ...206..883V}, which we revised to include fast and slow rotators in Paper VII (for photometric investigations see \citet{2011AdAst2011E..18L} and \cite{2012ApJS..198....2K}).

Additionally, the low values of S\'ersic indices for the bulges of fast rotators are characteristic of central light concentrations built from discs \citep[e.g. discy-bulges,][]{1993IAUS..153..209K,2005MNRAS.358.1477A}\footnote{These are sometimes referred to as pseudo-bulges  \citep[e.g.][]{2007MNRAS.381..401L, 2008AJ....136..773F}, in order to highlight their structural and presumably evolutionary differences from the classical bulges \citep{2004ARA&A..42..603K}. We, however find this terminology unnecessarily confusing as it encompasses structures with various morphologies, scales and potential origins.}. We remind the reader that we did not analyse barred galaxies and that our sample is devoid of spirals (and late-type galaxies in general). Also we have excluded from the fitting the central regions, while including higher resolution images could have an effect of decreasing the S\'ersic index \citep[e.g.][]{2003ApJ...582L..79B}. Nevertheless, it is clear from Figs.~\ref{f:hist} and~\ref{f:lambdaR} that bulges of  low S\'ersic index are typical among fast rotators and that their kinematics are disc-like, linking further the properties of early- and late-type galaxies. Similar results were reported recently by \citet{2012arXiv1204.5188F} for S0s and late-type galaxies. It is, however, also evident on Fig.~\ref{f:lambdaR} that there are fast rotators with disc-like kinematics and with bulges of high S\'ersic index, as well as fast rotators which are sufficiently well described with single components of low S\'ersic indices.

\subsubsection{Masses of discs}
\label{sss:mass}

Using dynamical masses from Paper XIX, we can estimate what mass fraction is in the exponential components. In calculating we assume that there is no difference in stellar populations between the bulge and the exponential components and that galaxies are well fitted by a single mass-to-light ratio in the dynamical models. With this caveat in mind and selecting galaxies with $D/T>0.05$, we find that the total mass in the exponential components is $\sim 4.12\times10^{12}$ M$_\odot$, or 27 per cent of the total mass of investigated galaxies. Selecting galaxies with $D/T<0.05$ and $n_b<3$, gives the total mass of $2.10\times10^{12}$ M$_\odot$ or 14 per cent of the total mass of investigated galaxies. Combining these two figures we find that $\sim41$ per cent of stellar mass in early-type galaxies is in discs or disc-like components. The rest is shared mostly between single component slow rotators and bulges of fast rotators. Note that we did not include here the contribution of the barred galaxies.

\subsection{Decomposition and classifications of early-type galaxies}
\label{ss:outlier}

\subsubsection{Hubble types and angular momentum}
\label{sss:hub}

\begin{figure}
        \includegraphics[width=\columnwidth]{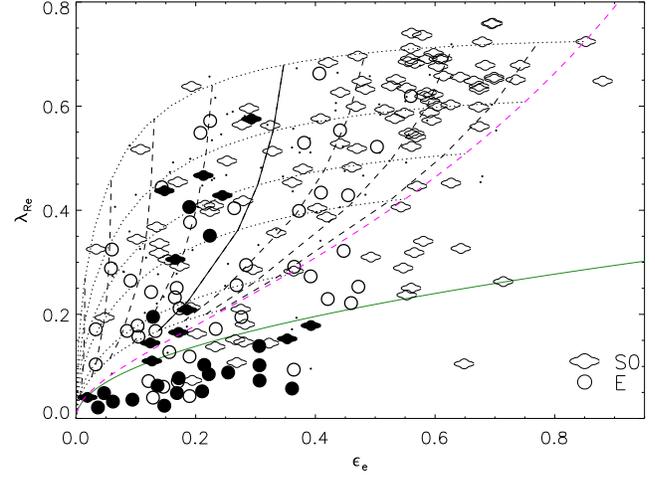}
\caption{\label{f:ES0} Distribution of elliptical (morphological type T $< -3.5$) and S0 (morphological type T $> -3.5$) galaxies in $\lambda_R$ versus $\epsilon$ diagram, as in Fig.~8 of Paper III. Solid symbols show ellipticals and S0s which are best fit with a single component S\'ersic function of a large index ($n>3$), and a decomposition of their profiles was not deemed necessary. As in Fig.~\ref{f:lambdaR}, the green line separates slow (below the line) from fast (above the line) rotators (Paper III), the dashed magenta line shows the edge-on view for ellipsoidal galaxies with anisotropy $\beta = 0.7 \times \epsilon$ from \citet{2007MNRAS.379..418C}, and dots are not-analysed barred \atlas galaxies. The dotted lines correspond to the location of galaxies with intrinsic ellipticities between 0.25 and 0.85 in steps of 0.1. The dashed lines show the location of galaxies originally on the magenta line as the inclination is varied in steps of 10\degr,  decreasing from the magenta line (90\degr) to the left. As a guide line, the line that was plotted solid corresponds for the inclination of 50\degr. The formulas to plot these lines can be found in \citet{2007MNRAS.379..418C}.}
\end{figure}

On Fig.~\ref{f:ES0} we repeat the $\lambda_R - \epsilon$ plot, with symbols differentiating between galaxies classified as ellipticals and S0s using morphological types from the HyperLeda catalog \citet{2003A&A...412...45P}. In Paper III we commented on the discrepancy between E/S0 and fast/slow rotator classifications. Here we want to compare our decomposition results with both of these approaches, and with solid symbols we plot those galaxies, which are sufficiently well described with a single S\'ersic profiles of a large index ($n_b>3$). 

There are 31 galaxies with that property, of which 20 are slow and 11 fast rotators. As fractions of the analysed slow and fast rotators, these galaxies make up 59  and 7 per cent, respectively. Based on their morphological classification, ellipticals best fit with a single component profiles of a large index are typically found under the green line defining the slow rotator class.  As a contrary, among the fast rotators, objects with the same structural properties are typically classified as S0s. Concentrating on the $\lambda_R>0.25$ region, there are such 7 galaxies, 2 classified as ellipticals (NGC\,0680 and NGC\,4486A) and 5 as S0s (NGC\,2695, NGC\,4753, NGC\,4459, NGC\,5869 and NGC\,3182, in order of decreasing $\lambda_R$). NGC\,0680 is characterised by having evidence for a major merger, with a series of shells, arcs and two plumes rich in HI \citep[][hereafter Paper IX]{2011MNRAS.417..863D}. A similar shell like structure is also visible in NGC\,5869 and in NGC\,4753. Although these galaxies have significant and ordered rotation in their inner regions, the outer regions seem not to be fully relaxed, possibly having multiple structural components which are not any better described with two than with one components. The light profile of NGC\,4486A is unfortunately contaminated by a bright star, nearly co-spatial with the nucleus of the galaxy, and we moved the inner fitting limit out  to 5\arcsec, which is comparable to the effective radius of this galaxy, and the fit is likely not robust. Other S0 galaxies either have dust (NGC\,4459 and NGC\,4753) or show significant wiggles in their profiles (NGC\,2695, NGC\,3182), which are not removed with a two component fits. 

Light profiles of fast rotators with $\lambda_R<0.25$ are different from the above mentioned galaxies. The four galaxies characterised by single components of high S\'ersic indices in this region are: NGC\,3607 (S0), NGC\,3193 (elliptical), NGC\,5485 (S0) and NGC\,3073 (S0). All galaxies except NGC\,5485 do not show strong evidence for an exponential profiles. A blind decomposition assigns between 0.03 and 0.08 of the light fraction to an exponential profile, but the fits are barely improved with respect to one component fits. All four galaxies are somewhat special, but NGC\,5485 is the most intriguing as this is the one of the two galaxies in the entire \atlas sample which shows a prolate rotation (around its major axis), coinciding with a dust disc in a polar configuration. Even though this galaxy has a significant exponential component, it is not possible to associate it to the observed rotation, and call this component a disc. 

Below the green line, most interesting are the galaxies that can be decomposed or have one component with a low S\'ersic index. There are 14 such objects (NGC\,4168, NGC\,3608, NGC\,5198, NGC\,4458, NGC\,5813, NGC\,3414, NGC\,7454, NGC\,4191, NGC\,4559, UGC03960, PGC050395, NGC\,1222, PGC28887 and NGC\,4690, in order of increasing  $\lambda_R$), 7 classified as S0 and 7 as Es. The profiles for these galaxies, except NGC\,4191 and NGC\,7454, require a significant fraction ($>0.2$) of the exponential components in their lights. NGC\,4191 and NGC\,4550 are $2\sigma$ galaxies, and their low S\'ersic indices are consistent with these galaxies being made of counter-rotating discs  \citep{1992ApJ...394L...9R,1992ApJ...400L...5R,2007MNRAS.379..418C,2011MNRAS.412L.113C}. NGC\,7454 and NGC\,5198 are galaxies with non-regular but featureless kinematics. Atypically for slow rotators, NGC\,5198 and UGC03960 have HI gas, in both cases in peculiar configurations \citep[][hereafter Paper XIII]{2012MNRAS.422.1835S}. The last five galaxies in this list are found close to the green line, and they are likely to be transitional objects in terms of $\lambda_R$. The other five galaxies have KDCs and possibly the exponential profiles could be associated with the stellar distributions forming the KDCs

\subsubsection{A transitional region in $\lambda_R$}
\label{sss:lr}

There seems to exist a transitional region between fast and slow rotators, and it can be broadly put to be between $0.1<\lambda_R < 0.25$. Almost all galaxies above this region can be considered disc dominated galaxies or at least galaxies with significant disk fractions. Below this region galaxies are typically, with a few exceptions, single component systems of high S\'ersic index. Within the region, however, there is a mix of objects, fast rotators with no and slow rotators with a significant fraction of light in exponential components. 

This region was also highlighted in the study of binary mergers by \citet[][hereafter Paper VI]{2011MNRAS.416.1654B}. There we found that slow rotator remnants of binary mergers (of 1:1 and 1:2 mass ratios) are typically found below this region. Above the region, however, is the area populated by fast rotators remnants of binary mergers, whose progenitors were on prograde orbits (prograde or retrograde motion of the main progenitor has a strong influence on the dynamical structure of the remnant). The transitional region itself is also populated by merger remnants, but this time remnants of re-mergers of galaxies that lie above or below this region (see Fig. 11 in Paper VI). Although these were non-cosmological mergers, their results highlight that this region will likely contain galaxies with special dynamical structures. 

Furthermore, part of this region is populated by galaxies seen at low inclination, while their edge on projections are on the dashed magenta line on Fig.~\ref{f:lambdaR} (see Fig.~1 of Paper III for the illustration of the projections in $\lambda_R - \epsilon$ diagram). This means that galaxies in this region could be a mix of two populations, oblate galaxies with discs projected at low inclinations and remnants of major mergers. In this respect the varied properties of light profiles of galaxies are no more surprising than their varied kinematic properties, and one could expect more surprises from galaxies in this region.

\subsubsection{Hubble types, angular momentum and decomposition results}
\label{sss:had}

\begin{table}
\caption{Median values and standard deviation of S\'ersic indices and D/T ratios for galaxies as classified by apparent shape or angular momentum. }
\label{t:ESO}
\begin{tabular}{c|cc|cc}
\hline
Classification& $\overline{D/T}$ &$ \sigma_{D/T}$ & $\overline{n_b}$ & $\sigma_{n_b}$\\
(1)& (2)&(3) &(4) &(5)\\
\hline
E        & 0.19 & 0.29 & 3.8 & 2.2\\
S0      & 0.37 & 0.39 & 1.4 & 1.0\\
\hline
SR      & 0.00 & 0.16 & 4.8 & 1.9\\
FR      & 0.41 & 0.36 & 1.7 & 1.3\\
\hline
\hline
E FR  & 0.32 & 0.28 & 2.7 & 2.1\\
S0 FR& 0.58 & 0.43 & 1.4 & 0.8\\
\hline
E SR  & 0.00 & 0.14 & 5.1 & 1.7\\
S0 SR& 0.00 & 0.19 & 4.1 & 2.4\\
\hline
\end{tabular}
\\
Note that a number of galaxies are single components systems with D/T=0. In these cases $n_b$ was the S\'ersic index of the single component.
\end{table}

In Table~\ref{t:ESO} we list the median values and the standard deviations of S\'ersic indices and D/T ratios, splitting the analysed galaxies into ellipticals and S0s, fast and slow rotators, as well as the combination of the two classification: fast rotating ellipticals (E FR), fast rotating S0 (S0 FR), slow rotating ellipticals (E SR) and slow rotating S0s (S0 SR). In terms of the decomposition parameters, both classifications give similar results, but fast -- slow division highlights more the differences between the objects with higher and lower D/T ratios and S\'ersic indices, than the standard Hubble classification. This is enhanced if we sort ellipticals and S0s depending on their angular momentum content. We can see that slow rotating ellipticals and S0s are structurally very similar, while fast rotating ellipticals and S0 show a certain range of properties, but they are rather very different from their slow rotating counterparts. As general conclusion of this section, based on Fig.~\ref{f:ES0} and Table~\ref{t:ESO} we stress that results of the decomposition are more closely related to the fast -- slow classification. They could be used to improve on the standard Hubble classification, but they cannot be used as a substitute for the kinematic classification. 

As a guideline, when stellar kinematics is not available, we recommend to use the following combination of criteria to select tentative fast and slow rotators: a D/T $> 0.05$ (a D/T $>0.1$ is also acceptable, depending on the confidence of the decomposition) for galaxies which need to be decomposed in (at least) two components, and $n<3$ for galaxies not requiring a decompositions. We stress that with this selection one can misclassify up to 40 per cent of slow rotators. 

The large spread of possible values for D/T ratios when elliptical/S0 classification is used, as well as for fast rotators is likely a manifestation of the inclination effects. In addition, the semi-analytic models of Paper VIII suggest that there are differences between fast rotators. In particular, there is a range of D/T ratios (as we confirm in Section~\ref{ss:dt}), where those with small ratios are likely to grow discs via cold accretion flows or grow bulges via minor mergers, while fast rotators with large D/T have exhausted their gas reservoirs (and can not replenish it) and live in dense environments resembling passively evolved spirals. In the following two sections we address these two issues, by investigating the influence of the inclination on our results and looking for differences among fast rotators.

\subsection{Inclination effects}
\label{ss:incl}

The change of D/T ratios or values of $n_b$ from the top right (mostly blue) corners of the panels in Fig.~\ref{f:lambdaR} to the bottom left (orange and red) corners could be caused by inclination effects. This is expected as ellipsoidal galaxies viewed edge-on, and having an anisotropy as found in \cite{2007MNRAS.379..418C}, lie on the dashed magenta line. Their projections due to varying inclinations are found to the left of this line (see Fig~\ref{f:ES0}), within the region inhabited by the majority of fast rotators, where the changes in D/T and $n_b$ are the most obvious. Given the known effects of the inclination on the ability to find discs in model galaxies \citep[e.g.][]{1990ApJ...362...52R,1996MNRAS.279..993G}, we can also expect that finding discs using the decomposition method will be affected as well. In order to gain a qualitative understanding of the effects of the inclination on the decomposition parameters we performed the following test. 

We selected two galaxies (NGC\,4621 and NGC\,5308), a galaxy with a weak and a strong disc (and small and large D/T ratios), respectively, which can be reasonably assumed to be close to edge on. We used the Multi-Gaussian Expansion (MGE) method \citep{1992A&A...253..366M, 1994A&A...285..723E} as implemented by \citet{2002MNRAS.333..400C} to parameterise their light distributions as a series of two-dimensional gaussians. Assuming the galaxies are seen edge-on, the MGE models specify the intrinsic shapes of these galaxies. The models were projected at a series of inclinations. Each of these models was then analysed in the same way as the original images: we extracted an azimuthally averaged light profile (letting the ellipse parameters free during the fit) and fitted the light profile as described in Section~\ref{ss:method} with a general S\'ersic and an exponential component. 

In Table~\ref{t:incl} we list the parameters of the decompositions of our MGE models. The results of this idealised analysis is that although there are some changes in the recovered parameters, they are systematic, but not large. The D/T fraction decreases as the viewing inclination approaches the face-on orientation, but the amplitude of the change is relatively small.  In addition, the change of $n_b$ and the sizes of the two components are also increasing, where the increase is more pronounced for the models with the smaller disc. 

\begin{table}
\caption{Inclination effect on the parameters of the decomposition }
\label{t:incl}
\begin{tabular}{cccccc}
\hline
name&Incliantion & D/T & $n_b$& R$_e$ & R$_s$\\
(1)& (2)&(3) &(4) &(5)&(6)\\
\hline
		&10& 0.72 & 1.56&  3.1&19.8\\
		&20& 0.73 & 1.49&  4.6&19.9\\
		&30& 0.74 & 1.45&  4.7&19.7\\
NGC\,5308&40& 0.77 & 1.39&  4.6&19.3\\
		&50& 0.79 & 1.33&  4.4&19.0\\
		&60& 0.82 & 1.24&  4.2&18.5\\
		&70& 0.85 & 1.10&  3.9&18.0\\
		&90& 0.88 & 0.87&  3.5&17.3\\
\hline
		&10& 0.17& 6.0& 58.5&30.8\\
		&20& 0.20& 5.6& 49.4&31.8\\
		&30& 0.17& 5.9& 56.7&29.9\\
NGC\,4621&40& 0.17& 5.7& 54.7&29.4\\
		&50& 0.18& 5.6& 52.3&28.8\\
		&60& 0.25& 5.0& 39.0&30.4\\
		&70& 0.27& 4.8& 35.3&30.7\\
		&90& 0.33& 4.4& 29.3&31.2\\
\hline
\end{tabular}
\\
\end{table}

\begin{figure}
        \includegraphics[width=\columnwidth]{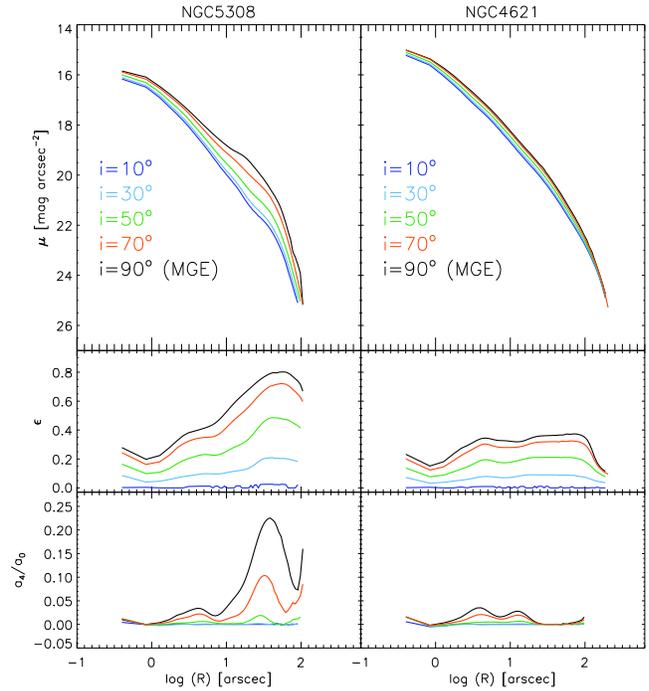}
\caption{\label{f:incl} {\it Top to bottom: } Surface brightness, flattening and disciness radial profiles for model galaxies with different fractions of light in the exponential components. {\it Left to right:} MGE models and their projections at 70\degr, 50\degr, 30\degr and 10\degr are based on NGC\,5308 (D/T$\sim0.8$) and NGC\,4621 (D/T$\sim0.35$). These galaxies were chosen as they are seen close to edge on and the intrinsic MGE model is considered to be seen at 90\degr. Colours on all panels correspond to models projected at different inclinations, as shown in the legend. Note that as the inclination decreases, the profiles of the corresponding model also decrease in the maximum amplitude. }

\end{figure}

The changes of the model D/T and $n_b$ with inclination can account for a change of at most 20-25\% in D/T and 1-1.5 in $n_b$ in Fig.~\ref{f:lambdaR}. The reason for this is likely in the systematics associated with the decomposition of the profiles. We illustrate this with  Fig.~\ref{f:incl}, where we show the radial profiles of the surface brightness, ellipticity and the disciness parameter \citep[e.g.][we plot the Fourier term, $a_4/a_0$, associated with the $\cos(4\theta)$ harmonics, normalised by the intensity]{1989A&A...217...35B}, for our two model galaxies seen at different inclinations (we show every other inclination for clarity). 

Looking at the edge-on case (90\degr) of the NGC\,5308 model, the disc component is clearly visible as a bump in the surface brightness profile at about $\log(R)=1.3$. The same bump is clearly associated with the rise in ellipticity and high $a_4/a_0$ which measures the disciness. At this inclination we can be sure that the recovered parameters indeed describe a disc. As the inclination decreases, the profiles also change. Ellipticity and disciness show a dramatic change, while the surface brightness changes less prominently, but the bump in the profile steadily decreases. These same changes are also visible for the models of NGC\,4621, but the differences at various inclinations are much smaller.

As demonstrated by \citet{1990ApJ...362...52R}, the disciness parameter looses its usefulness below an inclination of 50-60\degr. The differences in ellipticity between a bulge and a disc, if they existed in the first place, are erased below an inclination of 30-40\degr. The only signature of a disc, or, to be more precise, a necessity for another component, is visible in the light profile of the model such as NGC\,5308. The light profiles of the NGC\,4621 model, which had a relatively small disc, become less curved as the inclination is decreasing, and offer less hints for a need of a disc. In this model, below an inclination of 70\degr\, there is basically no clear photometric evidence for a disc. Our results are in agreement with \citet{1996MNRAS.279..993G}, who also note that only strong discs are visible at low inclinations. 

These examples show the dramatic effect of the inclination on the photometry and the observed shape of galaxies. Unless the disc is the dominant component, it will not be possible to recognise it below a certain inclination ($\sim50\degr$). A decomposition method might recover a certain amount of the disc at a low inclination in a galaxy such as represented by our model of NGC\,4621, but the confidence that this model could really be distinguished from a single component model, or that the exponential is really needed, is generally low. 

This should be taken into account when judging the decomposition results, including those presented here. Below an inclination of 50\degr, the photometric evidence for discs disappear and this might explain the large fraction of galaxies classified as ellipticals among fast rotators left of the line corresponding to this inclination (and above the magenta line) in Fig.~\ref{f:ES0}. It can also be used to explain why fast rotators with single component of high S\'ersic index are also found left of that line. Kinematic signatures of discs are more robust with respect to the changes in inclinations. The disc-like kinematics, found in nearly oblate axisymmetric objects (as well as bars) is visible at inclinations of 20\degr or even less \citep{2008MNRAS.390...93K}. Complex kinematics, on the other hand is a clear signature that the mass distribution is not favourable for the existence of discs.

\subsection{Two types of ETGs with discs}
\label{ss:2types}

The incidence of discs among slow rotators, large ranges of D/T ratios and S\'ersic indices (both $n$ and $n_b$) among fast rotators suggest there are sub-populations present among these galaxies. Additionally, different types of fast rotators are predicted by the semi-analytic models (Paper VIII). In this section we explore this by dividing galaxies in three bins, using both kinematic and photometric information on the disc components. The galaxies in the three bins can be described as having: {\it no discs}, {\it intermediate discs} or {\it dominant discs}. Following the results of Sections~\ref{ss:dt} and \ref{ss:lesson}, the selection of bins is made by requiring that galaxies are: 

\begin{itemize}

\item[i)] {\it No discs:} those slow rotators with $D/T <0.05$, $n_b > 3$ and not $2\sigma$ galaxies. This selection yields 20 objects (only slow rotators). 

\item[ii)] {\it Intermediate discs:}  those slow rotators which have $0.05 < D/T < 0.5$ or those that have $D/T < 0.05$, but $n_b < 3$, or those fast rotators which have $D/T<0.5$ and $n_b > 3$. No $2\sigma$ galaxies are taken in this bin. This selection yields 36 objects, including 9 slow rotators.

\item[iii)] {\it Dominant discs:} those slow and fast rotators with $D/T>0.5$, or those fast rotators with $D/T<0.5$ but $n_b<3$, and all (both fast and slow rotator) $2\sigma$. This selection yields 124 objects, including 5 slow rotators.

\end{itemize}

The {\it no disc} bin comprises slow rotators which do not have any signature (neither in the kinematics nor in the photometry) of disc-like components, and it is the most conservative estimate for non-existence of discs in early-type galaxies. We required $n_b>3$ (actually, for these galaxies $n_b$ is the global S\'ersic index, as they are all best fit with a single component) to remove the few galaxies with low S\'ersic index. As $2\sigma$ galaxies are made of two counter-rotating discs, or at least of two flattened families of counter-rotating orbits of high angular momentum \citep[for detailed dynamical models of $2\sigma$ galaxies see][]{2007MNRAS.379..418C}, these galaxies should be considered to have large disc contributions, even though their kinematics are not disc like. Therefore, we also removed all slow rotator $2\sigma$ galaxies.

\begin{figure*}
        \includegraphics[width=\textwidth]{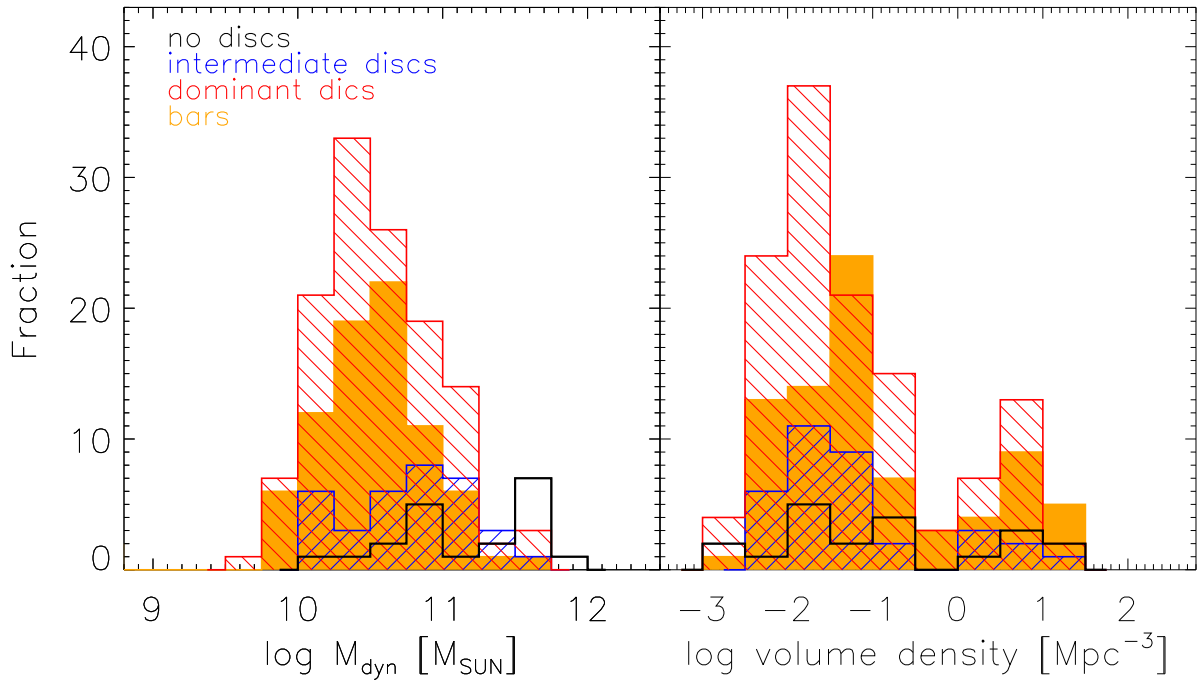}
\caption{\label{f:mass}  Distribution of \atlas galaxies of different disc content with respect to the total galaxy mass (left) and environment (right). In both panels galaxies are divided in three classes as specified in the legend (left panel) and in text (Section~\ref{ss:2types}) and we added all barred galaxies for which we did not attempt a decomposition. Open histogram shows {\it no discs}, red (left slanted) histogram shows {\it intermediate discs}, blue (right slanted) histogram {\it dominant discs} distributions and orange filled histogram shows barred galaxies. }
\end{figure*}
The {\it Intermediate discs} contain all galaxies which have some indications of discs, but these discs do not dominate the total light. This bin collects most of the slow rotators of typically higher $\lambda_R$ (for the range of $\lambda_R$ found among slow rotators; see open symbols on Fig~\ref{f:ES0}), and those fast rotators that have relatively small exponential discs and bulge components of high S\'ersic indices. The reason for this requirement is that a systems with a bulge component fit by a low S\'ersic index next to an exponential disc could be approximated as a double discs system or at least as being made of two disc-like components and should be excluded from this class. Again, no $2\sigma$ galaxies are taken in this bin. 

Finally, the {\it Dominant discs} bin gathers all remaining galaxies, including all remaining slow rotators with strong photometric disc contribution, all $2\sigma$ galaxies, and all fast rotators which either have a $D/T>0.5$ or $D/T> 0.5$ and $n_b>0.3$, for the same reason as explained in the previous paragraph. Given the previous results, it is not a surprise that most of our galaxies indeed fall in this group. 

We did not include barred galaxies as they were not analysed in this paper. However, if we were to include barred and ringed systems, it is likely that they would be split between {\it Dominant discs} and {\it Intermediate discs}, stronger barred systems probably contributing to the latter. In Fig.~\ref{f:mass}, which summarises the results of this section, we include barred galaxies in a separate bin for comparison with other three bins defined above. 

In Fig.~\ref{f:mass}, we present the mass and environment dependence for \atlas galaxies. We used mass estimates from Paper IX, and the density estimator from Paper VII (see Section~\ref{sss:mass}). As a measure of the environment, we use the volume density in Mpc$^{-3}$ of galaxies inside a sphere of a radius which includes ten nearest neighbours. Here we used the best distance estimates to get the three-dimensional distribution of galaxies (for more details see Paper VII). This density estimator is good to differentiate between cluster and field regions, or Virgo and non-Virgo densities in the \atlas sample. 

In both histograms shown on Fig.~\ref{f:mass} there is a substantial overlap between the bins, but a clear trend in mass can be seen on the left hand panel. The {\it Dominant discs} are typically found in lower mass systems (centred around $10^{10.3}$ M$_\odot$), the {\it Intermediate discs} in intermediate and more massive systems (centred around $10^{10.9}$ M$_\odot$), while the population of {\it No discs} dominates the most massive end of the distribution of \atlas galaxies (beyond $10^{11.5}$ M$_\odot$). Bars are distributed similarly like {\it Dominant discs}, and the K-S test gives a probability of 0.98 that these two distributions are drawn from the same parent sample. A contrary result is obtained if one compares the distribution of bars and {\it Intermediate discs} (K-S test probability is 0.003). This result is consistent with the observed distribution of galaxy properties on the mass -- size diagram and our interpretation of ETGs scaling relations \citet[][hereafter Paper XX]{2012arXiv1208.3523C}).

A more complex picture is evident in the right hand plot of the same figure which considered the environmental dependence. There is no major difference between fractions of different types of galaxies between Virgo (log(volume density) $> $0) and non-Virgo environments. Outside of Virgo, {\it Dominant discs} and {\it Intermediate discs} have similar distributions, while bars favour a bit more dense environments. Within Virgo, densest regions are favoured by {\it No disc} populations (as shown already in Paper VII), while {\it Intermediate discs} are found more towards the outskirts. Bars and {\it Dominant Discs} are found also in denser environments within the cluster, but bars tend to be more similarly distributed like {\it No disc} galaxies.

%
%
\section{Conclusions}
\label{s:conc}
 
In this work we performed a disc-bulge decomposition of \atlas galaxies with the aim to investigate the photometric evidence for discs in early-type galaxies, and to link them with our kinematic data. For this purpose we selected all (obviously) non-barred galaxies from our sample (180 galaxies out of 260, with 34 slow and 146 fast rotators), and performed a two component decomposition onto an exponential disc and a bulge described by the  S\'ersic function of a free index. We did not try to reproduce other components (i.e. bars and rings). The removal of the barred objects is justifiable as these galaxies are known to contain discs and they are found in fast rotators, therefore, the link between photometry and kinematics for these systems is clear, and we can not fit them accurately with our two component approach. We also performed a single component fits with a S\'ersic function and several tests with 1D and 2D decompositions methods (presented in the Appendix~\ref{A:choice}). The results of the fits are presented in Table~\ref{t:master}.

Before listing our main conclusion, we would like to highlight that global S\'ersic index is a poor estimate of galaxy morphology. It is widely used to differentiate between early- and late-type galaxies, but even when applied on a sample of only early-type galaxies it does not recover either the traditional Hubble classification based on the apparent shapes or the modern kinematic classification based on the specific angular momentum. Using the decomposition into a bulge and a disc does improve the agreement between morphological and kinematic classifications, but it is still not sufficiently good. While it can be used to highlight those objects which are likely consistent with being fast rotators and disc related (by assuming low S\'ersic index for light profiles requiring only a single component and D/T $>0.05$ for two component fits), it still fails in recognising slow rotators (or even galaxies commonly classified as ellipticals). This is of particular importance for higher redshift studies and studies of large samples of galaxies. 

Our main conclusions are: 

\begin{itemize}

\item Using the S\'ersic index alone (obtained by fitting a single S\'ersic function to the light profile) is not sufficient to distinguish between fast and slow rotators. The distribution of S\'ersic indices for slow and fast rotators are not drawn from the same sample, and typically fast rotators have low $n$ ($<3$). There is, however, a significant overlap of slow and fast rotators for $n>3$. Based on the \atlas sample of nearby early-type galaxies there is a 5 per cent chance that an object with $n<3$ is a slow rotator. For an object with $n>3$ there is, however, only a 22 per cent chance that it is a slow rotator. 

\item Single-component Sersic fits were adequate for 43 per cent of the analysed early-type galaxies (77 of 180 galaxies). The light profiles of other galaxies were better fit with two sub-components. The single-component galaxies do not contain a formal exponential component (with n=1), but 46 (of 77 or 59 per cent) of them have a low S\'ersic index ($n<3$), frequently around a value of 1. 

\item The exponential sub-components, or single-components with low S\'ersic indices ($n<3$), are found in the majority of early-type galaxies. We show that these components are present in galaxies with regular rotation, intermediate to high angular momentum and objects with $h_3 - V/\sigma$ anti-correlation typical for discs. Therefore, we associate exponential sub-components with discs. Similarly, single-components of low S\'ersic indices can be associated with discs (if $n\sim1$) and disc-like structures (for other n that are $<3$). 

\item  About 17 per cent of \atlas (early-type) galaxies (31 of 180 galaxies, or 12 per cent of 258 \atlas galaxies with good imaging, assuming here not analysed bars are disc related structures) do not have any evidence for discs or disc-like structures. 

\item About 41 per cent of the stellar mass of early-type galaxies is in discs or disc-like components. 

\item  Disc or disc-like components are typically found in fast rotators, while in some slow rotators the presence of exponential sub-components or single-components with low S\'ersic indices ($n<3$) could be related to structures made of more complex orbital families (with high angular momentum) allowed in non-axisymmetric potentials. These components are often related to kinematically distinct cores (KDCs). We note that one galaxy, NGC\,5485, has an exponential sub-component, but its orientation is perpendicular to the sense of rotation, and, hence, it can not be taken as an evidence for a disc. 

\item 24 of 34 (70 per cent) slow rotators are best fitted with single-components. Of these 4 have a low S\'ersic index ($<3$). Other slow rotators (10) have a substantial fraction of light in the exponential components. 

\item 93 of analysed 146 fast rotators (64 per cent) have exponential sub-components (discs). 42 of the remaining 53 fast rotators have single-components of low S\'ersic index ($<3$). There are only 11 fast rotators that do not show clear evidence for discs or disc like structures in their photometry. For some of these galaxies inclination effects could be the reason for not detecting the disc-like structures in photometry, some are recent merger remnants while rest are complex systems.

\item S\'ersic index of the bulge sub-component is smaller than 3 for 73 of 103 early-type galaxies, for which a two component fit was deemed necessary. The same is true for 70 objects if $n=2.5$ is used. It is not obvious that only secular evolution is responsible for build up of these sub-components.

\item There are trends between $D/T$ and $n_b$ with $\lambda_R$, such that for high $\lambda_R$, $D/T$ is high and $n_b$ is low, but there is no clear correlation. The S\'ersic index $n_{tot}$ from a single fit to galaxies does not correlate strongly with D/T ratio, as shown by other studies, or with  $\lambda_R$.

\item Decomposing those galaxies that require two components into discs and bulges improves the differentiation between fast and slow rotators compared to using a single component S\'ersic index. To a first approximation, it is possible to describe fast rotators as early-type galaxies with exponential discs (D/T $>0.05$) or, for single component S\'ersic fits, low $n$ ($n<3$). Similarly, slow rotators can be described as galaxies without exponential components and high $n$. We recommend this criteria when stellar kinematics is not available, but the correspondence is not 1:1, with a 7 per cent probability (11 of 146 analysed fast rotators) to miss a fast rotator and a 59 per cent probability (20 of 34 analysed slow rotators do not have disc-like components) to correctly recognise a slow rotator, implying that the decomposition can be used only as a guidance for classification. In general, kinematic analysis and classification based on the angular momentum content remains the best attempt to mitigate the influence of inclination effects. 

\item As noted previously by other authors, there is a significant dependance of photometric parameters on the inclination effects. Strong (exponential) disc signatures, however, can be seen in the light profiles even at low inclinations, while weak discs disappear sooner and are hard to detect below an inclination of $\sim50\degr$.

\item Disc dominated galaxies are typically the least massive, while galaxies with no tracers of discs are the most massive systems in the nearby Universe. Barred galaxies have a consistent distribution of mass as systems dominated by discs.

\item There is no strong relation between the environment and the amount of disc light and discs are found in all environments. At high densities there is a weak evidence that disc dominated systems are found in more denser regions than galaxies with smaller disc contributions. Barred galaxies are found at all densities, but typically in denser regions than dominant discs, and have a similar distribution like galaxies with no discs.

\end{itemize}

\vspace{+1cm}
\noindent{\bf Acknowledgements}\\

\noindent MC acknowledges support a Royal Society University Research Fellowship. MS acknowledges support from an STFC Advanced Fellowship ST/F009186/1. RMcD is supported by the Gemini Observatory, which is operated by the Association of Universities for Research in Astronomy, Inc., on behalf of the international Gemini partnership of Argentina, Australia, Brazil, Canada, Chile, the United Kingdom, and the United States of America. SK acknowledges support from the Royal Society Joint Projects Grant JP0869822. TN and MBois acknowledge support from the DFG Cluster of Excellence `Origin and Structure of the Universe'. PS is an NWO/Veni fellow. The research leading to these results has received funding from the European Community's Seventh Framework Programme (/FP7/2007-2013/) under grant agreement No 229517. This work was supported by the rolling grants `Astrophysics at Oxford' PP/E001114/1 and ST/H002456/1 and visitors grants PPA/V/S/2002/00553, PP/E001564/1 and ST/H504862/1 from the UK Research Councils. RLD acknowledges travel and computer grants from Christ Church, Oxford and support from the Royal Society in the form of a Wolfson Merit Award 502011.K502/jd. We acknowledge the usage of the HyperLeda database (http://leda.univ-lyon1.fr).This paper is based on observations obtained at the William Herschel Telescope, operated by the Isaac Newton Group in the Spanish Observatorio del Roque de los Muchachos of the Instituto de Astrof\'{\i}sica de Canarias.  Funding for the SDSS and SDSS-II was provided by the Alfred P. Sloan Foundation, the Participating Institutions, the National Science Foundation, the U.S. Department of Energy, the National Aeronautics and Space Administration, the Japanese Monbukagakusho, the Max Planck Society, and the Higher Education Funding Council for England. The SDSS was managed by the Astrophysical Research Consortium for the Participating Institutions.

%
%
\bibliographystyle{mn2e}

\begin{thebibliography}{120}
\expandafter\ifx\csname natexlab\endcsname\relax\def\natexlab#1{#1}\fi

\bibitem[{{Abazajian} {et~al}\mbox{.}(2009){Abazajian}, {Adelman-McCarthy},
  {Ag{\"u}eros}, {Allam}, {Allende Prieto}, {An}, {Anderson}, {Anderson}, \&
  {et al.}}]{2009ApJS..182..543A}
{Abazajian} K.~N. {et~al.}, 2009, \apjs, 182, 543

\bibitem[{{Aguerri} \& {Trujillo}(2002)}]{2002MNRAS.333..633A}
{Aguerri} J.~A.~L., {Trujillo} I., 2002, \mnras, 333, 633

\bibitem[{{Allen} {et~al}\mbox{.}(2006){Allen}, {Driver}, {Graham}, {Cameron},
  {Liske}, \& {de Propris}}]{2006MNRAS.371....2A}
{Allen} P.~D., {Driver} S.~P., {Graham} A.~W., {Cameron} E., {Liske} J., {de
  Propris} R., 2006, \mnras, 371, 2

\bibitem[{{Andredakis} {et~al}\mbox{.}(1995){Andredakis}, {Peletier}, \&
  {Balcells}}]{1995MNRAS.275..874A}
{Andredakis} Y.~C., {Peletier} R.~F., {Balcells} M., 1995, \mnras, 275, 874

\bibitem[{{Arnold} {et~al}\mbox{.}(1994){Arnold}, {de Zeeuw}, \&
  {Hunter}}]{1994MNRAS.271..924A}
{Arnold} R., {de Zeeuw} P.~T., {Hunter} C., 1994, \mnras, 271, 924

\bibitem[{{Athanassoula}(2005)}]{2005MNRAS.358.1477A}
{Athanassoula} E., 2005, \mnras, 358, 1477

\bibitem[{{Bacon} {et~al}\mbox{.}(2001){Bacon}, {Copin}, {Monnet}, {Miller},
  {Allington-Smith}, {Bureau}, {Carollo}, {Davies}, {Emsellem}, {Kuntschner},
  {Peletier}, {Verolme}, \& {de Zeeuw}}]{2001MNRAS.326...23B}
{Bacon} R. {et~al.}, 2001, \mnras, 326, 23

\bibitem[{{Balcells} {et~al}\mbox{.}(2003){Balcells}, {Graham},
  {Dom{\'{\i}}nguez-Palmero}, \& {Peletier}}]{2003ApJ...582L..79B}
{Balcells} M., {Graham} A.~W., {Dom{\'{\i}}nguez-Palmero} L., {Peletier} R.~F.,
  2003, \apjl, 582, L79

\bibitem[{{Barden} {et~al}\mbox{.}(2005){Barden}, {Rix}, {Somerville}, {Bell},
  {H{\"a}u{\ss}ler}, {Peng}, {Borch}, {Beckwith}, {Caldwell}, {Heymans},
  {Jahnke}, {Jogee}, {McIntosh}, {Meisenheimer}, {S{\'a}nchez}, {Wisotzki}, \&
  {Wolf}}]{2005ApJ...635..959B}
{Barden} M. {et~al.}, 2005, \apj, 635, 959

\bibitem[{{Bell} {et~al}\mbox{.}(2003){Bell}, {McIntosh}, {Katz}, \&
  {Weinberg}}]{2003ApJS..149..289B}
{Bell} E.~F., {McIntosh} D.~H., {Katz} N., {Weinberg} M.~D., 2003, \apjs, 149,
  289

\bibitem[{{Bender} {et~al}\mbox{.}(1994){Bender}, {Saglia}, \&
  {Gerhard}}]{1994MNRAS.269..785B}
{Bender} R., {Saglia} R.~P., {Gerhard} O.~E., 1994, \mnras, 269, 785

\bibitem[{{Bender} {et~al}\mbox{.}(1989){Bender}, {Surma}, {Doebereiner},
  {Moellenhoff}, \& {Madejsky}}]{1989A&A...217...35B}
{Bender} R., {Surma} P., {Doebereiner} S., {Moellenhoff} C., {Madejsky} R.,
  1989, \aap, 217, 35

\bibitem[{{Benson} {et~al}\mbox{.}(2007){Benson}, {D{\v z}anovi{\'c}}, {Frenk},
  \& {Sharples}}]{2007MNRAS.379..841B}
{Benson} A.~J., {D{\v z}anovi{\'c}} D., {Frenk} C.~S., {Sharples} R., 2007,
  \mnras, 379, 841

\bibitem[{{Binney} \& {Merrifield}(1998)}]{1998gaas.book.....B}
{Binney} J., {Merrifield} M., 1998, {Galactic astronomy}. Princeton, NJ,
  Princeton University Press

\bibitem[{{Blanton} {et~al}\mbox{.}(2003){Blanton}, {Hogg}, {Bahcall},
  {Baldry}, {Brinkmann}, {Csabai}, {Eisenstein}, \& {et
  al.}}]{2003ApJ...594..186B}
{Blanton} M.~R., {Hogg} D.~W., {Bahcall} N.~A., {Baldry} I.~K., {Brinkmann} J.,
  {Csabai} I., {Eisenstein} D., {et al.}, 2003, \apj, 594, 186

\bibitem[{{Bois} {et~al}\mbox{.}(2011){Bois}, {Emsellem}, {Bournaud},
  {Alatalo}, {Blitz}, {Bureau}, {Cappellari}, {Davies}, \& {et
  al.}}]{2011MNRAS.416.1654B}
{Bois} M. {et~al.}, 2011, \mnras, 416, 1654, Paper VI

\bibitem[{{Boroson}(1981)}]{1981ApJS...46..177B}
{Boroson} T., 1981, \apjs, 46, 177

\bibitem[{{Burstein}(1979)}]{1979ApJ...234..435B}
{Burstein} D., 1979, \apj, 234, 435

\bibitem[{{Byun} \& {Freeman}(1995)}]{1995ApJ...448..563B}
{Byun} Y.~I., {Freeman} K.~C., 1995, \apj, 448, 563

\bibitem[{{Caon} {et~al}\mbox{.}(1993){Caon}, {Capaccioli}, \&
  {D'Onofrio}}]{1993MNRAS.265.1013C}
{Caon} N., {Capaccioli} M., {D'Onofrio} M., 1993, \mnras, 265, 1013

\bibitem[{{Cappellari}(2002)}]{2002MNRAS.333..400C}
{Cappellari} M., 2002, \mnras, 333, 400

\bibitem[{{Cappellari}(2008)}]{2008MNRAS.390...71C}
{Cappellari} M., 2008, \mnras, 390, 71

\bibitem[{{Cappellari} {et~al}\mbox{.}(2007){Cappellari}, {Emsellem}, {Bacon},
  {Bureau}, {Davies}, {de Zeeuw}, {Falc{\'o}n-Barroso}, {Krajnovi{\'c}},
  {Kuntschner}, {McDermid}, {Peletier}, {Sarzi}, {van den Bosch}, \& {van de
  Ven}}]{2007MNRAS.379..418C}
{Cappellari} M. {et~al.}, 2007, \mnras, 379, 418

\bibitem[{{Cappellari} {et~al}\mbox{.}(2011{\natexlab{a}}){Cappellari},
  {Emsellem}, {Krajnovi{\'c}}, {McDermid}, {Scott}, {Verdoes Kleijn}, {Young},
  {Alatalo}, \& {et al.}}]{2011MNRAS.413..813C}
{Cappellari} M. {et~al.}, 2011{\natexlab{a}}, \mnras, 413, 813, Paper I

\bibitem[{{Cappellari} {et~al}\mbox{.}(2011{\natexlab{b}}){Cappellari},
  {Emsellem}, {Krajnovi{\'c}}, {McDermid}, {Serra}, {Alatalo}, {Blitz}, {Bois},
  \& {et al.}}]{2011MNRAS.416.1680C}
{Cappellari} M. {et~al.}, 2011{\natexlab{b}}, \mnras, 416, 1680, Paper VII

\bibitem[{{Cappellari} {et~al}\mbox{.}(2012{\natexlab{a}}){Cappellari},
  {McDermid}, {Alatalo}, {Blitz}, {Bois}, {Bournaud}, {Bureau}, {Crocker},
  {Davies}, {Davis}, {de Zeeuw}, {Duc}, {Khochfar}, {Krajnovic}, {Kuntschner},
  {Morganti}, {Naab}, {Oosterloo}, {Sarzi}, {Scott}, {Serra}, {Weijmans}, \&
  {Young}}]{2012arXiv1208.3523C}
{Cappellari} M. {et~al.}, 2012{\natexlab{a}}, ArXiv:1208.3523, Paper XX

\bibitem[{{Cappellari} {et~al}\mbox{.}(2012{\natexlab{b}}){Cappellari},
  {Scott}, {Alatalo}, {Blitz}, {Bois}, {Bournaud}, {Bureau}, {Crocker},
  {Davies}, {Davis}, {de Zeeuw}, {Duc}, {Khochfar}, {Krajnovic}, {Kuntschner},
  {McDermid}, {Morganti}, {Naab}, {Oosterloo}, {Sarzi}, {Serra}, {Weijmans}, \&
  {Young}}]{2012arXiv1208.3522C}
{Cappellari} M. {et~al.}, 2012{\natexlab{b}}, ArXiv:1208.3522, Paper XIX

\bibitem[{{Chen} {et~al}\mbox{.}(2010){Chen}, {C{\^o}t{\'e}}, {West}, {Peng},
  \& {Ferrarese}}]{2010ApJS..191....1C}
{Chen} C.-W., {C{\^o}t{\'e}} P., {West} A.~A., {Peng} E.~W., {Ferrarese} L.,
  2010, \apjs, 191, 1

\bibitem[{{Ciotti}(1991)}]{1991A&A...249...99C}
{Ciotti} L., 1991, \aap, 249, 99

\bibitem[{{Ciotti} \& {Bertin}(1999)}]{1999A&A...352..447C}
{Ciotti} L., {Bertin} G., 1999, \aap, 352, 447

\bibitem[{{Coccato} {et~al}\mbox{.}(2011){Coccato}, {Morelli}, {Corsini},
  {Buson}, {Pizzella}, {Vergani}, \& {Bertola}}]{2011MNRAS.412L.113C}
{Coccato} L., {Morelli} L., {Corsini} E.~M., {Buson} L., {Pizzella} A.,
  {Vergani} D., {Bertola} F., 2011, \mnras, 412, L113

\bibitem[{{C{\^o}t{\'e}} {et~al}\mbox{.}(2004){C{\^o}t{\'e}}, {Blakeslee},
  {Ferrarese}, {Jord{\'a}n}, {Mei}, {Merritt}, {Milosavljevi{\'c}}, {Peng},
  {Tonry}, \& {West}}]{2004ApJS..153..223C}
{C{\^o}t{\'e}} P. {et~al.}, 2004, \apjs, 153, 223

\bibitem[{{Courteau} {et~al}\mbox{.}(1996){Courteau}, {de Jong}, \&
  {Broeils}}]{1996ApJ...457L..73C}
{Courteau} S., {de Jong} R.~S., {Broeils} A.~H., 1996, \apjl, 457, L73

\bibitem[{{Davies} {et~al}\mbox{.}(1983){Davies}, {Efstathiou}, {Fall},
  {Illingworth}, \& {Schechter}}]{1983ApJ...266...41D}
{Davies} R.~L., {Efstathiou} G., {Fall} S.~M., {Illingworth} G., {Schechter}
  P.~L., 1983, \apj, 266, 41

\bibitem[{{de Jong}(1996)}]{1996A&AS..118..557D}
{de Jong} R.~S., 1996, \aaps, 118, 557

\bibitem[{{de Jong} {et~al}\mbox{.}(2004){de Jong}, {Simard}, {Davies},
  {Saglia}, {Burstein}, {Colless}, {McMahan}, \&
  {Wegner}}]{2004MNRAS.355.1155D}
{de Jong} R.~S., {Simard} L., {Davies} R.~L., {Saglia} R.~P., {Burstein} D.,
  {Colless} M., {McMahan} R., {Wegner} G., 2004, \mnras, 355, 1155

\bibitem[{{de Souza} {et~al}\mbox{.}(2004){de Souza}, {Gadotti}, \& {dos
  Anjos}}]{2004ApJS..153..411D}
{de Souza} R.~E., {Gadotti} D.~A., {dos Anjos} S., 2004, \apjs, 153, 411

\bibitem[{{de Vaucouleurs}(1959)}]{1959HDP....53..311D}
{de Vaucouleurs} G., 1959, Handbuch der Physik, 53, 311

\bibitem[{{de Zeeuw} {et~al}\mbox{.}(2002){de Zeeuw}, {Bureau}, {Emsellem},
  {Bacon}, {Carollo}, {Copin}, {Davies}, {Kuntschner}, {Miller}, {Monnet},
  {Peletier}, \& {Verolme}}]{2002MNRAS.329..513D}
{de Zeeuw} P.~T. {et~al.}, 2002, \mnras, 329, 513

\bibitem[{{D'Onofrio}(2001)}]{2001MNRAS.326.1517D}
{D'Onofrio} M., 2001, \mnras, 326, 1517

\bibitem[{{Duc} {et~al}\mbox{.}(2011){Duc}, {Cuillandre}, {Serra},
  {Michel-Dansac}, {Ferriere}, {Alatalo}, {Blitz}, {Bois}, \& {et
  al.}}]{2011MNRAS.417..863D}
{Duc} P.-A. {et~al.}, 2011, \mnras, 417, 863, Paper IX

\bibitem[{{Emsellem} {et~al}\mbox{.}(2011){Emsellem}, {Cappellari},
  {Krajnovi{\'c}}, {Alatalo}, {Blitz}, {Bois}, {Bournaud}, {Bureau}, \& {et
  al.}}]{2011MNRAS.414..888E}
{Emsellem} E. {et~al.}, 2011, \mnras, 414, 888, Paper III

\bibitem[{{Emsellem} {et~al}\mbox{.}(2007){Emsellem}, {Cappellari},
  {Krajnovi{\'c}}, {van de Ven}, {Bacon}, {Bureau}, {Davies}, {de Zeeuw},
  {Falc{\'o}n-Barroso}, {Kuntschner}, {McDermid}, {Peletier}, \&
  {Sarzi}}]{2007MNRAS.379..401E}
{Emsellem} E. {et~al.}, 2007, \mnras, 379, 401

\bibitem[{{Emsellem} {et~al}\mbox{.}(2004){Emsellem}, {Cappellari}, {Peletier},
  {McDermid}, {Geacon}, {Bureau}, {Copin}, {Davies}, {Krajnovi{\' c}},
  {Kuntschner}, {Miller}, \& {de Zeeuw}}]{2004MNRAS.352..721E}
{Emsellem} E. {et~al.}, 2004, \mnras, 352, 721

\bibitem[{{Emsellem} {et~al}\mbox{.}(1994){Emsellem}, {Monnet}, \&
  {Bacon}}]{1994A&A...285..723E}
{Emsellem} E., {Monnet} G., {Bacon} R., 1994, \aap, 285, 723

\bibitem[{{Erwin} {et~al}\mbox{.}(2008){Erwin}, {Pohlen}, \&
  {Beckman}}]{2008AJ....135...20E}
{Erwin} P., {Pohlen} M., {Beckman} J.~E., 2008, \aj, 135, 20

\bibitem[{{Faber} {et~al}\mbox{.}(1997){Faber}, {Tremaine}, {Ajhar}, {Byun},
  {Dressler}, {Gebhardt}, {Grillmair}, {Kormendy}, {Lauer}, \&
  {Richstone}}]{1997AJ....114.1771F}
{Faber} S.~M. {et~al.}, 1997, \aj, 114, 1771

\bibitem[{{Fabricius} {et~al}\mbox{.}(2012){Fabricius}, {Saglia}, {Fisher},
  {Drory}, {Bender}, \& {Hopp}}]{2012arXiv1204.5188F}
{Fabricius} M.~H., {Saglia} R.~P., {Fisher} D.~B., {Drory} N., {Bender} R.,
  {Hopp} U., 2012, ArXiv e-prints

\bibitem[{{Ferrarese} {et~al}\mbox{.}(2006){Ferrarese}, {C{\^o}t{\'e}},
  {Jord{\'a}n}, {Peng}, {Blakeslee}, {Piatek}, {Mei}, {Merritt},
  {Milosavljevi{\'c}}, {Tonry}, \& {West}}]{2006ApJS..164..334F}
{Ferrarese} L. {et~al.}, 2006, \apjs, 164, 334

\bibitem[{{Ferrarese} {et~al}\mbox{.}(1994){Ferrarese}, {van den Bosch},
  {Ford}, {Jaffe}, \& {O'Connell}}]{1994AJ....108.1598F}
{Ferrarese} L., {van den Bosch} F.~C., {Ford} H.~C., {Jaffe} W., {O'Connell}
  R.~W., 1994, \aj, 108, 1598

\bibitem[{{Fisher} \& {Drory}(2008)}]{2008AJ....136..773F}
{Fisher} D.~B., {Drory} N., 2008, \aj, 136, 773

\bibitem[{{Freeman}(1970)}]{1970ApJ...160..811F}
{Freeman} K.~C., 1970, \apj, 160, 811

\bibitem[{{Gadotti}(2008)}]{2008MNRAS.384..420G}
{Gadotti} D.~A., 2008, \mnras, 384, 420

\bibitem[{{Gadotti}(2009)}]{2009MNRAS.393.1531G}
{Gadotti} D.~A., 2009, \mnras, 393, 1531

\bibitem[{{Gadotti} \& {S{\'a}nchez-Janssen}(2012)}]{2012MNRAS.423..877G}
{Gadotti} D.~A., {S{\'a}nchez-Janssen} R., 2012, \mnras, 423, 877

\bibitem[{{Gerhard}(1993)}]{1993MNRAS.265..213G}
{Gerhard} O.~E., 1993, \mnras, 265, 213

\bibitem[{{Gerhard} \& {Binney}(1996)}]{1996MNRAS.279..993G}
{Gerhard} O.~E., {Binney} J.~J., 1996, \mnras, 279, 993

\bibitem[{{Gradshteyn} {et~al}\mbox{.}(2000){Gradshteyn}, {Ryzhik}, {Jeffrey},
  \& {Zwillinger}}]{2000tisp.book.....G}
{Gradshteyn} I.~S., {Ryzhik} I.~M., {Jeffrey} A., {Zwillinger} D., 2000, {Table
  of Integrals, Series, and Products}

\bibitem[{{Graham} {et~al}\mbox{.}(1996){Graham}, {Lauer}, {Colless}, \&
  {Postman}}]{1996ApJ...465..534G}
{Graham} A., {Lauer} T.~R., {Colless} M., {Postman} M., 1996, \apj, 465, 534

\bibitem[{{Graham}(2001)}]{2001AJ....121..820G}
{Graham} A.~W., 2001, \aj, 121, 820

\bibitem[{{Graham} \& {Driver}(2005)}]{2005PASA...22..118G}
{Graham} A.~W., {Driver} S.~P., 2005, \pasa, 22, 118

\bibitem[{{Graham} {et~al}\mbox{.}(2003){Graham}, {Erwin}, {Trujillo}, \&
  {Asensio Ramos}}]{2003AJ....125.2951G}
{Graham} A.~W., {Erwin} P., {Trujillo} I., {Asensio Ramos} A., 2003, \aj, 125,
  2951

\bibitem[{{Hoyos} {et~al}\mbox{.}(2011){Hoyos}, {den Brok}, {Verdoes Kleijn},
  {Carter}, {Balcells}, {Guzm{\'a}n}, {Peletier}, \& {et
  al.}}]{2011MNRAS.411.2439H}
{Hoyos} C., {den Brok} M., {Verdoes Kleijn} G., {Carter} D., {Balcells} M.,
  {Guzm{\'a}n} R., {Peletier} R., {et al.}, 2011, \mnras, 411, 2439

\bibitem[{{Hubble}(1922)}]{1922ApJ....56..162H}
{Hubble} E.~P., 1922, \apj, 56, 162

\bibitem[{{Hubble}(1926)}]{1926ApJ....64..321H}
{Hubble} E.~P., 1926, \apj, 64, 321

\bibitem[{{Hubble}(1936)}]{1936RNeb..........H}
{Hubble} E.~P., 1936, Yale University Press

\bibitem[{{Jeans}(1929)}]{1929Jeans}
{Jeans} J.~H., 1929, Cambridge University Press

\bibitem[{{Jedrzejewski}(1987)}]{1987MNRAS.226..747J}
{Jedrzejewski} R.~I., 1987, \mnras, 226, 747

\bibitem[{{Johnston} {et~al}\mbox{.}(2012){Johnston}, {Arag{\'o}n-Salamanca},
  {Merrifield}, \& {Bedregal}}]{2012MNRAS.422.2590J}
{Johnston} E.~J., {Arag{\'o}n-Salamanca} A., {Merrifield} M.~R., {Bedregal}
  A.~G., 2012, \mnras, 422, 2590

\bibitem[{{Jorgensen} \& {Franx}(1994)}]{1994ApJ...433..553J}
{Jorgensen} I., {Franx} M., 1994, \apj, 433, 553

\bibitem[{{Kent}(1985)}]{1985ApJS...59..115K}
{Kent} S.~M., 1985, \apjs, 59, 115

\bibitem[{{Khochfar} {et~al}\mbox{.}(2011){Khochfar}, {Emsellem}, {Serra},
  {Bois}, {Alatalo}, {Bacon}, {Blitz}, {Bournaud}, \& {et
  al.}}]{2011MNRAS.417..845K}
{Khochfar} S. {et~al.}, 2011, \mnras, 417, 845, Paper VIII

\bibitem[{{Kormendy}(1977)}]{1977ApJ...217..406K}
{Kormendy} J., 1977, \apj, 217, 406

\bibitem[{{Kormendy}(1993)}]{1993IAUS..153..209K}
{Kormendy} J., 1993, in IAU Symposium, Vol. 153, Galactic Bulges, {Dejonghe}
  H., {Habing} H.~J., eds., p. 209

\bibitem[{{Kormendy} \& {Bender}(2012)}]{2012ApJS..198....2K}
{Kormendy} J., {Bender} R., 2012, \apjs, 198, 2

\bibitem[{{Kormendy} {et~al}\mbox{.}(2009){Kormendy}, {Fisher}, {Cornell}, \&
  {Bender}}]{2009ApJS..182..216K}
{Kormendy} J., {Fisher} D.~B., {Cornell} M.~E., {Bender} R., 2009, \apjs, 182,
  216

\bibitem[{{Kormendy} \& {Kennicutt}(2004)}]{2004ARA&A..42..603K}
{Kormendy} J., {Kennicutt}, Jr. R.~C., 2004, \araa, 42, 603

\bibitem[{{Krajnovi{\'c}} {et~al}\mbox{.}(2008){Krajnovi{\'c}}, {Bacon},
  {Cappellari}, {Davies}, {de Zeeuw}, {Emsellem}, {Falc{\'o}n-Barroso},
  {Kuntschner}, {McDermid}, {Peletier}, {Sarzi}, {van den Bosch}, \& {van de
  Ven}}]{2008MNRAS.390...93K}
{Krajnovi{\'c}} D. {et~al.}, 2008, \mnras, 390, 93

\bibitem[{{Krajnovi{\'c}} {et~al}\mbox{.}(2006){Krajnovi{\'c}}, {Cappellari},
  {de Zeeuw}, \& {Copin}}]{2006MNRAS.366..787K}
{Krajnovi{\'c}} D., {Cappellari} M., {de Zeeuw} P.~T., {Copin} Y., 2006,
  \mnras, 366, 787

\bibitem[{{Krajnovi{\'c}} {et~al}\mbox{.}(2011){Krajnovi{\'c}}, {Emsellem},
  {Cappellari}, {Alatalo}, {Blitz}, {Bois}, {Bournaud}, {Bureau}, \& {et
  al.}}]{2011MNRAS.414.2923K}
{Krajnovi{\'c}} D. {et~al.}, 2011, \mnras, 414, 2923, Paper II

\bibitem[{{Lackner} \& {Gunn}(2012)}]{2012MNRAS.421.2277L}
{Lackner} C.~N., {Gunn} J.~E., 2012, \mnras, 421, 2277

\bibitem[{{Landsman}(1993)}]{1993adass...2..246L}
{Landsman} W.~B., 1993, in ASP Conf. Ser. 52: Astronomical Data Analysis
  Software and Systems II, pp. 246--+

\bibitem[{{Laurikainen} {et~al}\mbox{.}(2007){Laurikainen}, {Salo}, {Buta}, \&
  {Knapen}}]{2007MNRAS.381..401L}
{Laurikainen} E., {Salo} H., {Buta} R., {Knapen} J.~H., 2007, \mnras, 381, 401

\bibitem[{{Laurikainen} {et~al}\mbox{.}(2009){Laurikainen}, {Salo}, {Buta}, \&
  {Knapen}}]{2009ApJ...692L..34L}
{Laurikainen} E., {Salo} H., {Buta} R., {Knapen} J.~H., 2009, \apjl, 692, L34

\bibitem[{{Laurikainen} {et~al}\mbox{.}(2011){Laurikainen}, {Salo}, {Buta}, \&
  {Knapen}}]{2011AdAst2011E..18L}
{Laurikainen} E., {Salo} H., {Buta} R., {Knapen} J.~H., 2011, Advances in
  Astronomy, 2011

\bibitem[{{Laurikainen} {et~al}\mbox{.}(2010){Laurikainen}, {Salo}, {Buta},
  {Knapen}, \& {Comer{\'o}n}}]{2010MNRAS.405.1089L}
{Laurikainen} E., {Salo} H., {Buta} R., {Knapen} J.~H., {Comer{\'o}n} S., 2010,
  \mnras, 405, 1089

\bibitem[{{MacArthur} {et~al}\mbox{.}(2003){MacArthur}, {Courteau}, \&
  {Holtzman}}]{2003ApJ...582..689M}
{MacArthur} L.~A., {Courteau} S., {Holtzman} J.~A., 2003, \apj, 582, 689

\bibitem[{{Markwardt}(2009)}]{2009ASPC..411..251M}
{Markwardt} C.~B., 2009, in Astronomical Society of the Pacific Conference
  Series, Vol. 411, Astronomical Society of the Pacific Conference Series,
  {D.~A.~Bohlender, D.~Durand, \& P.~Dowler}, ed., pp. 251--+

\bibitem[{{McIntosh} {et~al}\mbox{.}(2005){McIntosh}, {Bell}, {Rix}, {Wolf},
  {Heymans}, {Peng}, {Somerville}, {Barden}, {Beckwith}, {Borch}, {Caldwell},
  {H{\"a}u{\ss}ler}, {Jahnke}, {Jogee}, {Meisenheimer}, {S{\'a}nchez}, \&
  {Wisotzki}}]{2005ApJ...632..191M}
{McIntosh} D.~H. {et~al.}, 2005, \apj, 632, 191

\bibitem[{{M{\'e}ndez-Abreu} {et~al}\mbox{.}(2008){M{\'e}ndez-Abreu},
  {Aguerri}, {Corsini}, \& {Simonneau}}]{2008A&A...478..353M}
{M{\'e}ndez-Abreu} J., {Aguerri} J.~A.~L., {Corsini} E.~M., {Simonneau} E.,
  2008, \aap, 478, 353

\bibitem[{{Monnet} {et~al}\mbox{.}(1992){Monnet}, {Bacon}, \&
  {Emsellem}}]{1992A&A...253..366M}
{Monnet} G., {Bacon} R., {Emsellem} E., 1992, \aap, 253, 366

\bibitem[{{Mor\'e} {et~al}\mbox{.}(1980){Mor\'e}, {Garbow}, \&
  {Hillstrom}}]{1980ANL.....80.74M}
{Mor\'e} J.~J., {Garbow} B.~S., {Hillstrom} K.~E., 1980, User Guide for
  MINPACK-1, Argonne National Laboratory Report ANL-80-74, 74

\bibitem[{{Naab} \& {Trujillo}(2006)}]{2006MNRAS.369..625N}
{Naab} T., {Trujillo} I., 2006, \mnras, 369, 625

\bibitem[{{Paturel} {et~al}\mbox{.}(2003){Paturel}, {Petit}, {Prugniel},
  {Theureau}, {Rousseau}, {Brouty}, {Dubois}, \&
  {Cambr{\'e}sy}}]{2003A&A...412...45P}
{Paturel} G., {Petit} C., {Prugniel} P., {Theureau} G., {Rousseau} J., {Brouty}
  M., {Dubois} P., {Cambr{\'e}sy} L., 2003, \aap, 412, 45

\bibitem[{{Peng} {et~al}\mbox{.}(2002){Peng}, {Ho}, {Impey}, \&
  {Rix}}]{2002AJ....124..266P}
{Peng} C.~Y., {Ho} L.~C., {Impey} C.~D., {Rix} H.-W., 2002, \aj, 124, 266

\bibitem[{{Pignatelli} {et~al}\mbox{.}(2006){Pignatelli}, {Fasano}, \&
  {Cassata}}]{2006A&A...446..373P}
{Pignatelli} E., {Fasano} G., {Cassata} P., 2006, \aap, 446, 373

\bibitem[{{Reynolds}(1920)}]{1920MNRAS..80..746R}
{Reynolds} J.~H., 1920, \mnras, 80, 746

\bibitem[{{Rix} \& {White}(1990)}]{1990ApJ...362...52R}
{Rix} H., {White} S.~D.~M., 1990, \apj, 362, 52

\bibitem[{{Rix} {et~al}\mbox{.}(1992){Rix}, {Franx}, {Fisher}, \&
  {Illingworth}}]{1992ApJ...400L...5R}
{Rix} H.-W., {Franx} M., {Fisher} D., {Illingworth} G., 1992, \apjl, 400, L5

\bibitem[{{Rubin} {et~al}\mbox{.}(1992){Rubin}, {Graham}, \&
  {Kenney}}]{1992ApJ...394L...9R}
{Rubin} V.~C., {Graham} J.~A., {Kenney} J.~D.~P., 1992, \apjl, 394, L9

\bibitem[{{Saglia} {et~al}\mbox{.}(1997){Saglia}, {Bertschinger}, {Baggley},
  {Burstein}, {Colless}, {Davies}, {McMahan}, \&
  {Wegner}}]{1997ApJS..109...79S}
{Saglia} R.~P., {Bertschinger} E., {Baggley} G., {Burstein} D., {Colless} M.,
  {Davies} R.~L., {McMahan}, Jr. R.~K., {Wegner} G., 1997, \apjs, 109, 79

\bibitem[{{Sandage}(2005)}]{2005ARA&A..43..581S}
{Sandage} A., 2005, \araa, 43, 581

\bibitem[{{Sandage} {et~al}\mbox{.}(1970){Sandage}, {Freeman}, \&
  {Stokes}}]{1970ApJ...160..831S}
{Sandage} A., {Freeman} K.~C., {Stokes} N.~R., 1970, \apj, 160, 831

\bibitem[{{Scarlata} {et~al}\mbox{.}(2007){Scarlata}, {Carollo}, {Lilly},
  {Sargent}, {Feldmann}, {Kampczyk}, {Porciani}, \& {et
  al.}}]{2007ApJS..172..406S}
{Scarlata} C., {Carollo} C.~M., {Lilly} S., {Sargent} M.~T., {Feldmann} R.,
  {Kampczyk} P., {Porciani} C., {et al.}, 2007, \apjs, 172, 406

\bibitem[{{Schombert} \& {Bothun}(1987)}]{1987AJ.....93...60S}
{Schombert} J.~M., {Bothun} G.~D., 1987, \aj, 93, 60

\bibitem[{{Scorza} {et~al}\mbox{.}(1998){Scorza}, {Bender}, {Winkelmann},
  {Capaccioli}, \& {Macchetto}}]{1998A&AS..131..265S}
{Scorza} C., {Bender} R., {Winkelmann} C., {Capaccioli} M., {Macchetto} D.~F.,
  1998, \aaps, 131, 265

\bibitem[{{Scott} {et~al}\mbox{.}(2012){Scott}, {Alatalo}, {Blitz}, {Bois},
  {Bournaud}, {Bureau}, {Crocker}, {Davies}, {Davis}, {de Zeeuw}, {Duc},
  {Emsellem}, {Khochfar}, {Krajnovi{\'c}}, {Kuntschner}, {McDermid},
  {Morganti}, {Naab}, {Oosterloo}, {Sarzi}, {Serra}, {Weijmans}, \&
  {Young}}]{2012Scott}
{Scott} N. {et~al.}, 2012, in preparation, Paper XXI

\bibitem[{{Serra} {et~al}\mbox{.}(2012){Serra}, {Oosterloo}, {Morganti},
  {Alatalo}, {Blitz}, {Bois}, {Bournaud}, {Bureau}, , \& {et
  al.}}]{2012MNRAS.422.1835S}
{Serra} P. {et~al.}, 2012, \mnras, 422, 1835, Paper XIII

\bibitem[{{S\'ersic}(1968)}]{1968adga.book.....S}
{S\'ersic} J.~L., 1968, {Atlas de galaxias australes}. Cordoba, Argentina:
  Observatorio Astronomico, 1968

\bibitem[{{Shen} {et~al}\mbox{.}(2003){Shen}, {Mo}, {White}, {Blanton},
  {Kauffmann}, {Voges}, {Brinkmann}, \& {Csabai}}]{2003MNRAS.343..978S}
{Shen} S., {Mo} H.~J., {White} S.~D.~M., {Blanton} M.~R., {Kauffmann} G.,
  {Voges} W., {Brinkmann} J., {Csabai} I., 2003, \mnras, 343, 978

\bibitem[{{Simard} {et~al}\mbox{.}(2009){Simard}, {Clowe}, {Desai},
  {Dalcanton}, {von der Linden}, {Poggianti}, {White}, \& {et
  al.}}]{2009A&A...508.1141S}
{Simard} L., {Clowe} D., {Desai} V., {Dalcanton} J.~J., {von der Linden} A.,
  {Poggianti} B.~M., {White} S.~D.~M., {et al.}, 2009, \aap, 508, 1141

\bibitem[{{Simard} {et~al}\mbox{.}(2011){Simard}, {Mendel}, {Patton},
  {Ellison}, \& {McConnachie}}]{2011ApJS..196...11S}
{Simard} L., {Mendel} J.~T., {Patton} D.~R., {Ellison} S.~L., {McConnachie}
  A.~W., 2011, \apjs, 196, 11

\bibitem[{{Simard} {et~al}\mbox{.}(2002){Simard}, {Willmer}, {Vogt},
  {Sarajedini}, {Phillips}, {Weiner}, {Koo}, {Im}, {Illingworth}, \&
  {Faber}}]{2002ApJS..142....1S}
{Simard} L. {et~al.}, 2002, \apjs, 142, 1

\bibitem[{{Spitzer} \& {Baade}(1951)}]{1951ApJ...113..413S}
{Spitzer}, Jr. L., {Baade} W., 1951, \apj, 113, 413

\bibitem[{{Statler}(1991)}]{1991AJ....102..882S}
{Statler} T.~S., 1991, \aj, 102, 882

\bibitem[{{Trujillo} {et~al}\mbox{.}(2004){Trujillo}, {Erwin}, {Asensio Ramos},
  \& {Graham}}]{2004AJ....127.1917T}
{Trujillo} I., {Erwin} P., {Asensio Ramos} A., {Graham} A.~W., 2004, \aj, 127,
  1917

\bibitem[{{van den Bergh}(1976)}]{1976ApJ...206..883V}
{van den Bergh} S., 1976, \apj, 206, 883

\bibitem[{{van der Marel} \& {Franx}(1993)}]{1993ApJ...407..525V}
{van der Marel} R.~P., {Franx} M., 1993, \apj, 407, 525

\bibitem[{{Weinzirl} {et~al}\mbox{.}(2009){Weinzirl}, {Jogee}, {Khochfar},
  {Burkert}, \& {Kormendy}}]{2009ApJ...696..411W}
{Weinzirl} T., {Jogee} S., {Khochfar} S., {Burkert} A., {Kormendy} J., 2009,
  \apj, 696, 411

\bibitem[{{Yoon} {et~al}\mbox{.}(2011){Yoon}, {Weinberg}, \&
  {Katz}}]{2011MNRAS.414.1625Y}
{Yoon} I., {Weinberg} M.~D., {Katz} N., 2011, \mnras, 414, 1625

\end{thebibliography}

%
%

\appendix

\section{Choosing the fitting method}
\label{A:choice}

\begin{figure*}
  	      \includegraphics[width=0.35\textwidth]{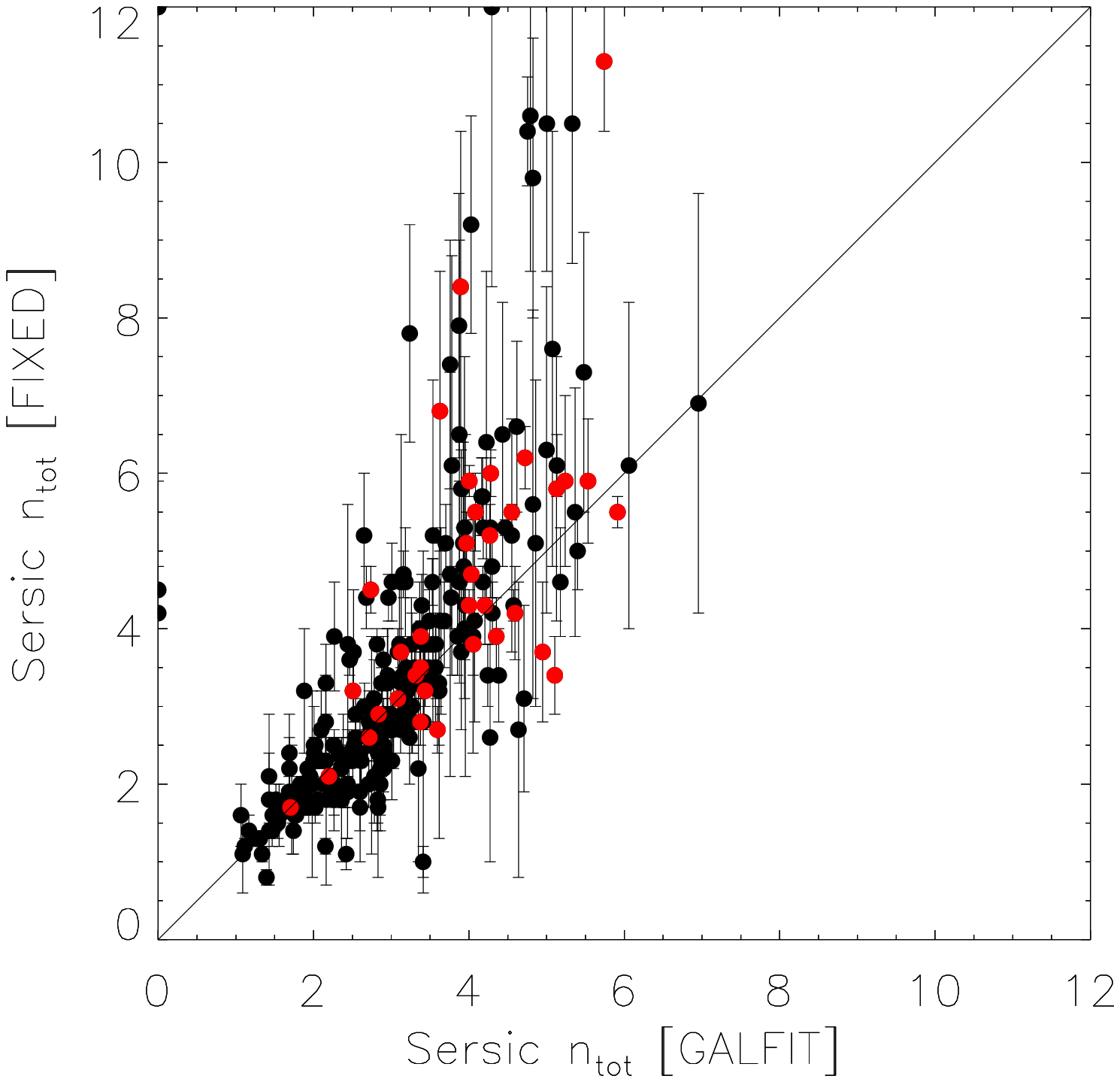}
 	      \includegraphics[width=0.35\textwidth]{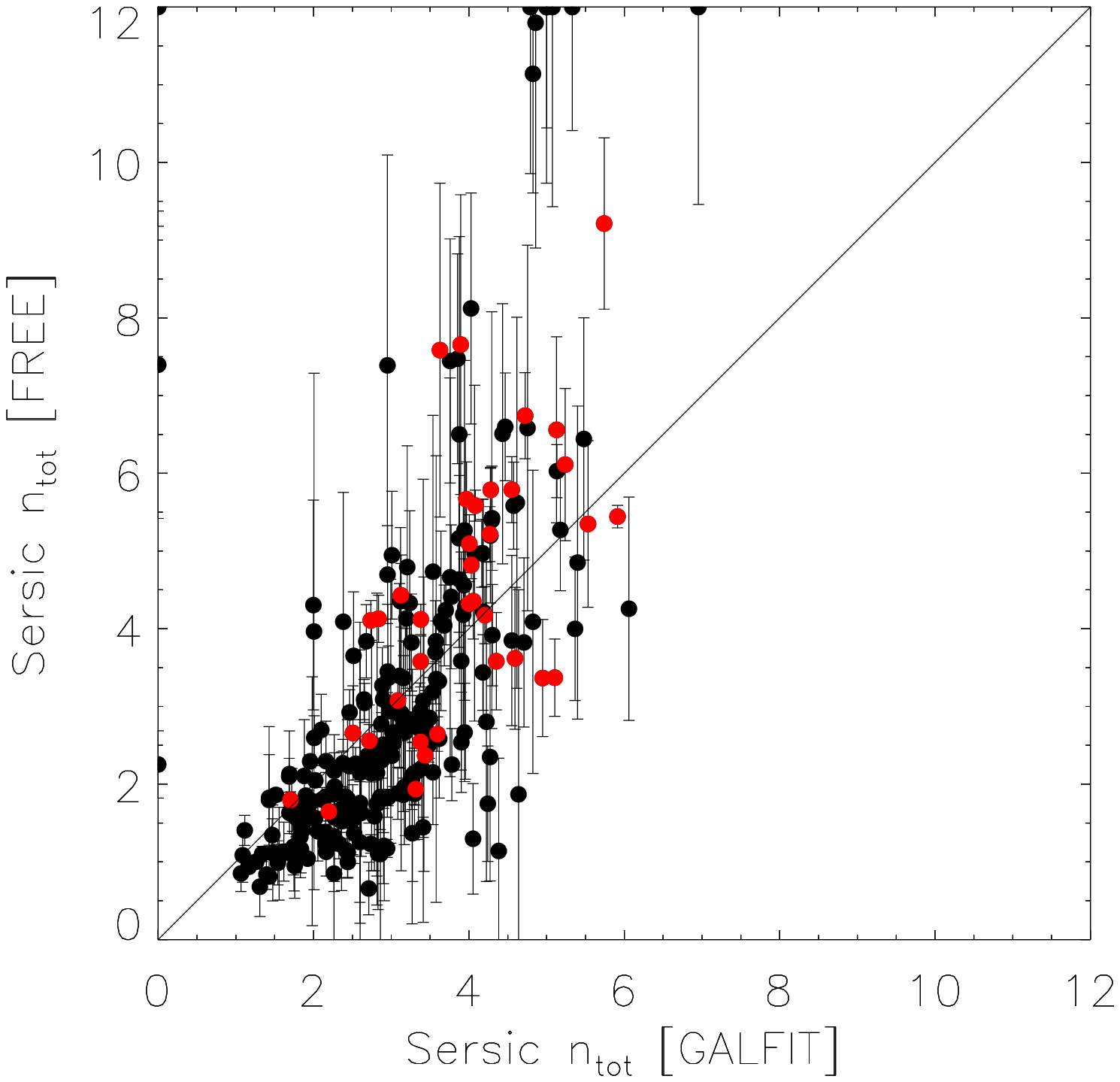}
                \includegraphics[width=0.35\textwidth]{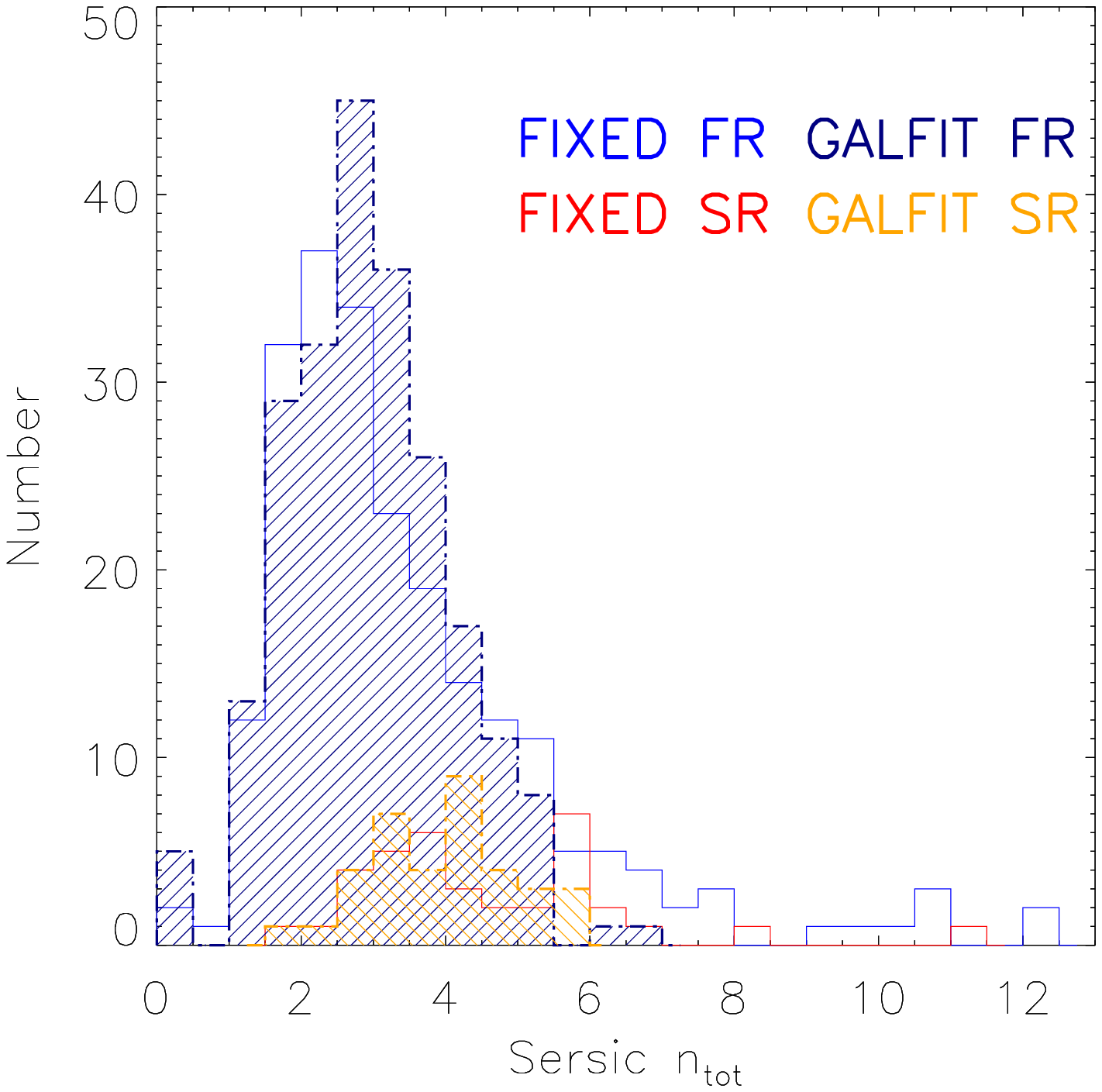}
                \includegraphics[width=0.35\textwidth]{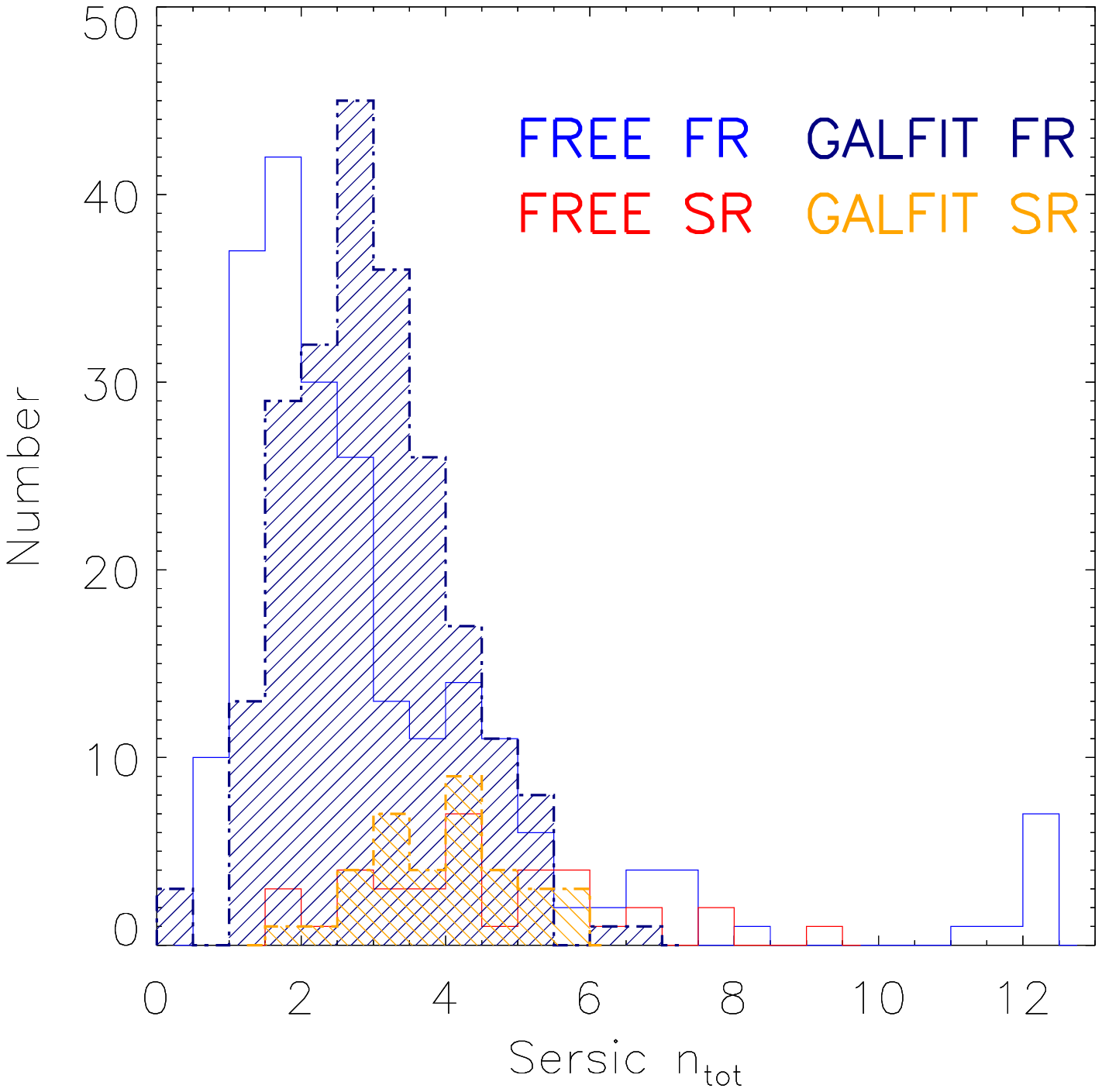}
\caption{\label{f:galfit1} Comparison of S\'ersic indices using our method (in 1D) and GALFIT (in 2D) on \atlas galaxies. {\it Left:} comparison of GALFIT results and single S\'ersic component fits to light profiles obtained by azimuthally averaging along fixed ellipses (described as FIXED in the legend). {\it Right:} comparison of GALFIT results and single S\'ersic component fits to light profiles obtained by azimuthally averaging along free ellipses (described as FREE in the legend). Top row panels show a direct comparison for objects, while bottom row panels show histograms of respective distributions. On top panels, slow rotators are shown with red symbols.  On bottom panels, GALFIT results are shown with hatched histograms, while 1D results with open histograms, and colours relate to the separation into fast and slow rotators, as shown in the legend. }
\end{figure*}

As mentioned in Section~\ref{ss:1or2}, there are various methods which can be used to parametrically describe a light distribution of a galaxy. The availability of computing power made techniques working in two-dimension (2D) widely used in the recent years, which are especially better suited for working with spatially poorly resolved galaxies at higher redshifts. Our method of choice, however, was to fit one dimensional (1D) light profiles obtained by azimuthally averaging along ellipses, because this approach allowed for a uniform and a systematic treatment of early-type galaxies with and without discs. In particular, in the case of one component fits we used profiles azimuthally averaged along ellipses with fixed position angle and flattening, while in the case of two component decomposition we used profiles azimuthally averaged along best fitting ellipses, where the ellipse fitting program was allowed to vary the position angle and flattening of the ellipses. 

There are, however, different approaches with regard to what is the best suited 1D light profile for the decomposition. For example, one could take major axis cuts \citep[e.g][]{1977ApJ...217..406K, 1979ApJ...234..435B, 2008AJ....136..773F, 2012ApJS..198....2K}, major and minor axis cuts \citep[e.g][]{1985ApJS...59..115K,1996A&AS..118..557D} or azimuthally averaged light profiles \citep[e.g][]{1981ApJS...46..177B, 1997ApJS..109...79S, 2002MNRAS.333..633A, 2003ApJ...594..186B,2003ApJ...582..689M, 2006MNRAS.369..625N}. While azimuthally averaging increases the signal-to-noise ratio and removes local irregularities, the argument against this procedure is that, unless the galaxy is seen directly face-on, the mixing of the disc and bulge components is such that the radial light profile becomes ambiguous, i.e. azimuthally averaging mixes the contributions of the disc and the bulge. \citet{2012MNRAS.423..877G} point out this problem of averaging along isophotes in an edge-on galaxy, but remark also that it is less an issue for other inclinations. As our galaxies are seen at (random) range of inclinations, and we desired a uniform approach to all galaxies, we did not change the extraction of 1D profiles. We, however, made a test by extracting light profiles along the major axes and while we found some differences, they do not change our results and conclusions. 

In this appendix we want to understand the origin of differences between our 1D and a 2D approach. Our wish is not to weigh relative merits of these two approaches, but to quantify the differences one can expect between them. As our choice of 2D decomposition algorithm we use GALFIT \citep{2002AJ....124..266P}.

\subsection{One component fits}
\label{AA:1comp}

We first run GALFIT to fit a single S\'ersic function to our images. As a preparation of the images before running GALFIT, we estimated the sky levels and determined the centre for each galaxy. Furthermore, we created error images based on Poisson noise and seeing images using the same average seeing as given in Section~\ref{ss:method}. As initial values for position angle and flattening of the galaxies we used values from Paper II, which are the same as used for 1D single component fits. The final values for the ellipse parameters returned by GALFIT are very similar to Paper II values. The rms for ellipticites is 0.063 and for position angles $3.67\degr$, which are both consistent with errors estimated in Paper II. The comparison with the single component 1D fits described in Section~\ref{s:1comp} are shown on the left panel of Fig.~\ref{f:galfit1}, which shows the distribution of the S\'ersic indices. For completeness we also show results of the 1D fits to light profiles obtained by azimuthally averaging along free ellipses on the right panel of Fig.~\ref{f:galfit1}. Note that these latter results come from the fits which were used to judge whether a decomposition is necessary or a single component is sufficient to describe the light profile (see Section~\ref{ss:method}).

There is a general similarity between the 1D and 2D results when 1D light profiles are obtained by azimuthally averaging along fixed ellipses. The rms of the difference of these two estimates is $\sim0.8$, and there is a trend for some galaxies to have larger $n_{tot}$ values using our method, but the difference of the medians of the two distributions is 0.08. The non-symmetric shape of the distributions is clearly seen on the bottom panels with histograms.

Comparison of the 2D results with those in 1D using the free ellipses is shown on the right-hand panels. There are two notable properties: the spread around the one-to-one line is larger (rms of $\sim1.1$) than in the case using fixed ellipses and there is a trend such that 1D $n_{tot}$ are smaller than 2D values (median difference of -0.35) when 2D $n_{tot} < 4$. 

The cause for the better agreement of 2D results and fits to 1D profiles obtained by azimuthally averaging along fixed ellipses can be understood if galaxies are divided into fast and slow rotators. When free ellipses are used, distributions of S\'ersic indices for fast rotators in 1D and 2D cases are different (lower right panel of Fig.~\ref{f:galfit1}). Distributions for slow rotators are, however, quite similar. In this work we show that fast rotators, unlike slow rotators, can be decomposed into two components (Sections~\ref{ss:dt} and \ref{ss:lesson}) of typically different ellipticities. Fitting a single component to light profiles extracted along fixed or free ellipses will give different results as the light profiles themselves differ. As we are fitting one component, it is reasonable to ignore the changes in ellipticities and extract light profiles along the fixed ellipses. For galaxies that may show strong variations in ellipticity (or position angle) due to their triaxial structure (and not existence of multiple components), this approach might not be the most optimal. These objects are typically slow rotators and do not require a decomposition in two components. We compared the results of the fits from free and fixed ellipse models and found only three galaxies that have $n_{tot}$ different for 1 or more between these two cases (NGC4486 -- does not show any signature of rotation, NGC5576 and NGC7454 -- both galaxies have peculiar and non-regular velocity maps). Hence, as there is only a handful of such objects in our sample and for the sake of uniformity we fit them as all other objects. This closely resembles what is done in 2D (galaxy is assumed to have fixed position angle and ellipticity) and explains the similarity of the results with these two methods.

\begin{figure}
              \includegraphics[width=\columnwidth]{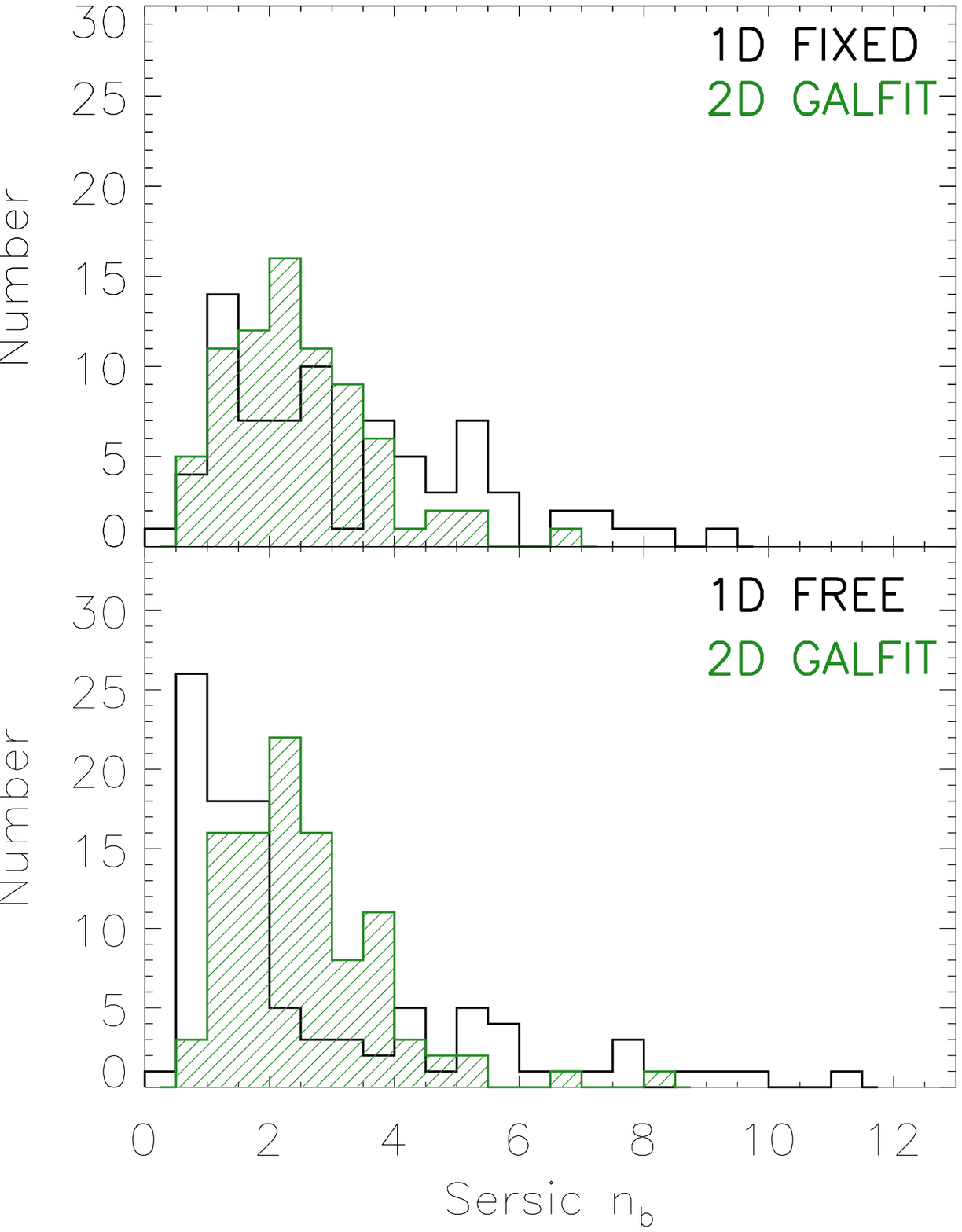}
\caption{\label{f:galfit2}  Comparison of bulge S\'ersic indices obtained by decomposing \atlas galaxies into a bulge and a disc component, using GALFIT (hatched histograms on all panels) and fitting 1D light profiles (black histograms on all panels) extracted by azimuthally averaging along fixed ellipses (top; described as 1D FIXED in the legend) and free (bottom panel; described as 1D FREE in the legend). Only galaxies that required two components in 1D fits are shown, which explains the difference between the bottom panel and histograms in Fig.~\ref{f:hist}.}
\end{figure}

\subsection{Two components fits}
\label{AA:2comp}

We also run GALFIT to decompose the images in free S\'ersic and exponential components, and we decomposed 1D profiles obtained by azimuthally averaging along fixed ellipses using the same 1D algorithm as in the main text (see Section~\ref{ss:method}). The results of this exercise are shown on Fig.~\ref{f:galfit2}, where we compare S\'ersic indices of the disc and bulge components for these three methods (2D GALFIT, 1D along free and fixed ellipses). Before running GALFIT, images were prepared as in the case of single component fitting, but this time we fix in GALFIT the position angle and flattening of the exponential components, while these parameters were left free for the bulge components. The parameters were fixed to the values in Paper II (these are the same values used to fix the parameters of the ellipses when extracting 1D light profiles). On Fig.~\ref{f:galfit2} we include only those objects which required two components in 1D (free or fixed, respectively) fits.

\begin{figure}
              \includegraphics[width=\columnwidth]{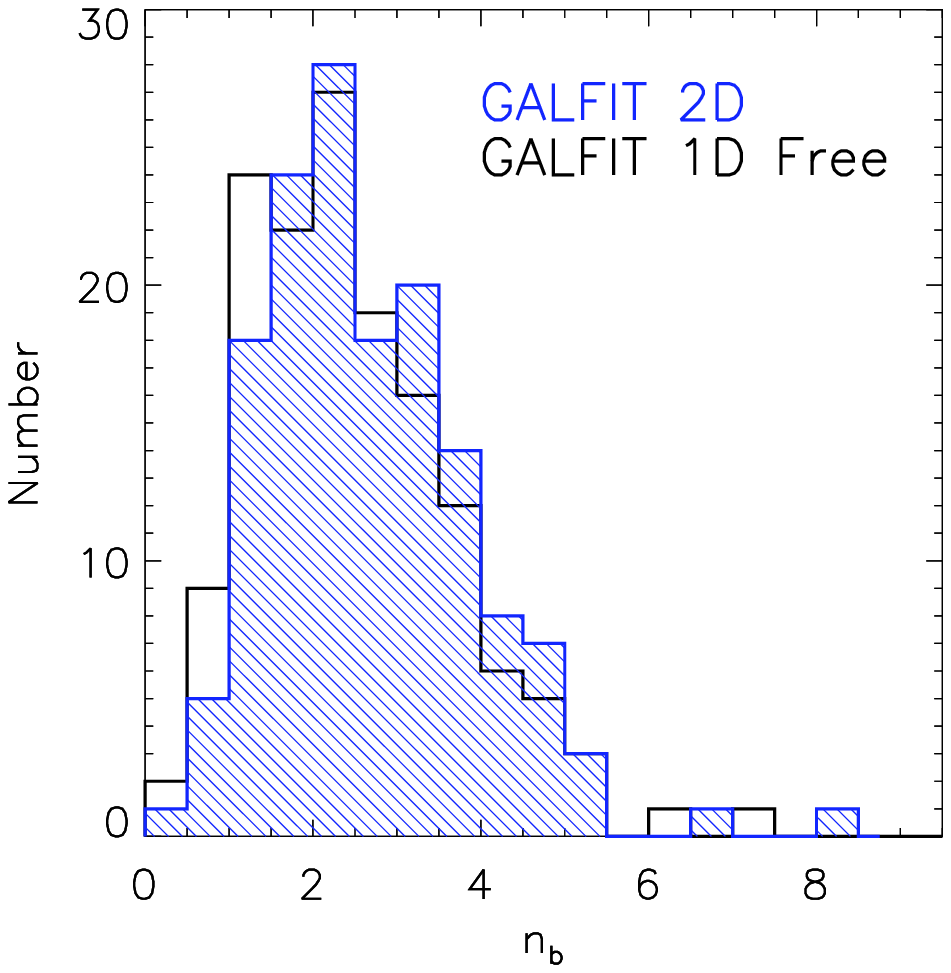}
\includegraphics[width=\columnwidth]{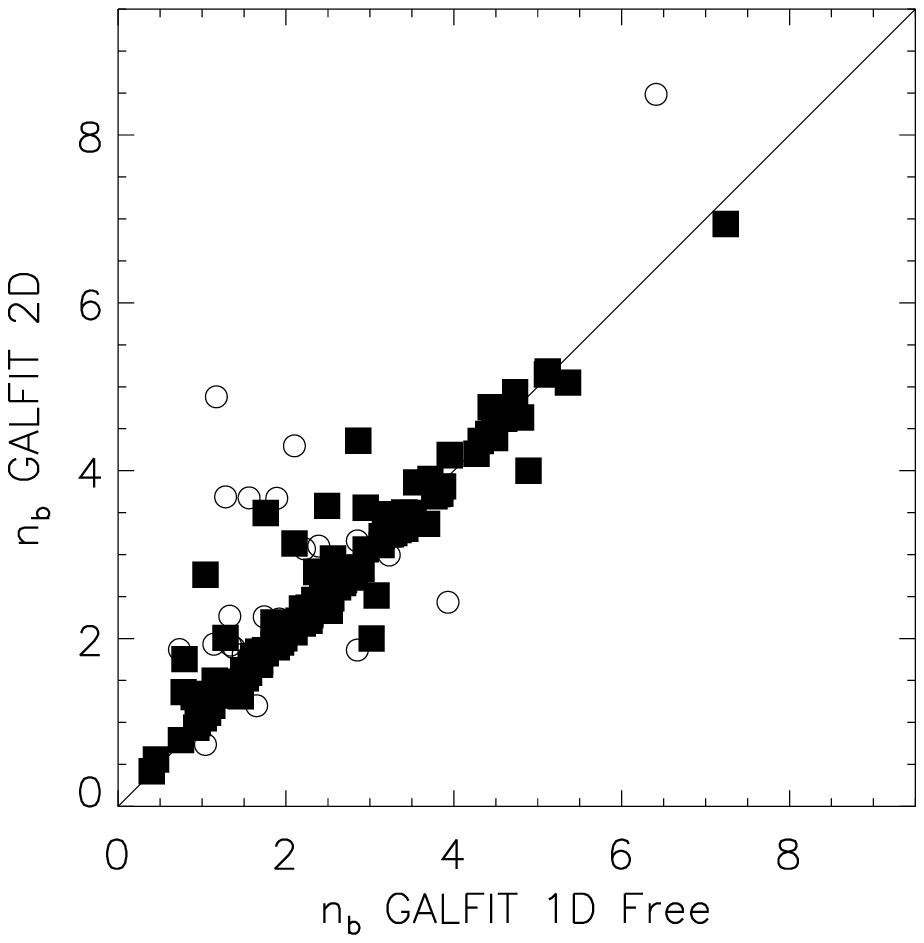}
              
\caption{\label{f:gmodel}  {\it Top:} Distribution of  S\'ersic indices obtained by using GALFIT to fit the same sample of ATLAS$^{\rm3D}$ galaxies as in the main text (hatched blue histogram) and fitting 1D light profiles extracted by azimuthally averaging along free ellipses of the same GALFIT models. {\it Bottom} Comparison of individual values of S\'ersic indices. Open circles are models for which our 1D algorithm automatically returned the best fit with only one (free S\'ersic) component, while solid squares are galaxies decomposed into a disc and a bulge.}
\end{figure}

Again, there are differences between 1D and 2D approaches and between light profiles extracted from free and fixed ellipses. The differences are more pronounced between 1D free and 2D methods. The trend is the same as seen in the case of fitting only one component to the light profile: the 1D free $n_b$ are smaller than the 2D $n_b$ for about 1--2 units, and the 1D distribution of $n_b$ is asymmetric, while the 2D distribution is more symmetric. Bulge S\'ersic indices of 1D fixed ellipse fits are more similar to 2D results, although they span a larger range of values. Note that we run fits on the 1D fixed profiles within the same fitting range as for 1D free profiles, which sometimes might not be optimal.

The difference between results obtained by GALFIT and 1D light profiles extracted along free ellipses warrants a further test of the 1D fitting method, specifically, can 1D methods recover parameters of model galaxies? For this purpose we use our GALFIT two component models to extract light profiles along azimuthally averaged ellipses of free parameters. The extraction was done in the same way as for galaxy images using kinemetry. These profiles were then fitted with our 1D algorithm. The only significant difference with the fits to the real galaxies was that we used a fixed range for all galaxies, between 2.5\arcsec and the radius at which the intensity of the models was equal to one (i.e no special fitting ranges for individual galaxies). This was possible as GALFIT models are made of only two components (e.g. no nuclear or halo components, only bulge and a disc). We also excluded all models for which GALFIT predicted bulge or disc sizes of less then 2.5\arcsec and $n_b$ smaller than 0.3, as these are 1D fit boundary conditions. 

The comparison is shown in Fig.~\ref{f:gmodel}. The top panel shows the two distributions of the S\'ersic index $n_b$, while the bottom panel shows a more direct comparison between individual values for each galaxy. The two distributions are not identical, but are generally similar. On the bottom panel, we highlight with open circles those models for which our 1D algorithm returned the best fit with only one (free S\'ersic) component (i.e. for the fitting range and the starting parameters the algorithm found the best fit solution with a single component model). These cases are typically the largest outliers and give an estimate of the systematic errors involved related to the choice of initial conditions and the fitting range used. If they are excluded from the comparison, the rms of the difference in $n_b$ is 0.18 and the Kolmogorov-Smirnov test gives the probability of $\sim80$\% that the data are drawn from the same distribution. 

The results of this test suggest that the 1D fitting method used in the main text can recover the structural parameters of the models, fully justifying our approach. The differences between the methods presented in this Appendix point out large systematic uncertainties associated with the photometric decomposition, which are much larger than any statistical errors due to noise in the data. In the case of the 1D fits, the most dominant contributors are the methods used to extract the profiles (e.g. along fixed or free ellipses) and the fitting range. This should be kept in mind when comparing S\'ersic parameters obtained with different methods and approaches.

\section{Comparison with literature data}
\label {C:litcomp}
  
\begin{figure*}
              \includegraphics[width=\textwidth]{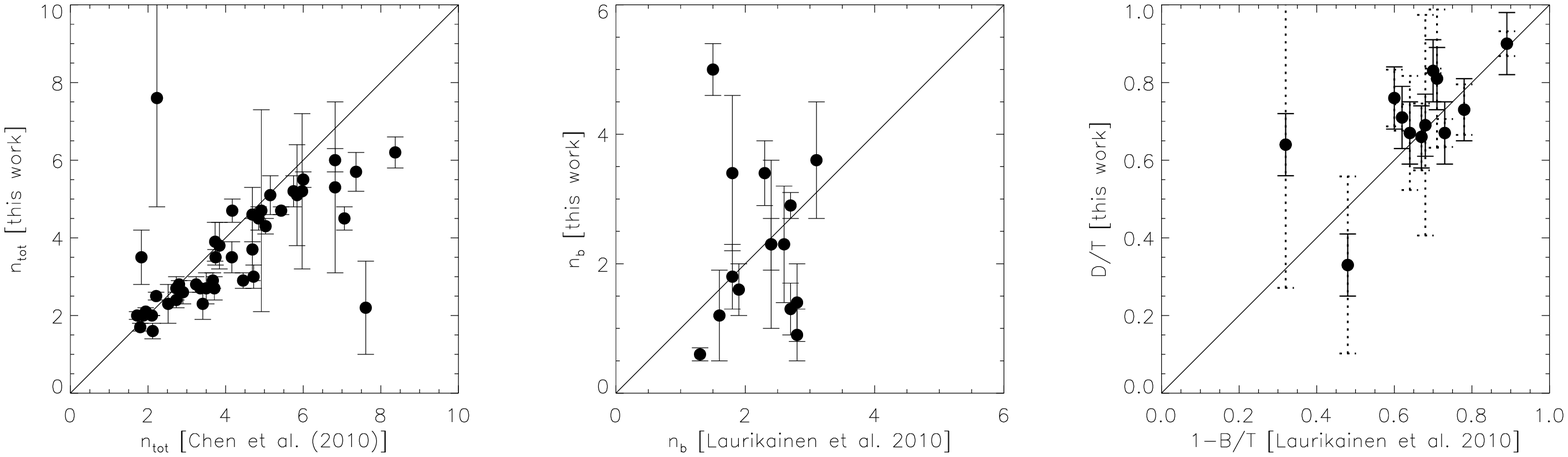}
\caption{\label{f:B1} Comparison of obtained S\'ersic indices and B/T ratios with the literature data for selected galaxies. {\it Left:} Comparison of  single component S\'ersic indices for galaxies in common with the ACSVCS \citet{2010ApJS..191....1C}. {\it Middle:} Comparison of the bulge S\'ersic indices for galaxies in common with \citet{2010MNRAS.405.1089L}. {\it Right:} Comparison of our B/T ratios with those of \citet{2010MNRAS.405.1089L}. Dashed error bars are individual Monte Carlo uncertainties, while thick error bars correspond to the median error (of 0.08) for all objects with D/T$>0$. In all panels the straight line is one-to-one relation. }
\end{figure*}

A comparison of the results of this work with published data faces to problems: there are not many studies that fit in a comparable way (i.e. decomposition into a free S\'ersic and an exponential functions), and  the number of galaxies in common is typically small. Studies which consider a similar set of nearby galaxies often use a parametrisation into a de Vaucouleurs and an exponential profiles \citep[e.g][]{1985ApJS...59..115K} or decompose galaxies in more than just two components \citep[e.g.][]{2012ApJS..198....2K}. 

We have selected two studies with which we have a relatively large overlap of objects. For the comparison of the single S\'ersic fits we use the results of the ACSVCS \citep{2004ApJS..153..223C} survey of Virgo galaxies presented in \citet{2010ApJS..191....1C}. A number of these galaxies are also present in \citet{2009ApJS..182..216K} and the authors show a general agreement between these two studies, hence we use only the larger ACSVCS sample. The comparison is shown in the left hand panel of Fig.~\ref{f:B1}. There are 44 galaxies in common and there is a generally good agreement between the values of the S\'ersic indices with an rms of 0.7. The two strongest outliers are NGC4267 and NCG4377 (above and below the one-to-one relation, respectively), for which the fits are poor, possibly due to bar/ring structures. Generally, at larger values of $n_{tot}$ the deviations increase in the sense that \citet{2010ApJS..191....1C} values are systematically larger. This can partially be explained by the fact that they use the HST imaging and exclude only the region within the break radius, which is for galaxies in common typically smaller than our 2\farcs5 inner limit.

For the comparison of our decomposition results we used the comprehensive study of S0 and spiral galaxies by \citet{2010MNRAS.405.1089L}. There are 23 galaxies in common (S0s), but in the right hand panel of Fig.~\ref{f:B1} we compare only 16. Of the seven discarded galaxies two were decomposed in more than two components, while other two objects were not decomposed by  \citet{2010MNRAS.405.1089L}. Also, three objects did not warrant the decomposition by our approach. There is a considerably larger spread between these two data sets (rms $\sim1.2$) than for the single S\'ersic fits comparison, but excluding two largest outliers on each side of the one-to-one relation, the remaining points are in a general agreement within uncertainties. 

A similar conclusion is achieved by looking at the comparison of D/T ratios. We converted our D/T ratios into B/T=1-D/T, in order to make use of B/T values from \citet{2010MNRAS.405.1089L}. We stress, however, a "bulge" may not necessarily be the same in these two studies, as \citet{2010MNRAS.405.1089L} decompose some galaxies in more than two components. Still, within our nominal (median) error of 0.08 in D/T, our results agree. The two largest outliers (NGC4694 and NGC5493, above and below the one-to-one relation, respectively) illustrate the difference in achieved results when using different methods.   \citet{2010MNRAS.405.1089L} decomposed both galaxies with more than two components, also using Ferrers functions for the possible bar component in NGC5493. We find large variations in possible B/T (or D/T) for both galaxies (one of the biggest in the sample) which indicate the complex nature of these systems. 

The comparisons of Fig.~\ref{f:B1} are encouraging, given that the fits are done with different methods and on different data. The role of systematic errors is hard to estimate in these studies, but should not be removed from consideration. As an example of possible systematic effects arising from the different methods applied on different samples, we compare our results with the results of two studies which analysed statistically large samples. The first one is a comparison with \citet{2009MNRAS.393.1531G}. That work analyses about 1000 galaxies between $0.02< z< 0.07$, selected in a similar mass range (M$_\star > 10^{10}$ M$_\odot$, but typically M$_\star <5\times10^{11}$ M$_\odot$), but with $q>0.9$. As the author notes, the latter selection is likely introducing a bias, as it is selecting galaxies that are more round, brighter and more concentrated. 

\begin{figure}
              \includegraphics[width=\columnwidth]{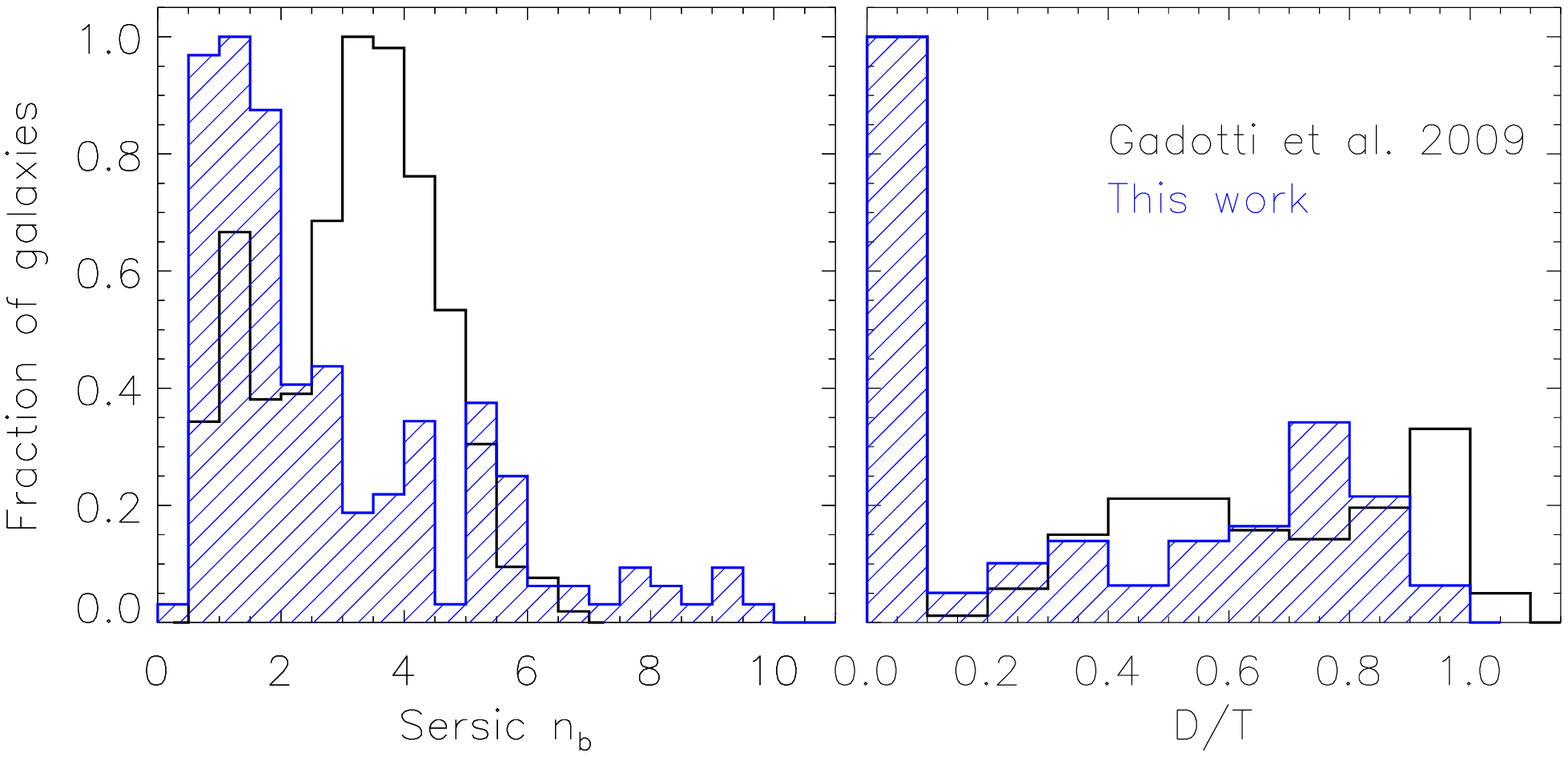}
               \includegraphics[width=\columnwidth]{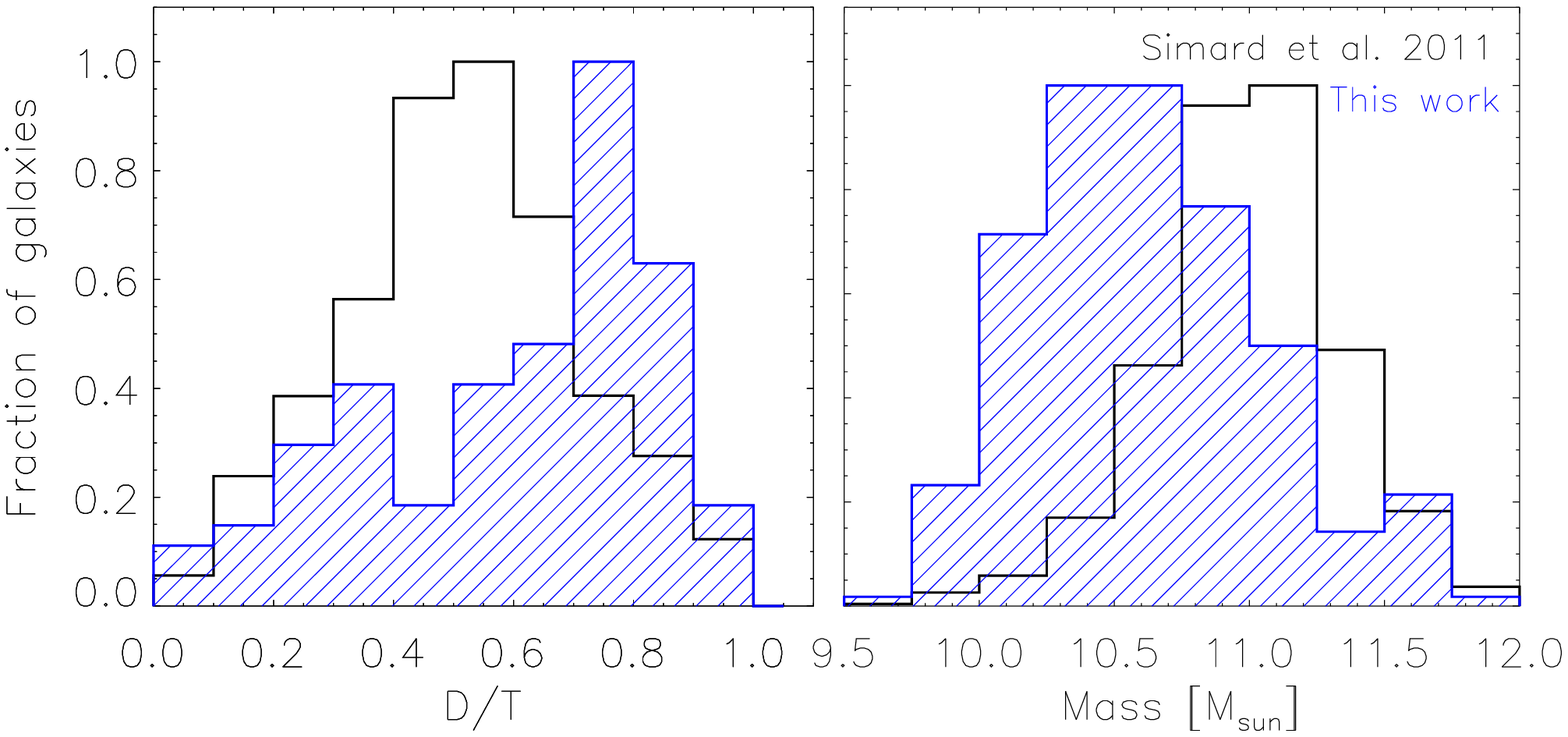}
\caption{\label{f:B2}  {\it Top:} Comparison with \citet{2009MNRAS.393.1531G} focusing on the bulge S\'ersic indices ({\it left}) and D/T ({\it right}). {\it Bottom:}  Comparison off D/T ratios ({\it left}) and stellar mass distribution ({\it right}) of a subsample of galaxies selected from with \citet{2011ApJS..196...11S}.  In all panels our data are shown with hatched (blue) histograms, but note that in the comparison with  \citet{2011ApJS..196...11S} we used only objects with D/T$>0$. All histograms are normalised to peak values.}
\end{figure}

On top panels of Fig.~\ref{f:B2}, we plot only the sub-sample of unbarred galaxies from  \citet{2009MNRAS.393.1531G}, as well as our results.   Most striking is the disparity of the $n_b$ distributions, our being smaller for about a value of 2, which is somewhat larger (but not inconsistent) than what we found in Sec.~\ref{AA:2comp}. In our sample, mostly slow rotators have larger indices, and it is possible that the mentioned bias introduced some excess of $n\sim4$ galaxies in \citet{2009MNRAS.393.1531G} sample. The distribution of D/T ratios, however, is rather similar. Both studies find a large number of galaxies with no exponential components (they are classified as ellipticals in \citet{2009MNRAS.393.1531G}, while in our case these are mostly slow rotators, but also fast rotators with small $n$), and a large spread of D/T values. 

We also compared our results with a recent study of \citet{2011ApJS..196...11S} who analyse more than a million of SDSS galaxies. From their catalogue we selected a set of objects trying to match the general properties of our sample (i.e. local early-types of a similar mass) and we looked for galaxies that can be decomposed into bulge and disc systems. Specifically, this meant we looked for objects with redshift below 0.1, ellipticity below 0.85, stellar mass in the range $9.7 < \log$(M$_\star$)$< 12$ M$_\odot$ (calculated from colours using \citet{2003ApJS..149..289B}), image smoothness parameter S2 $\le 0.075$ \citep{2009A&A...508.1141S}, and equivalent width of [OII] $<5$\AA. From these galaxies we further selected those that had P$_{pS} < 0.32$. P$_{pS}$ is the F-statistics probability that the decomposition into a bulge and a disc is not preferred to a single S\'ersic fit (low values mean that objects could be considered genuine two component systems).  

As \citet{2011ApJS..196...11S} note, the quality of imaging was typically insufficient to determine bulge S\'ersic indices, and there were no statistically significant differences between their n$_b=4$ and free n$_b$ models, Therefore, we do not compare the S\'ersic indices, but focus on the comparison of the disc fractions (obtained, however, from free $n_b$ models). On bottom panels of Fig.~\ref{f:B2} we show the comparison of D/T ratios of our sample and a sample selected as mentioned above (by using D/T=1-B/T to convert their B/T). We plot only our galaxies which could be decomposed in two components. A notable difference between these two samples is that our D/T values peak at a higher value. A possible explanation for this disparity is offered by the right hand panel comparing the actual mass distributions. As much as we tried to reproduce our sample by selecting galaxies form the much larger \citet{2011ApJS..196...11S} sample, the mass distributions are offset: the sample selected by the above criteria is dominated by galaxies just above $10^{11}$ M$_\odot$, while our sample is dominated by objects of $5\times10^{10}$ M$_\odot$.

\section{Decomposition properties of ATLAS$^{\rm 3D}$ galaxies}
\label{B:master}

\clearpage
\begin{deluxetable}{l  c  c  c   c  c  c  c  c  c  c  c  c  c  c  c  c  c  c  c  c}
\tablewidth{0pt}
\tabcolsep=4pt
\tabletypesize{\small}
\tablecaption{Fitting parameters for ATLAS$^{\rm 3D}$ galaxies.}
\tablehead{ 
\colhead{Name} &
\colhead{$\mu_{tot}$} &
\colhead{$R_{e,tot}$} &
\colhead{$n_{tot}$} &
\colhead{$\mu_b$} &
\colhead{$R_{e,b}$} &
\colhead{$n_b$} &
\colhead{$q_b$} &
\colhead{$\mu_d$} &
\colhead{$R_d$} &
\colhead{$q_d$} &
\colhead{D/T} \\
\colhead{} &
\colhead{mag/$\arcsec^2$} &
\colhead{$\arcsec$} &
\colhead{} &
\colhead{mag/$\arcsec^2$} &
\colhead{$\arcsec$} &
\colhead{} &
\colhead{} &
\colhead{mag/$\arcsec^2$} &
\colhead{$\arcsec$} &
\colhead{} &
\colhead{}\\
\colhead{(1)} &
\colhead{(2)} &
\colhead{(3)} &
\colhead{(4)} &
\colhead{(5)} &
\colhead{(6)} &
\colhead{(7)} &
\colhead{(8)} &
\colhead{(9)} &
\colhead{(10)} &
\colhead{(11)} &
\colhead{(12)} &
}
\startdata

    IC0560&     21.32 $\pm$      0.12&      16.5 $\pm$       1.0&       2.8 $\pm$       0.4&     19.39 $\pm$      0.58&       2.9 $\pm$       1.2&       0.9 $\pm$       0.4&      0.71&     19.58 $\pm$      0.06&      10.8 $\pm$       0.5&      0.47&      0.81 $\pm$     0.227&\\
    IC0598&     20.22 $\pm$      0.05&      12.9 $\pm$       0.3&       2.3 $\pm$       0.1&     20.57 $\pm$      0.07&      13.7 $\pm$       0.4&       1.8 $\pm$       0.2&      0.41&      0.00 $\pm$      0.00&       0.0 $\pm$       0.0&      0.00&      0.00 $\pm$     0.000&\\
    IC0676&     22.06 $\pm$      0.20&      24.7 $\pm$       2.8&       2.7 $\pm$       0.5&      0.00 $\pm$      0.00&       0.0 $\pm$       0.0&       0.0 $\pm$       0.0&      0.00&      0.00 $\pm$      0.00&       0.0 $\pm$       0.0&      0.00&      0.00 $\pm$     0.000&\\
    IC0719&     20.03 $\pm$      0.04&      14.5 $\pm$       0.2&       1.3 $\pm$       0.0&     20.00 $\pm$      0.06&      14.7 $\pm$       0.3&       1.0 $\pm$       0.1&      0.29&      0.00 $\pm$      0.00&       0.0 $\pm$       0.0&      0.00&      0.00 $\pm$     0.000&\\
    IC0782&     22.66 $\pm$      0.33&      22.3 $\pm$       4.4&       3.2 $\pm$       0.8&      0.00 $\pm$      0.00&       0.0 $\pm$       0.0&       0.0 $\pm$       0.0&      0.00&      0.00 $\pm$      0.00&       0.0 $\pm$       0.0&      0.00&      0.00 $\pm$     0.000&\\
    IC1024&     20.56 $\pm$      0.10&      15.2 $\pm$       0.7&       1.2 $\pm$       0.1&     20.45 $\pm$      0.12&      14.4 $\pm$       0.9&       1.4 $\pm$       0.2&      0.36&      0.00 $\pm$      0.00&       0.0 $\pm$       0.0&      0.00&      0.00 $\pm$     0.000&\\
    IC3631&     21.26 $\pm$      0.10&      12.4 $\pm$       0.6&       1.8 $\pm$       0.2&     20.89 $\pm$      0.17&      11.5 $\pm$       1.3&       1.1 $\pm$       0.4&      0.83&      0.00 $\pm$      0.00&       0.0 $\pm$       0.0&      0.00&      0.00 $\pm$     0.000&\\
\enddata
\tablecomments{\\
Column (1): The Name is the principal designation from LEDA, which is used as standard designation.\\
Column (2): Effective surface brightness of the single S\'ersic fit in mag arcsec$^{-2}$.\\
Column (3): Effective radius of the  single S\'ersic fit in arcsec. \\
Column (4):  S\'ersic index of the single S\'ersic fit.\\
Column (5): Effective surface brightness of the bulge component in mag arcsec$^{-2}$.\\
Column (6): Effective radius of the bulge in arcsec. \\
Column (7): S\'ersic index of the bulge component. \\
Column (8): Flattening of the bulge component. \\
Column (9): Effective surface brightness of the exponential component in mag arcsec$^{-2}$.\\
Column (10): Effective radius of the exponential component in arcsec.  \\
Column (11):  Flattening of the exponential component.  \\
Column (12): Disk-to-Total light ratio. \\
Note that Columns (2)-(4) are results of one component S\'ersic fits to profiles azimuthally averaged along fixed ellipses, while columns (5)-(11) are results of two components fits to profiles azimuthally averaged along free ellipses. This explains the difference between ($\mu_{tot}$, $R_{tot}$, $n_{tot}$) and ($\mu_b$, $R_b$, $n_b$) when two component fit was not required ($\mu_d=0$, $R_d=0$).\\
$^\star$ - no available r-band imaging.
$^{\dagger}$ -- single component fit did not converge (parameters at boundary conditions)
$^{\ddagger}$ -- two component fit did not converge (parameters at boundary conditions)
}
\label{t:master}
\end{deluxetable}


\label{lastpage}

\end{document}